\documentclass[a4paper, 12pt, one column, aas_macros]{article}

\usepackage[english]{babel}
\usepackage[utf8x]{inputenc}
\usepackage[T1]{fontenc}
\usepackage{aas_macros}
\usepackage{color}
\usepackage{xcolor}
\usepackage{graphicx}
\usepackage{epstopdf}
\usepackage{caption}
\usepackage{soul}
\usepackage{authblk}
\usepackage{subcaption}

\graphicspath{ {./figure/} }

\newcommand{\oc}[1]{\textcolor{black}{#1}}

\usepackage[top=1.3cm, bottom=2.0cm, outer=2.5cm, inner=2.5cm, heightrounded,
marginparwidth=1.5cm, marginparsep=0.4cm, margin=2.5cm]{geometry}

\usepackage[colorlinks=true, linkcolor=blue, urlcolor=blue, citecolor=blue]{hyperref} 
\usepackage{amsmath}  
\usepackage{amsfonts} %
\usepackage{amssymb}  %

\usepackage[authoryear]{natbib}
\setcitestyle{authoryear,open={(},close={)}}

\title{Modal Force Partitioning - A Method for Determining the Aerodynamic Loads for Decomposed Flow Modes with Application to Aeroacoustic Noise}
\author[1]{Suryansh Prakhar}
\author[1]{Jung-Hee Seo}
\author[1,\footnote{Email address for correspondence: \href{mailto:mittal@jhu.edu}{\texttt{mittal@jhu.edu}}}]{Rajat Mittal}
\affil[1]{Department of Mechanical Engineering, Johns Hopkins University, Baltimore, MD, USA}

\begin{document}
\date{}
\maketitle

\begin{abstract}
Aerodynamic loads play a central role in many fluid dynamics applications, and we present a method for identifying the structures (or modes) in a flow that make dominant contributions to the time-varying aerodynamic loads in a flow. The method results from the combination of the force partitioning method (Menon and Mittal, \emph{J. Fluid Mech.}, 907:A37, 2021)  and modal decomposition techniques such as Reynolds decomposition, triple decomposition, and proper orthogonal decomposition, and is applied here to three distinct flows - two-dimensional flows past a circular cylinder and an airfoil, and the three-dimensional flow over a revolving rectangular wing. We show that \oc{the} force partitioning method applied to modal decomposition of velocity fields results in complex, and difficult to interpret inter-modal interactions. We therefore propose and apply modal decomposition directly to the $Q$-field associated with these flows. The variable $Q$ is a non-linear observable that is typically used to identify vortices in a flow, and we find that the direct decomposition of $Q$ leads to results that are more amenable to interpretation. We also demonstrate that this modal force partitioning can be extended to provide insights into the far-field aeroacoustic loading noise of these flows.
\end{abstract}

\section{Introduction}
Fluid flows often contain a range of spatial and temporal features, and since the early work of Reynolds
\citep{Reynolds_1895}, decomposing a flow into modes has been a well-established technique for identifying and distinguishing between distinct features and for reducing the dimensionality of flows \citep{st_modal}. \cite{Reynolds_1895} proposed one of the first modal decompositions in fluid dynamics; that of a turbulent flow into a time-mean and a fluctuation about the mean, a decomposition that was later referred to as the ``Reynolds decomposition'' and formed the basis of the Reynolds Averaged Navier Stokes (RANS) approach to turbulence modeling.
Approximately 50 years ago, \cite{hussain_mechanics_1970} introduced the ``triple decomposition'' of a flow field which decomposes the unsteady component of a flow into ``coherent'' and ``incoherent'' components, and this technique has been used extensively, particularly for flows that contain a dominant low-frequency time-scale \citep{trip_decomp_ex}. 

The Proper Orthogonal Decomposition (POD) method was first introduced in fluid dynamics by Lumley in 1967 \citep{lumley1967structure}, and it allowed for the decomposition of a flow into an infinite set of orthogonal eigenfunctions or modes. The objective of POD is to identify the dominant modes in the flow and to reduce the dimensionality of the flow, and this method also become extensively employed in the analysis of fluid flows \citep{rochuon_pod,zhao_pod}. The Fourier transform may be considered to be yet another fundamental modal decomposition that has been applied extensively to fluid flows and has even formed the basis of a class of discretization methods \citep{canuto_spectral_2006}.

The rise of data-enabled techniques and machine-learning in the last two decades has led to an explosion of interest and activity in modal decomposition techniques in fluid dynamics \citep{st_modal} and several new modal decomposition techniques such as dynamic mode decomposition (DMD), spectral POD (SPOD) and resolvents \citep{schmid_dmd,schmidt_spod,brunton_resolvent} have appeared on the scene.  Almost all modal decomposition techniques are applied to the velocity field and the modal decomposition can be expressed as
\begin{equation}
{\bf u}({\bf x},t)={\bf u_0}({\bf x})+{\bf u_1}({\bf x},t)+... +{\bf u_i}({\bf x},t)+... +{\bf u_N}({\bf x},t).
\label{umodal}
\end{equation}
where the terms on the right represent the modes. In most decomposition techniques\oc{,} the first term on the right-hand-side is the time-mean flow and $N$ can range from $N=1$ (as for Reynolds decomposition) to an arbitrarily large number (as for Fourier or POD modal decompositions). 

\begin{figure}[ht]
    \centering
 \begin{subfigure}[b]{0.3\textwidth}
      \centering
         \includegraphics[width=\textwidth]{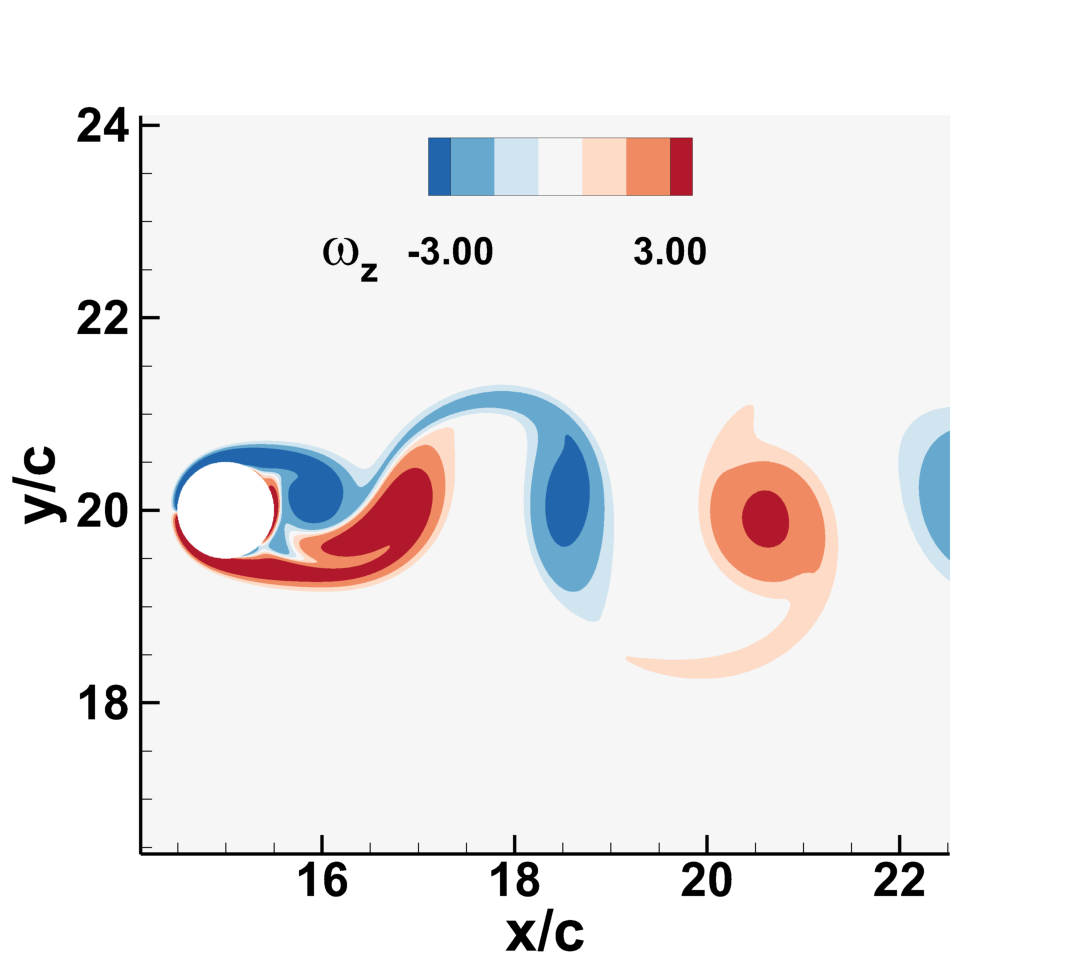}
         \caption{}
         \label{fig:cc:rd:TotFlow:a}
     \end{subfigure}
  \hfill
     \begin{subfigure}[b]{0.3\textwidth}
         \centering
         \includegraphics[width=\textwidth]{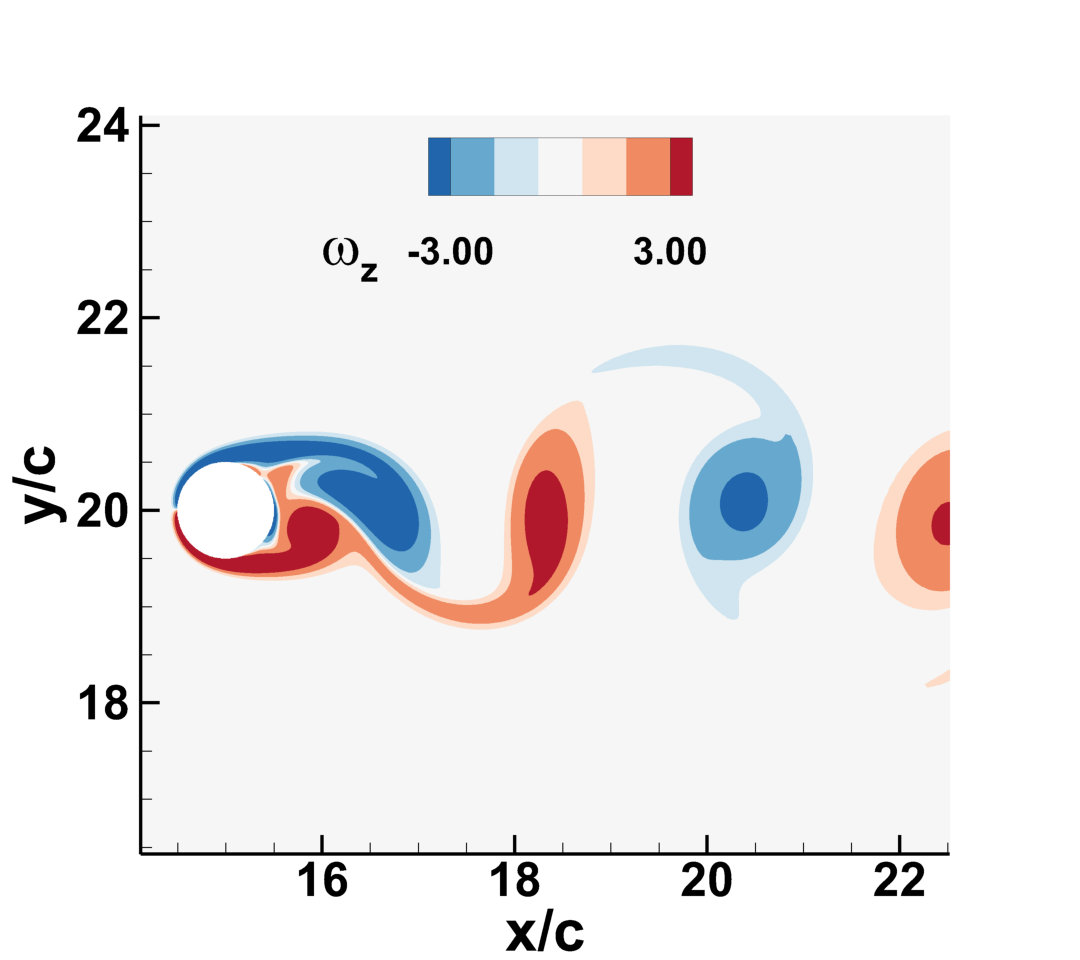}
         \caption{}
         \label{fig:cc:rd:TotFlow:b}
     \end{subfigure}
   \hfill
     \begin{subfigure}[b]{0.28\textwidth}
         \centering
         \includegraphics[width=1\textwidth]{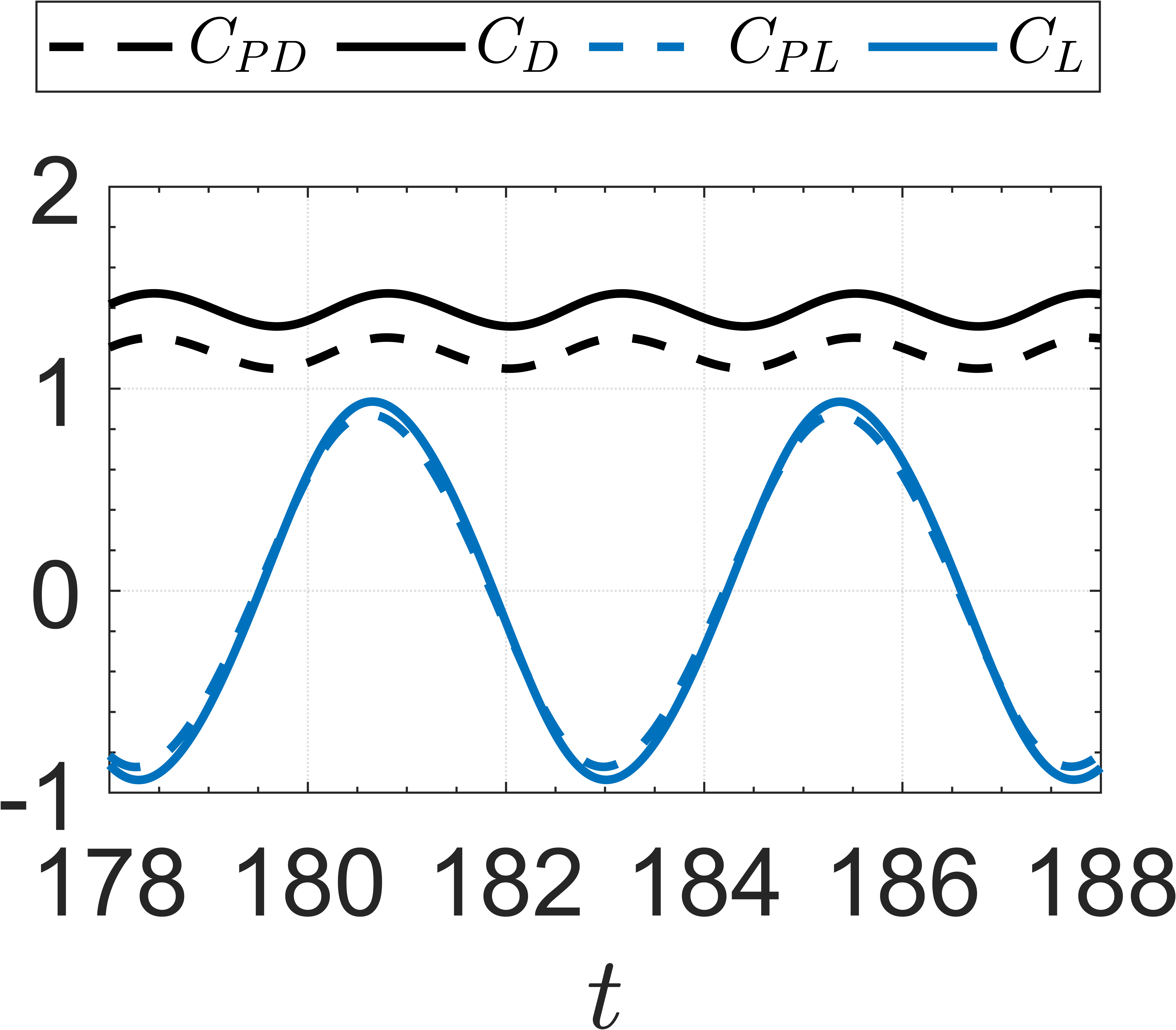}
         \caption{}
         \label{fig:cc:rd:TotFlow:c}
     \end{subfigure}
        \caption{(a) and (b) Spanwise vorticity corresponding to flow past circular cylinder at Reynolds number of 300 showing the shedding of the vortices in the wake at two instances. (c) Time variation of coefficients of drag and lift (pressure induced and total) for the circular cylinder.}
\label{fig:cc:rd:TotFlow}
\end{figure}

\begin{figure}[ht]
    \centering
     \begin{subfigure}[b]{0.24\textwidth}
         \centering
         \includegraphics[width=\textwidth]{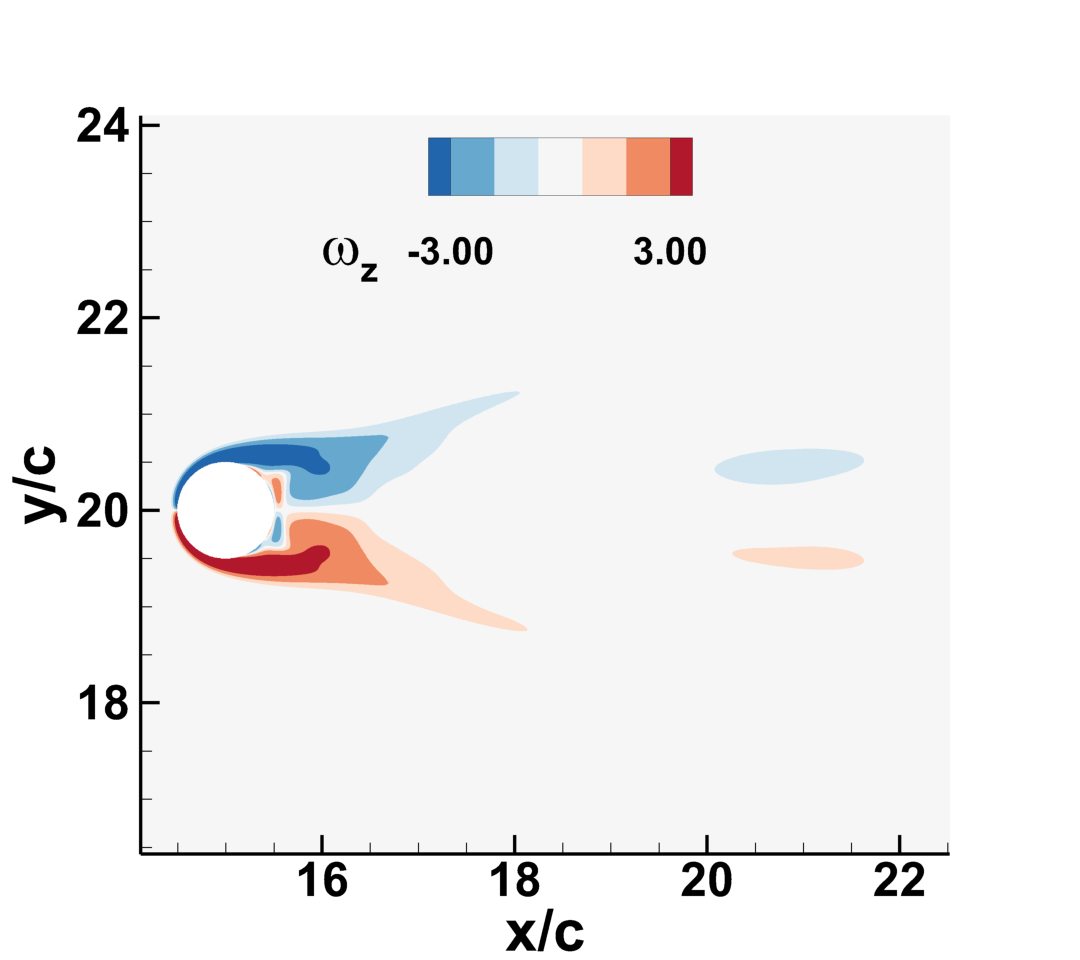}
         \caption{$\omega_z(\bf{u_0})$}
         \label{fig:cc:pod:omega:d}
     \end{subfigure}
     \hfill
 \begin{subfigure}[b]{0.24\textwidth}
      \centering
         \includegraphics[width=\textwidth]{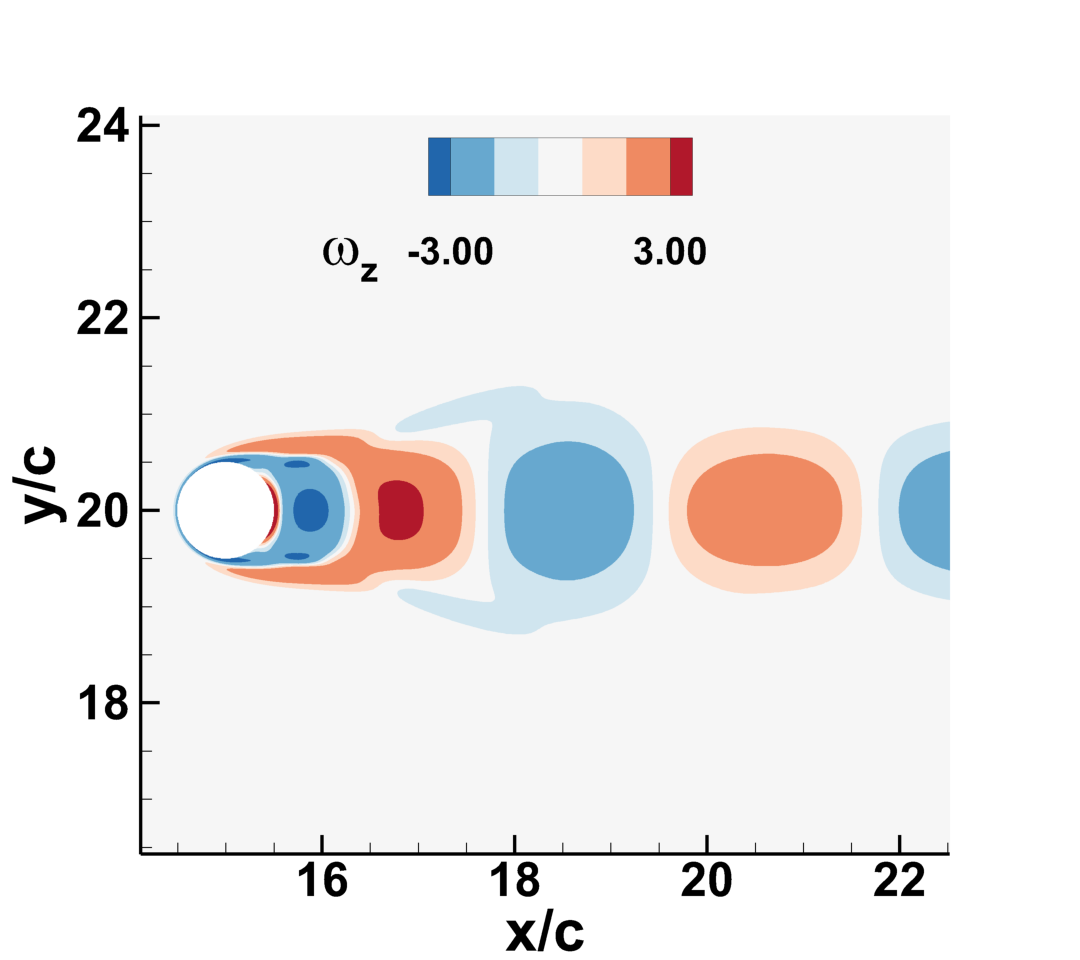}
         \caption{$\omega_z(\bf{u_1})$}
         \label{fig:cc:pod:omega:a}
     \end{subfigure}
  \hfill
     \begin{subfigure}[b]{0.24\textwidth}
         \centering
         \includegraphics[width=\textwidth]{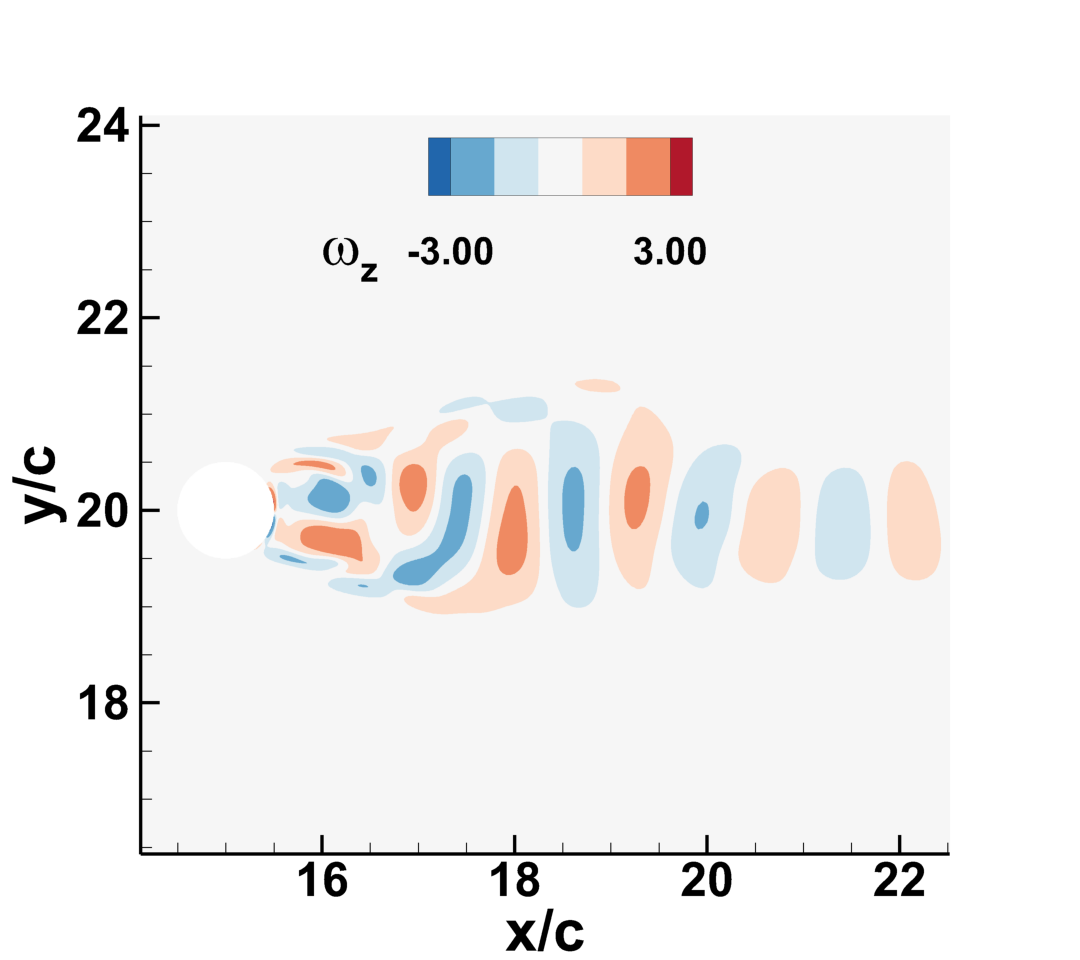}
         \caption{$\omega_z(\bf{u_2})$}
         \label{fig:cc:pod:omega:b}
     \end{subfigure}
   \hfill
     \begin{subfigure}[b]{0.24\textwidth}
         \centering
         \includegraphics[width=\textwidth]{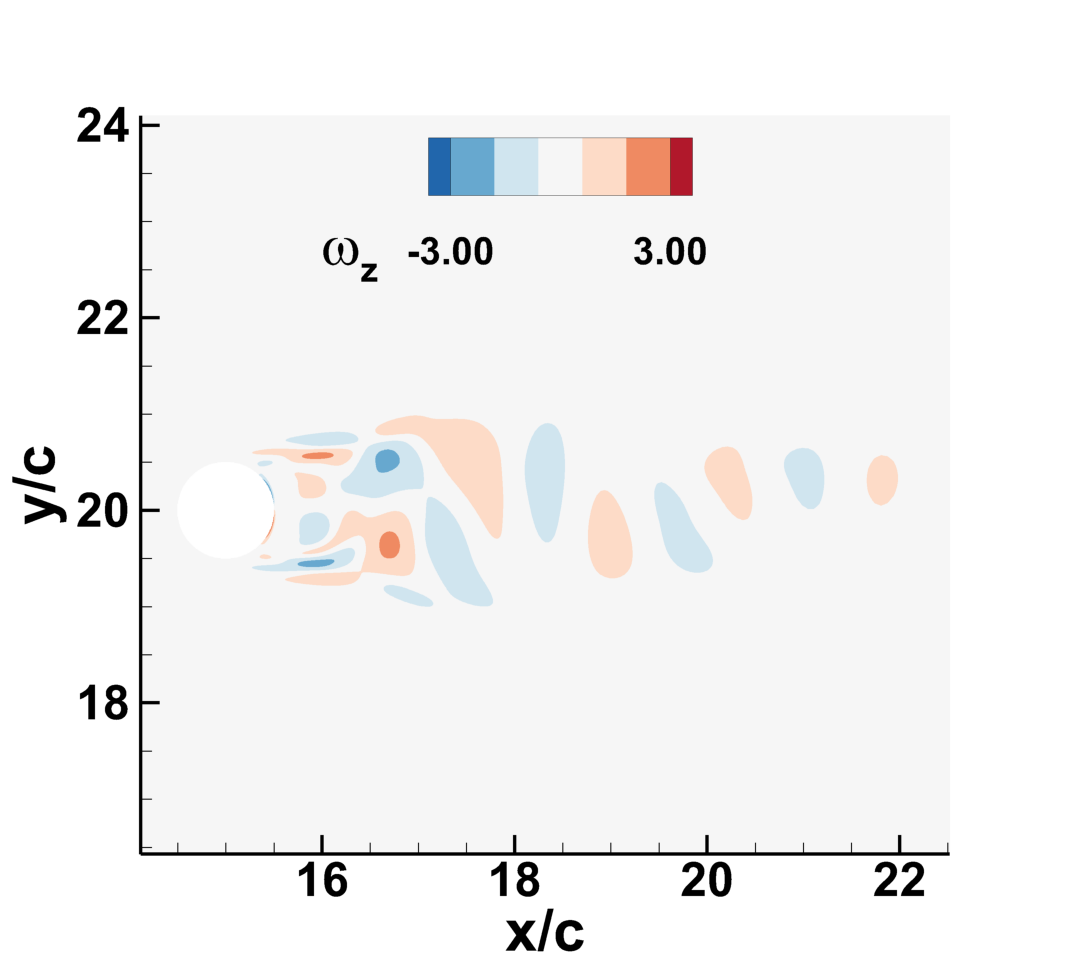}
         \caption{$\omega_z(\bf{u_3})$}
         \label{fig:cc:pod:omega:c}
     \end{subfigure}
     \hfill
     \begin{subfigure}[b]{0.6\textwidth}
         \centering
         \includegraphics[width=\textwidth]{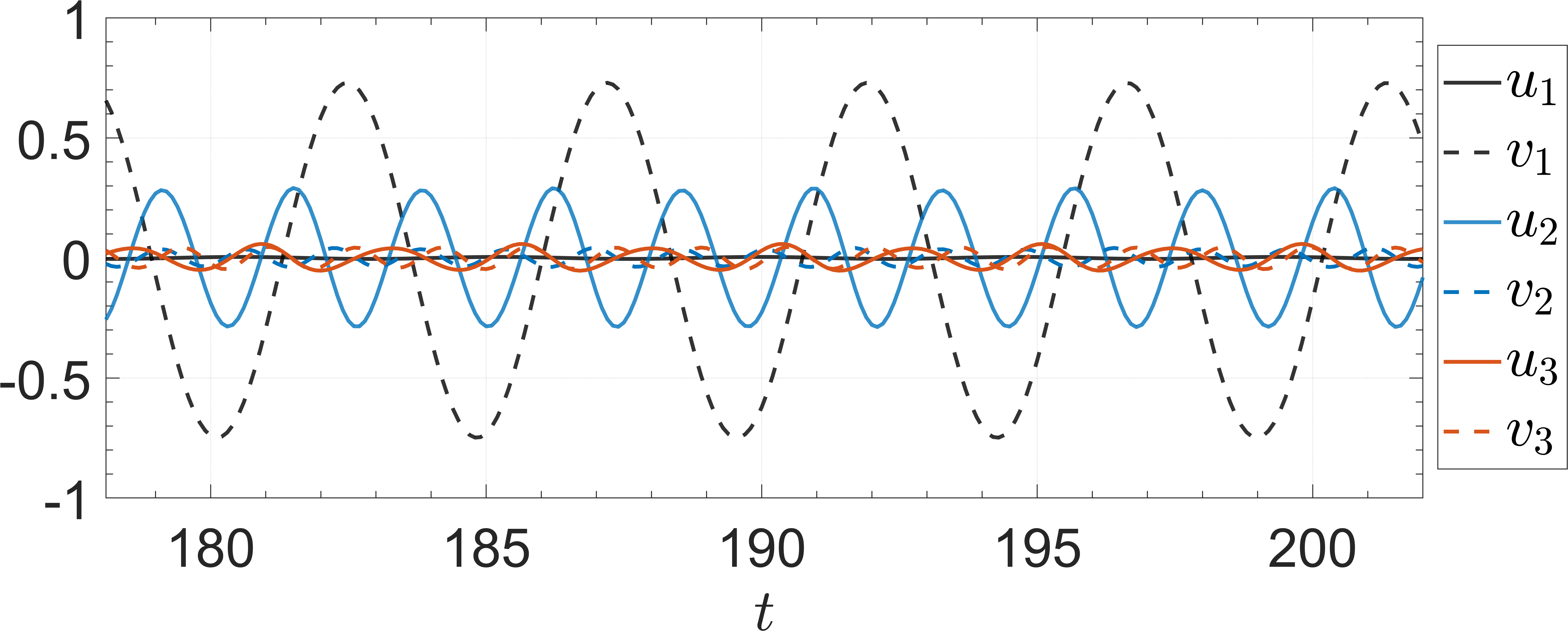}
         \caption{}
         \label{fig:cc:pod:omega:e}
     \end{subfigure}
        \caption{Proper orthogonal decomposition (POD) applied to the cylinder flow. Spanwise vorticity for (a) the mean mode (Mode-0), (b) Mode-1, (c) Mode-2 and (d) Mode-3. (e) Time variation of the streamwise and lateral velocity at a downstream distance of $d$ from the center of the cylinder on the wake centerline.}
\label{fig:cc:pod:omega}
\end{figure}

As an example, we present the POD of the flow past a circular cylinder, which is a quintessential vortex-dominated flow. Figure \ref{fig:cc:rd:TotFlow:a} shows the vortices for flow past a circular cylinder at Re=300 and figure \ref{fig:cc:rd:TotFlow:c} shows the time variation of the total lift and drag force coefficients, as well as the pressure components of these two coefficients. Figure \ref{fig:cc:pod:omega} shows the mean and the first 3 POD modes obtained from a modal decomposition of the velocity field. These 4 modes together constitute 94.2\% of the total variance in this flow and each mode has a distinct topology. For instance, Mode-1 is symmetric about the wake centerline, whereas Mode-2 and Mode-3 are not. Figure \ref{fig:cc:pod:omega:e} shows the time-variation of the velocity at one selected point in the wake for each of these three modes and it is clear that the time-variation of these modes is also quite distinct. Thus, the POD modes provide important information about the dominant space-time characteristics of the flow. 

The modes that are generated from POD (as well as from other decomposition methods) are however, purely kinematic in their description, i.e., they are designed for an examination of the topological characteristics of the flow field, but they do not directly provide any information on the \emph{dynamics} of the flow. In particular, we do not have the ability to determine the contributions that these modes make to the fluid dynamic forces on the submerged body. Such a capability would be extremely useful in several applications, including the following:
\begin{itemize}
\item determining the contribution of these modes to fluid dynamic loads (lift, drag, moments, etc.). Aerodynamic loads are key in a majority of scientific investigations and engineering applications that involve fluid flow and being able to quantify the aerodynamic loads associated with these modes could provide important insights into the flow physics associated with the generation of these forces. For instance, the symmetry properties of Mode-1 indicate that it contributes to drag but not to the lift force on the cylinder, Mode-2 and Mode-3 will contribute to the lift and the drag on the cylinder. However, we have no means of quantifying these contributions.
\item Understanding the contribution of decomposed modes to flow-induced vibration and flutter (FIVF). FIVF is determined by the phasing between the fluid dynamic loads and the movement of the structure, and since each mode typically has a distinct temporal profile, the ability to estimate the time-variation of the loads associated with each mode would provide a unique ability to analyze FIVF;
\item Analysis of the sources of flow-noise. Unsteady pressure loading of immersed surfaces serves as a source of flow noise \citep{Howe_book} in applications ranging from aeronautics \citep{farassat_noise} and marine engineering \citep{ship_acoustic} to bio-medicine \citep{bailoor_heart} and zoology \citep{zoo_acoustic}. Thus, correlating the sources flow noise to distinct flow modes could be very insightful; and
\item Developing physics-driven strategies for flow control. Effective flow control for applications such as drag reduction, lift/thrust enhancement, reduction/enhancement of FIVF, reduction in flow noise, could be empowered by understanding the contribution of modes to the relevant fluid dynamic loads induced by these modes. Control strategies could then be designed to systematically target those modes that make dominant contributions to these loads.
\end{itemize}

There have been previous some attempts to determine the fluid dynamic loads associated with the modal decomposition of flows. For instance, \citep{mittal1995effect} decomposed the velocity field past a circular cylinder obtained from a 3D spanwise homogeneous simulation into a spanwise averaged component and the remnant three-dimensional mode as follows $\bold{u}(x,y,z,t) = \bold{\bar{u}}^{2D}(x,y,t) + \bold{u}^{3D}(x,y,z,t)$ and then solved the pressure-Poisson equation for each mode to partition the contributions of these two modes on the pressure drag over the cylinder. This allowed them to determine the contributions that 3D flow features make to the total drag and lift forces on the cylinder and pinpoint the cause for the over-prediction of drag in 2D simulations of these flows.  

\cite{aghaei2022contributions} deployed the force-partitioning method (FPM) \citep{zhang2015centripetal,menon2021a,menon2021c} to decompose and analyze the pressure-induced drag for turbulent flow over rough walls. More details of the force partitioning method, which is based on the earlier work of \cite{quartapelle1983force}\oc{, \cite{wu_fpm}, \cite{chang1992potential}, \cite{howe_fpm}, and \cite{zhang2015centripetal},} will be provided in section \ref{sec:fpm} since it is central to the current paper, but it suffices for now to point out that FPM enables the partitioning of the pressure drag over the roughness elements into a component due to flow vorticity (the so-called ``vortex-induced'' component) and a component due to the viscous diffusion of momentum. The analysis was performed on data from direct numerical simulations of turbulent channel flows, at frictional Reynolds number of Re$_\tau$ = 500, with cubic and sand-grain roughened bottom walls. The results from these simulations showed that the vortex-induced pressure drag is the largest contributor (more than 50\%) to the total drag (which is the sum of pressure and shear drag) on the rough walls. A Reynolds decomposition of the flow into mean and fluctuation components was also performed and the contributions of these two modes on the mean drag were estimated using FPM and found to be nearly equal.

\cite{zhu2023force} applied the force and moment partitioning method (FMPM) of \cite{menon2021a} to experimental data for a NACA 0012 wing undergoing sinusoidal pitching in quiescent water. The velocity field was obtained from 2D PIV measurements at the central spanwise plane of the foil. The data was phase-averaged and the FMPM was applied to this phase-averaged component. The FMPM analysis enabled the separation of the pitching-moment contributions from the leading-edge and trailing-edge vortices, and their ratio was found to match empirical correlations.

\cite{Chiu2023} explored the influence of coherent structures on aerodynamic forces by using spectral proper orthogonal decomposition (SPOD) and ``force representation theory.'' This force representation theory is based on \cite{chang1992potential}, which is also based on the ideas of \cite{quartapelle1983force}. \cite{Chiu2023} found for instance that the large vortex structure in the zeroth frequency mode of the first SPOD mode, significantly impacted the lift and drag by inducing strong suction-side flow. This work will be discussed further in the paper since it has connections to the current work.

Finally, \cite{Seo2023} introduced a data-driven method for predicting vortex-induced sound from time-resolved velocimetry data and applied to flow through the slat of a multi-element high-lift airfoil. Time-resolved particle image velocimetry provided velocity fields in the slat-cove region of the airfoil, and these were reconstructed using rank-one modes from spectral proper orthogonal decomposition (SPOD). The pressure force and resulting dipole sound were calculated using force (\cite{zhang2015centripetal,menon2021a,menon2021c}) and acoustic partitioning (\cite{Seo2022}) methods (FAPM), involving volume integrals of the velocity gradient tensor's second invariant and geometry-dependent influence fields. Comparisons with measured sound data indicated that while shear layer modes contribute to tonal noise, their interaction with other wing components also generates significant flow noise.

Our current work has been motivated by the problem of aeroacoustic noise from small drones that operate at low tip Mach numbers. As drones are transforming several industries such as transportation, healthcare, vaccine delivery \citep{ziplinecovid}, rescue operations \citep{dronerescue}, food delivery, etc., the noise they produce during flight is one of the major factors limiting their wide-scale use in several of these applications \citep{DroneNoiseRev}. This makes reducing noise from drone rotors an important problem. However, the aeroacoustic noise from drones consists primarily of \oc{thickness noise and loading noise}. Thickness noise is relatively easy to estimate, but the latter is connected with the pressure loading of the blade, and these depend on the intrinsic unsteadiness in blade loading vector due to rotation, flow separation and blade vortex interaction \citep{lowmachaerobook}. Studies on drone rotor blades indicate complex vortical topologies and dynamics \citep{apsrotor,droneflowstruct} and all these flow structures present in the flow potentially affect the surface pressure on the blade. This makes the problem of understanding the causality of noise sources quite difficult. The ability to identify key flow features (or modes) that contribute most to the blade loading noise could enable us to pinpoint aspects in the shape and operation of these blades that could reduce the noise.

In the present study, we demonstrate the application of the modal force-partitioning method to estimate the pressure loading generated by the modes in the flow that result from various modal decompositions. We apply this methodology to  three widely used modal decomposition techniques: the Reynolds decomposition where any flow quantity can be split into a mean part and a fluctuating component, the triple decomposition (\cite{tripDecompSource}) which further splits the fluctuating part into coherent and non-coherent components, and finally the proper orthogonal decomposition (POD) (\cite{lumleyPOD,chatterPOD}) which segments the flow into modes that are orthogonal and arranged by their decreasing energy content.  This modal force-partitioning method (mFPM) is applied first to a canonical case of flow past a 2D circular cylinder and the case of 2D flow past NACA 0015 airfoil to understand fundamental mechanism for unsteady force and sound generation. We use these cases to propose a flow observable that provides a clearer identification of modes that make dominant contributions to the unsteady pressure loading on the immersed bodies. The method is then applied to the flow over a rotating blade to investigate the dominant flow structures associated with generation of loading noise.

\section{Methodology}
\subsection{Flow Solver}
The flow simulations are done by using a sharp-interface immersed boundary method based, incompressible Navier-Stokes solver called ViCar3D\citep{mittal2008,seo2011}. In this solver, the body is represented using unstructured triangular elements which is immersed in a non-uniform Cartesian grid with a discrete-forcing scheme coupled with ghost cells inside the body that can apply boundary conditions precisely on the body surface. The fractional step method of Van-Kan \citep{vankan} is used to split the momentum equation into an advection-diffusion equation and a pressure Poisson equation. The advection-diffusion equation is discretized using second-order Adam-Bashforth scheme for the convective term and implicit Crank-Nicholson for viscous terms, while the pressure Poisson equation is solved using gradient descent method. This solver has been validated for complex 2D and 3D cases which can be found in the references \oc{\citep{mittal2008,mittal_origin_2023,new_ibm}}.

\subsection{Force Partitioning Method (FPM)}
\label{sec:fpm}
In aerodynamics, the force due to vortices plays an important role but it is difficult to compute them because the pressure field is an elliptic variable which is calculated by solving a Laplace equation and hence total pressure force is continuously being influenced by all the flow features. The force partitioning method (FPM, \citep{menon2021a,menon2021b,menon2021c}) helps to overcome this problem and allows us to decompose the pressure forces into four components.  These components are the forces generated due to vortices (vortex-induced force), movement/acceleration of the body (we have called this the ``kinematic force''), acceleration of body or fluid (added mass force) and force due to viscous effects (viscous force). This is done by taking the velocity field calculated before and projecting it onto the following Laplace equation:
\begin{equation}
    \nabla^2\phi^{(i)}=0,
    \label{eqn:phi}
\end{equation}
with boundary conditions,
\begin{equation}
    \nabla\phi^{(i)} \cdot {\bf n}=  \begin{cases}
    n_i & \text{on $B$},\\
    0 & \text{on $\Sigma$}.
  \end{cases}
    \label{eqn:phi:bc}
\end{equation}
Here, $\phi$ is the ``influence field'' and $i=$1, 2 and 3 corresponds to x, y and z direction. $n_i$ represent normal direction vector, $B$ is boundary of the immersed surface and $\Sigma$ is the domain boundary.

\begin{figure}[ht]
\centering
\includegraphics[width=0.7\textwidth]{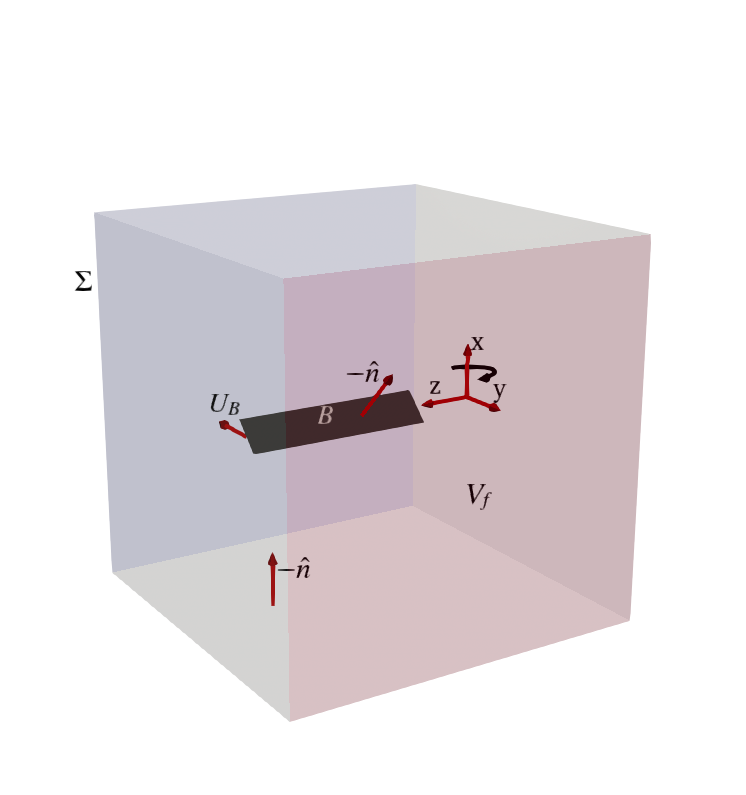}
\caption{FPM schematic (not to scale) for the revolving wing with the origin shown at the center of revolution.}
\label{FPMschematic}
\end{figure}
Then the Navier-Stokes equations,

\begin{equation}
\frac{\partial {\bf u}}{\partial t} + {\bf u}\cdot \nabla {\bf u}  = -\frac{1}{\rho}\nabla P +\nu \nabla^2 {\bf u} ,
\label{momeqn}
\end{equation}
 is projected onto the gradient of field of influence potential, $\phi$. Rearranging and integrating over the fluid domain gives,
\begin{equation}
    -\int_{V_f}\bold{\nabla} P \cdot \bold{\nabla} \phi^{(i)} dV= \int_{V_f} \rho \left[\frac{\partial {\bf u}}{\partial t} + {\bf u} \cdot \nabla {\bf u} - \nu \nabla^2 {\bf u} \right]\cdot \nabla\phi^{(i)} dV .
\end{equation}
The above equation can be simplified as \oc{(refer Appendix \ref{appendix:fpm})} 
\begin{equation}
\begin{aligned}
    \underbrace{\int_B P n_i dS}_{F^{(i)}} =
    \underbrace{-\int_B\left(  \rho \frac{D{\bf u}}{Dt} \cdot {\bf n} \right) \phi^{(i)} dS}_{F^{(i)}_B} - 
    \underbrace{\int_\Sigma\left( \rho \frac{D{\bf u}}{Dt} \cdot {\bf n} \right)  \phi^{(i)}  dS}_{F^{(i)}_O} +\\
    -    \underbrace{\int_{V_f} 2 \rho Q \phi^{(i)} dV}_{F^{(i)}_Q} +\underbrace{\int_{B+\Sigma} \left( \mu \nabla^2 {\bf u \cdot n} \right) \phi^{(i)} dS}_{F^{(i)}_\mu} 
    \label{fpmEqn}
\end{aligned}
\end{equation}
where the term on the LHS corresponds to the pressure loading on the body and terms on RHS corresponds to force due to $F_{B}$ - the acceleration reaction due to the motion of the body; $F_{O}$ the added mass force due to material acceleration of the flow on the outer boundary; ${F_Q}$ the vortex-induced force (VIF); and ${F_\mu}$ - the force due to viscous diffusion of momentum and  respectively. 
Here $Q$ is from the so-called $Q$-criterion \citep{Qcrit}, defined as,
\begin{equation}
Q=\frac{1}{2} (|| {\bf \Omega}||^2- || {\bf S}||^2) \equiv -\frac{1}{2}{\bf \nabla} \cdot ({\bf u}\cdot {\bf \nabla u})\,\,\, ,
\label{Qcriteqn}
\end{equation}
where ${\bf S}$ and ${\bf \Omega}$ are symmetric and anti-symmetric components of velocity gradient tensor respectively.
\oc{The derivation and application of FPM to several flows can be found in the references \citep{menon2021a,menon2021b,menon2021c}. \oc{It should be noted that the above partitioning is \emph{exact} and in principle, all of the terms in the partitioning can be estimated from the simulation data (see \cite{zhang2015centripetal} and \cite{menon2021c}).} For the moderate to high Reynolds number flows that are the subject of the current paper, the pressure force due to viscous momentum diffusion effects are generally small. For instance, for the airfoil case in section \ref{sec:airfoil}, the viscous diffusion term contributes less than 4\% to the total pressure lift.} Furthermore, $F_{B}$ is identically zero for stationary bodies and for the rotor rotating at constant angular velocity, \oc{the magnitude of this term is quite} negligible. Finally, since the outer domain is placed far from the body and the incoming flow is steady in all the cases studied here, $F_{O}$ is also negligible. \oc{Thus, assuming the application of this method for such flows, we proceed with the following approximation for the current study:}
\begin{equation}
\begin{aligned}
   F^{(i)} \approx F^{(i)}_Q = - \int_{V_f} 2 \rho Q \phi^{(i)} dV = \int_{V_f} {f^{(i)}_Q} dV
    \label{VIF}
\end{aligned}
\end{equation}
where ${f^{(i)}_Q}= - 2 \rho Q \phi^{(i)} $ is the vortex-force density for the pressure force in the $i^\textrm{th}$ direction. Note that the force due to individual vortices can be calculated by taking appropriate regions for the above integral (see \cite{menon2021c}) such as over individual vortices. 

\subsection{Modal Force Partitioning Method (mFPM)}
\label{sec:mFPM}
For Reynolds decomposition and $N=1$, ${\bf u_0}$ is the mean mode calculated as,
\begin{equation}
    {\bf u_0}({\bf x})=\frac{1}{T} \int_{t_0}^{t_0+T} {\bf u}({\bf x},t) dt \,\,\, \oc{.}
    \label{eqn:u0mean}
\end{equation}
${\bf u_1}({\bf x},t)$ is the fluctuation about the mean calculated using,
\begin{equation}
    {\bf u_1}({\bf x},t)= {\bf u}({\bf x},t)- {\bf u_0}({\bf x}) \,\,\, .
\end{equation}
For triple decomposition, $N=2$ with the mean velocity calculated using \oc{equation} \ref{eqn:u0mean}. ${\bf u_1}({\bf x},t)$ and ${\bf u_2}({\bf x},t)$ are the ``coherent'' and ``incoherent'' models of the flow. For $M$ period of data with a intrinsic time-period of $\tau$, these are calculated using \citep{triple_ref},
\oc{
\begin{equation}
    {\bf u_1}({\bf x},t)= \frac{1}{M} \sum_{n=0}^{M}  {\bf u}({\bf x},t+n\tau)- {\bf u_0}({\bf x}) \,\,\, ,
\end{equation}}
and 
\begin{equation}
    {\bf u_2}({\bf x},t)= {\bf u}({\bf x},t)- {\bf u_0}({\bf x}) -  {\bf u_1}({\bf x},t) \,\,\, .
\end{equation}
 
 For POD, $i=1,N$ correspond to the $N$ POD modes of the flow. We first subtract the mean flow, and as before, this is designated as $\bold{u}_0(\bold{x})$. In the snapshot POD \oc{\citep{sirovich}} approach employed here, the fluctuating velocity \oc{(}${\bf u}({\bf x},t)- {\bf u_0}({\bf x})$\oc{)} at $m$ selected grid points at a given time-step is arranged in a column vector and these vectors for $N$ sequential time-steps stacked to form $W$, a $(m \times N)$ matrix \citep{PodTutorial,parPODformula}. Here $m$ may be the entire grid or a subset of the grid points. The POD is obtained via the singular value decomposition (SVD) algorithm where $W$ is decomposed as
 \begin{equation}
     W=U \Sigma V^T\,\,\, ,
 \end{equation}
where $U$ is spatial eigenvector matrix with dimension ($m,N$), $\Sigma$ is a diagonal eigenvalue matrix of the POD and $V$ is temporal eigenvector matrix with dimension ($N,N$). Following standard techniques for efficiently computing this SVD via an eigenvalue problem for the temporal modes, the POD of the velocity field can be expressed as 
\begin{equation}
    {\bf u}({\bf x},t)= {\bf u_0}({\bf x}) + \sum_{m=1}^N {\bf u}_m ({\bf x},t)
\end{equation}
where 
\begin{equation}
    {\bf u}_m({\bf x},t) =\Sigma^{-1}_m \sum_{n=1}^{N} V(n,m) W(:,n)
\end{equation}

Once obtained, any given modal decomposition can be substituted into the expression for $Q$ (Eq. \ref{Qcriteqn}) to obtain
\begin{equation}
Q ( {\bf x},t)= -\frac{1}{2}  \sum_{m=0}^N \sum_{n=0}^N {\bf \nabla} \cdot \left(  {\bf u}_m \cdot  \nabla {\bf u}_n  \right) = \sum_{m=0}^N \sum_{n=0}^N \hat{Q}_{mn}\,\,\, ,
\label{Qcriteqn_two}
\end{equation}
where,
\begin{equation}
\oc{\hat{Q}_{mn}= -\frac{1}{2} {\bf \nabla} \cdot \left( {\bf u}_m \cdot {  \nabla {\bf u}_n } \right)\,\,.} 
\end{equation}
The vortex-induced pressure force can then be computed from the modal decomposition of the velocity field by the following expression:
\begin{equation}
    F^{(i)} \approx F^{(i)}_Q= \sum_{m=0}^N \sum_{n=0}^N {F}^{(i)}_{\hat{Q}_{mn}} \,\,\, 
    \label{eqn:fq}
\end{equation}
where 
\begin{equation}
   {F}^{(i)}_{\hat{Q}_{mn}}   =  -2\rho \int  \hat{Q}_{mn} \phi^{(i)} dV
    \label{eqn:Fmn}
\end{equation}
Thus, the total pressure induced force is the sum of intra-modal ($m=n$) and inter-modal ($m \neq n$) interactions between the modes of the flow, and these interactions can be estimated directly from the modal decomposition of the velocity field. It should be noted that $\hat{Q}_{nm}$ is symmetric in that $\hat{Q}^{(i)}_{mn} \equiv \hat{Q}^{(i)}_{nm}$ and therefore \oc{${F}^{(i)}_{\hat{Q}_{mn}}={F}^{(i)}_{\hat{Q}_{nm}}$}. Thus for these inter-modal ($m \neq n$) effects we always plot (or quantify) $2 \times \hat{Q}_{nm}$ and $2 \times {F}_{\hat{Q}_{mn}}$ to account for both contributions simultaneously.

\subsection{Force and Acoustic Partitioning Method (FAPM)}
\label{sec:fapm}
An extension of force partitioning method is the force and acoustic partitioning method (FAPM, \cite{Seo2022,Seo2023}). In this method we use the \oc{Ffowcs Williams} - Hawkings equation (FW-H, \oc{\cite{FWHorigpaper, zorumski_compact_noise,fwheqn2})} written for low surface Mach number, far field noise and a compact source as (see \cite{Seo2022} for details), 
\begin{equation}
    p^{'}=\frac{1}{4\pi}\left( \frac{{\bf \dot{F} \cdot r}}{cr^2} +\frac{{\bf F\cdot r}}{r^3} \right)_{t-\frac{r}{c}}.
    \label{APMeqn}
\end{equation}
 Here, $p^{'}$ is total loading noise, ${\bf F}$ is force and ${\bf r}$ is vector from point source to the point where the loading noise is computed. Plugging the force components from equation \ref{fpmEqn} into equation \ref{APMeqn} will give following noise components:
 \begin{equation}
     p^{'}=p_{B}^{'}+p_{O}^{'}+p_{\mu}^{'}+p_{Q}^{'}  \,\, ,
 \end{equation}
corresponding to noise due to blade acceleration, noise due to freestream unsteadiness, viscous diffusion induced loading noise and vortex-induced loading noise. For the results presented here, loading noise due to vortices will be the most dominant component as per the reasoning provided in the previous section and this is given by,
\begin{equation}
    p_{Q}^{'}=\frac{1}{4\pi}\left( \frac{{\bf \dot{F}_Q }}{cr^2} +\frac{{\bf F_Q}}{r^3} \right)_{t-\frac{r}{c}} \cdot {\bf r} \,\, \oc{.}
    \label{APMfqeqn}
\end{equation}
We now substitute the expression for the modal contribution of the vortex-induced force from equation \ref{eqn:fq} into the above equation and obtain the corresponding aeroacoustic noise associated with each of those intra-modal and inter-modal interaction as follows:
\begin{equation}
p'_{\hat{Q}_{mn}}=\frac{1}{4\pi r^2}\left[\left(\frac{1}{c} \frac{\partial}{\partial t}+\frac{1}{r}\right) {\bf{F}}_{\hat{Q}_{mn}} \right]_{(t-\frac{r}{c})} \cdot {\bf r}\,\, \oc{.}
\label{speqn}
\end{equation}

\section{Results}
In this section, we describe the results for three distinct cases with sequentially increasing complexity: 2D flow past a circular at a Reynolds number of 300; 2D flow past a NACA 0015 airfoil at a Reynolds number of 2500; and finally, 3D flow past an aspect-ratio of 5 revolving rectangular rotor blade at a tip-based Reynolds number of 3300. \oc{The grid convergence study for all these cases are shown in Appendix \ref{appendix:gc}.}

\subsection{Flow past a circular cylinder at Re=300}
\subsubsection{Reynolds decomposition}
\label{sec:pod:uv:cyl}

\begin{figure}
    \centering
     \begin{subfigure}[b]{0.28\textwidth}
         \centering
         \includegraphics[width=\textwidth]{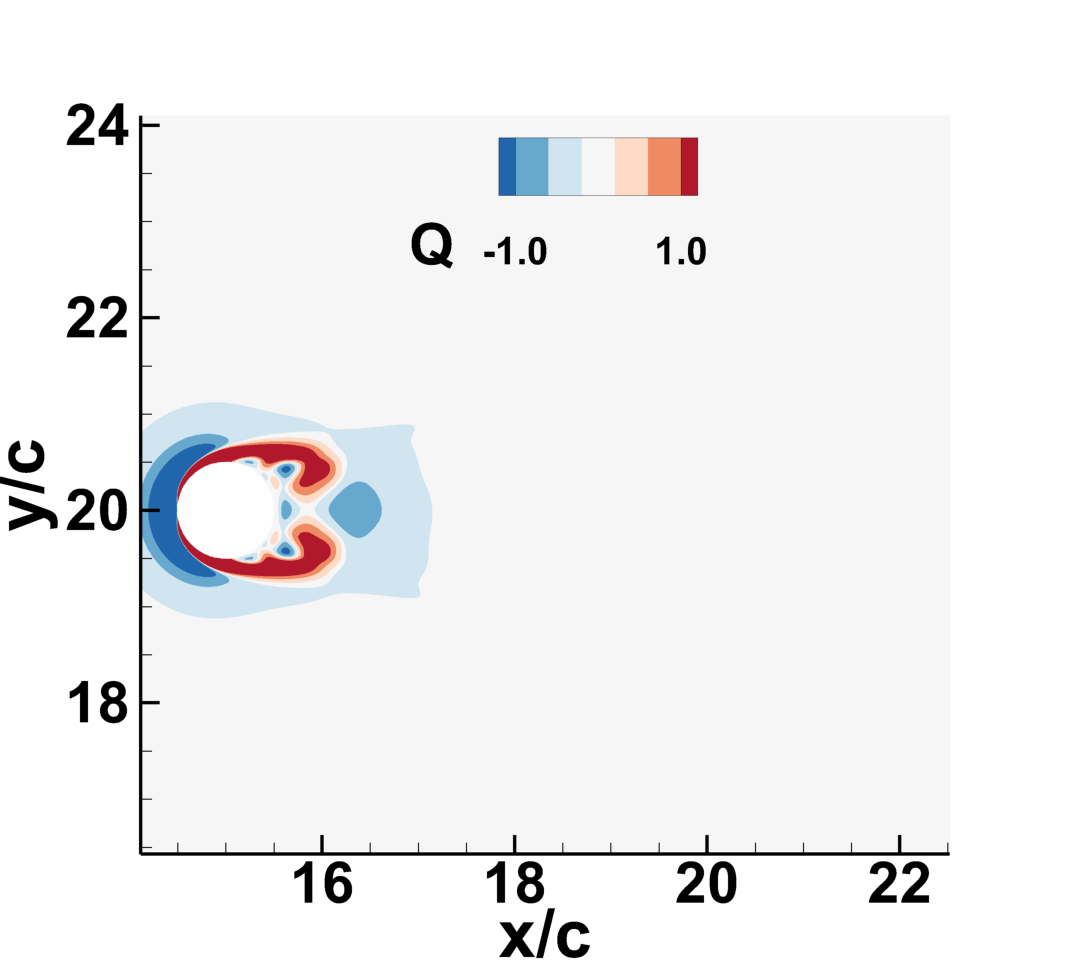}
         \caption{$\hat{Q}_{00}$}
         \label{fig:cc:ReD:QandfQ:a}
     \end{subfigure}
      \hfill
 \begin{subfigure}[b]{0.28\textwidth}
      \centering
         \includegraphics[width=\textwidth]{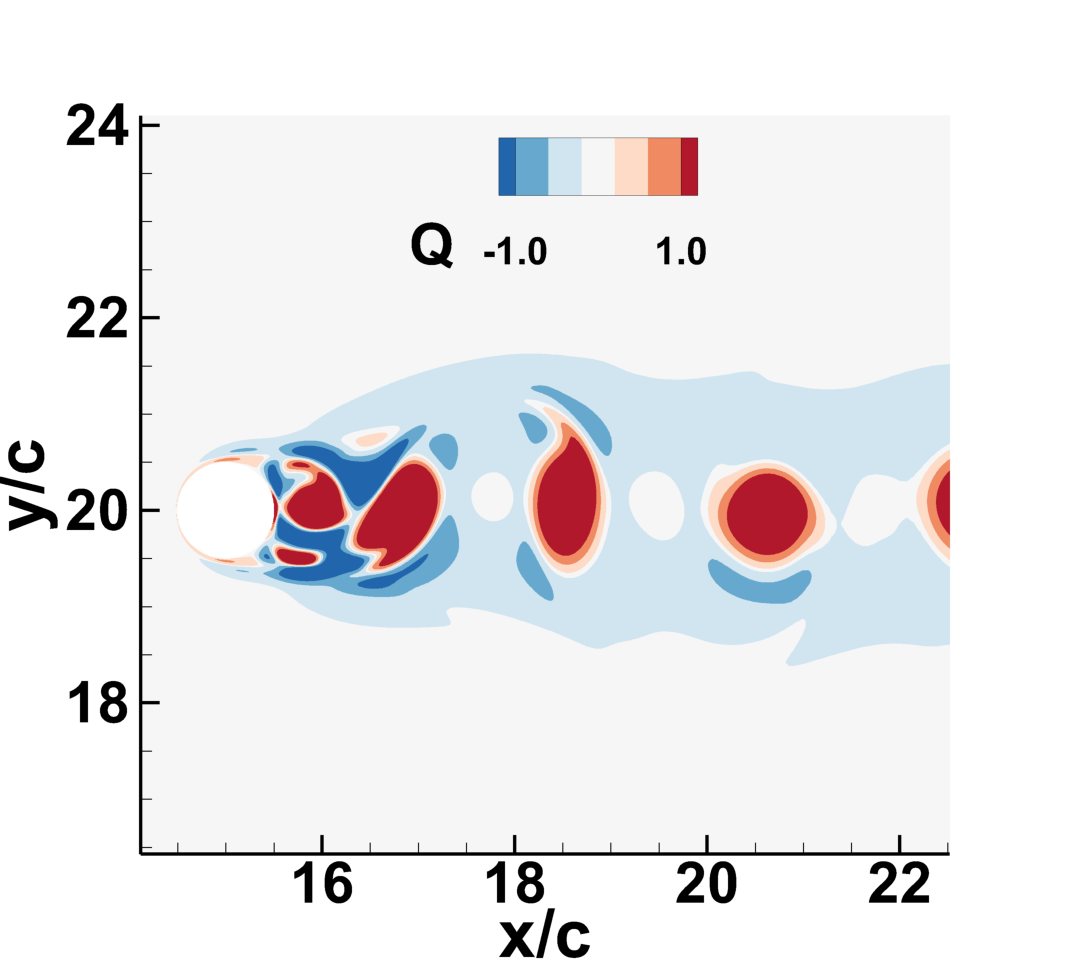}
         \caption{$\hat{Q}_{11}$}
         \label{fig:cc:ReD:QandfQ:b}
     \end{subfigure}
  \hfill
     \begin{subfigure}[b]{0.28\textwidth}
         \centering
         \includegraphics[width=\textwidth]{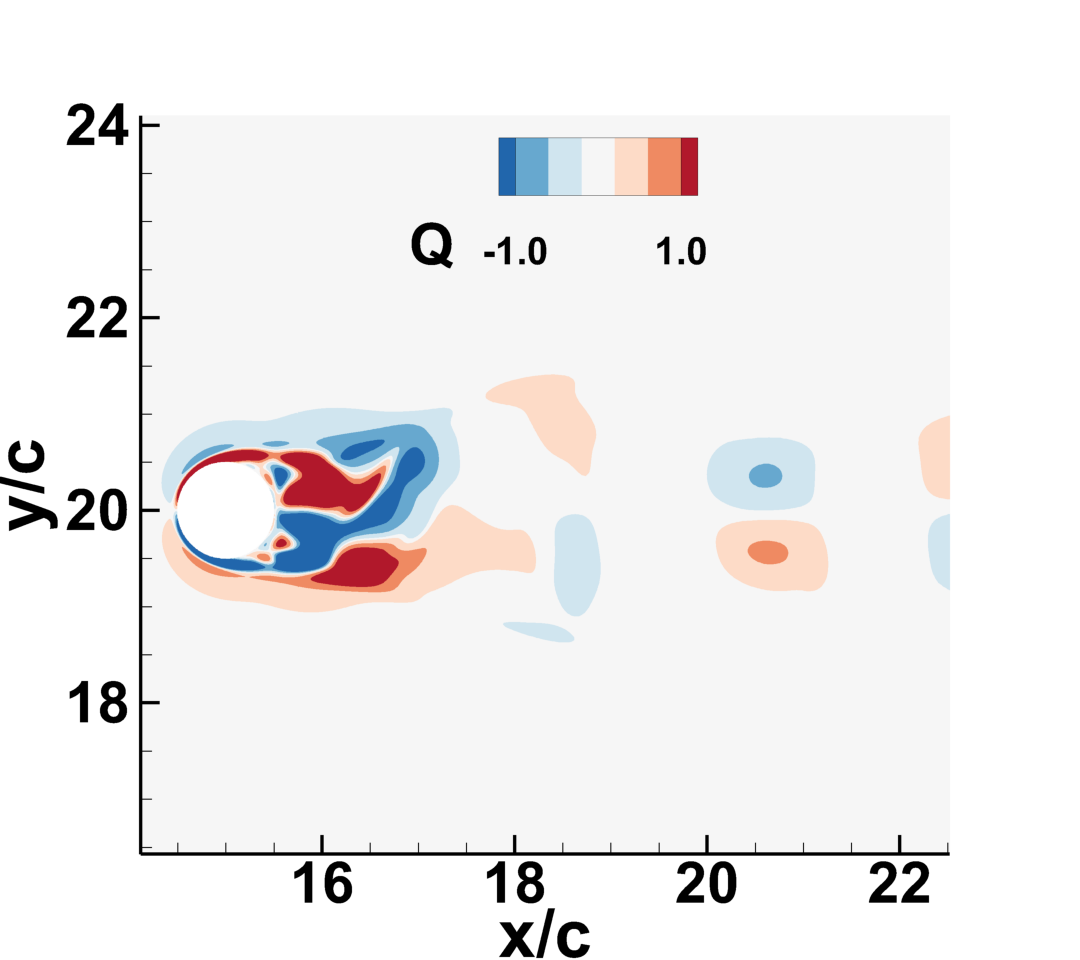}
         \caption{$\hat{Q}_{01}$}
         \label{fig:cc:ReD:QandfQ:c}
     \end{subfigure}
        \hfill
     \begin{subfigure}[b]{0.28\textwidth}
         \centering
         \includegraphics[width=\textwidth]{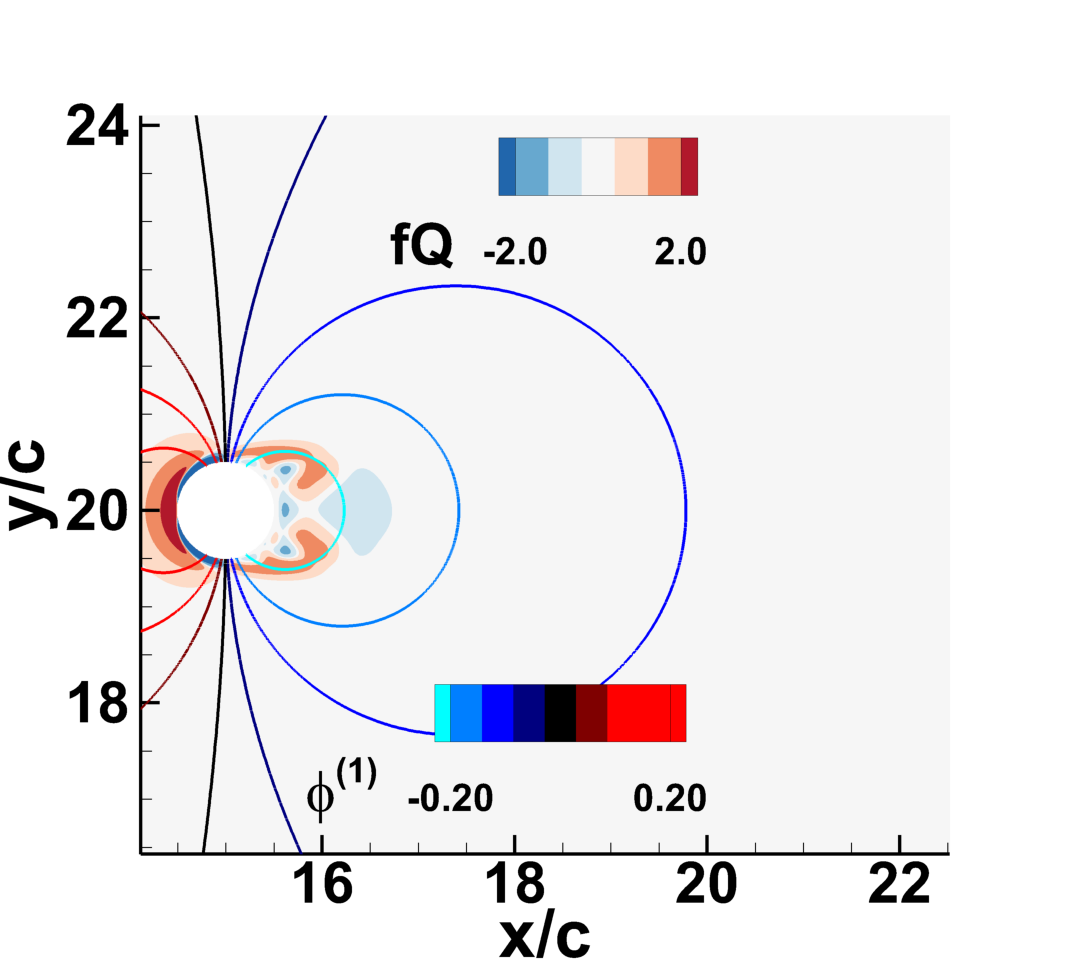}
         \caption{$f_{\hat{Q}_{00}}^{(1)}$}
         \label{fig:cc:ReD:QandfQ:d}
     \end{subfigure}
        \hfill
     \begin{subfigure}[b]{0.28\textwidth}
         \centering
         \includegraphics[width=\textwidth]{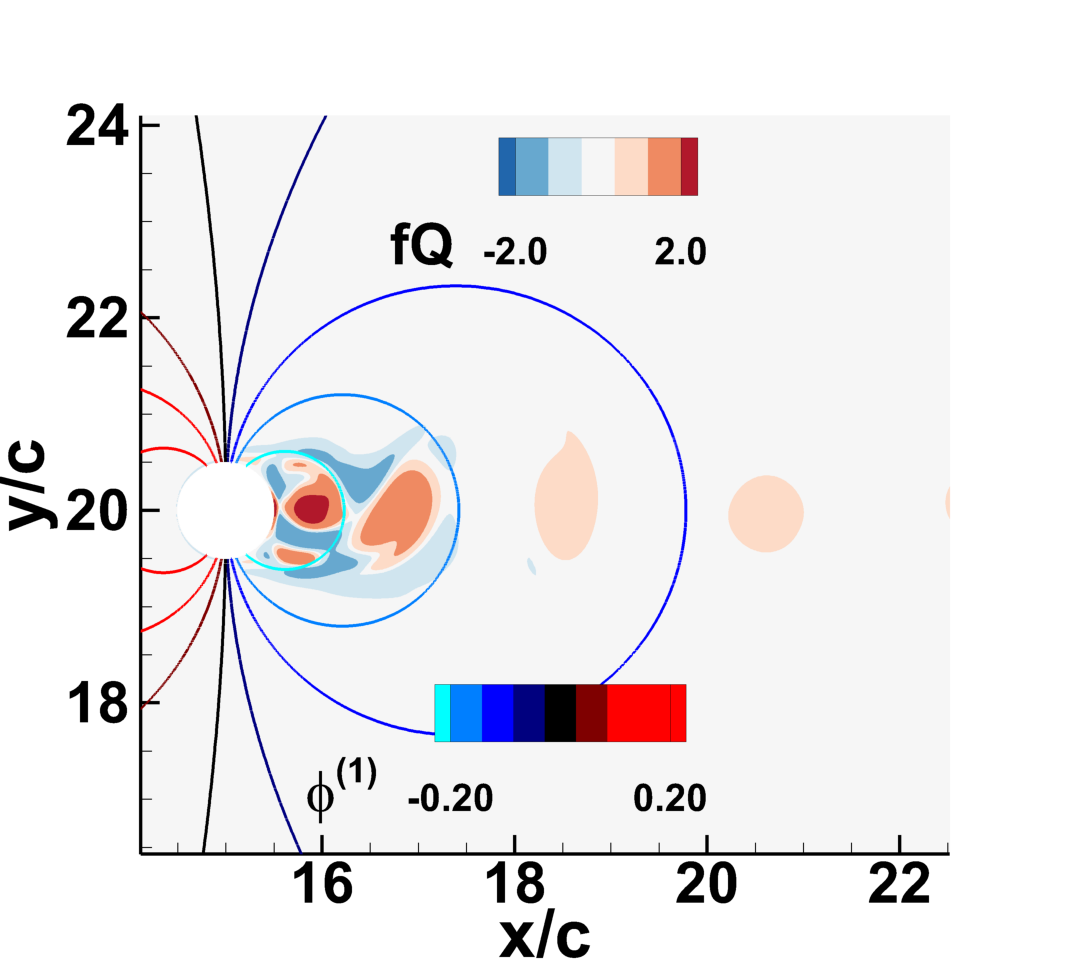}
         \caption{$f_{\hat{Q}_{11}}^{(1)}$}
         \label{fig:cc:ReD:QandfQ:e}
     \end{subfigure}
     \hfill
     \begin{subfigure}[b]{0.28\textwidth}
         \centering
         \includegraphics[width=\textwidth]{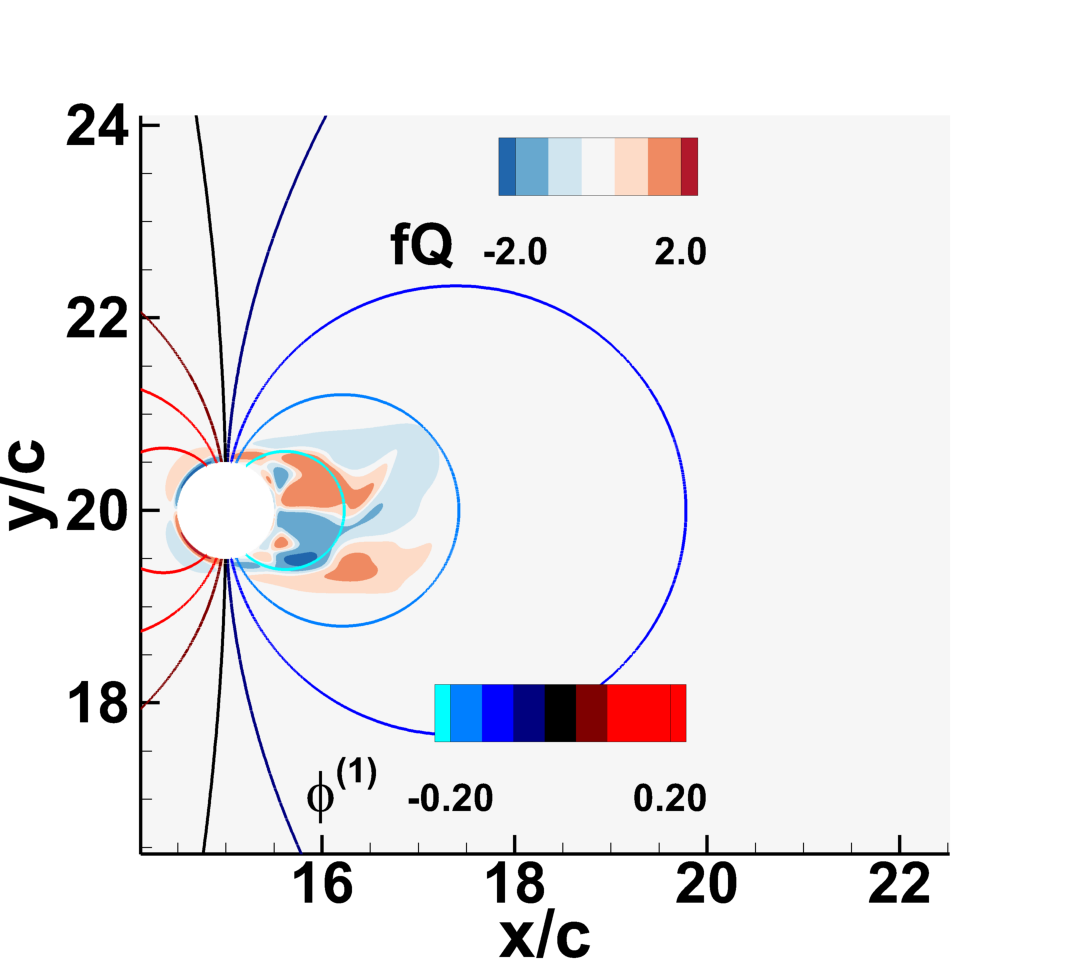}
         \caption{$f_{\hat{Q}_{01}}^{(1)}$}
         \label{fig:cc:ReD:QandfQ:f}
     \end{subfigure}
  \hfill
     \begin{subfigure}[b]{0.28\textwidth}
         \centering
         \includegraphics[width=\textwidth]{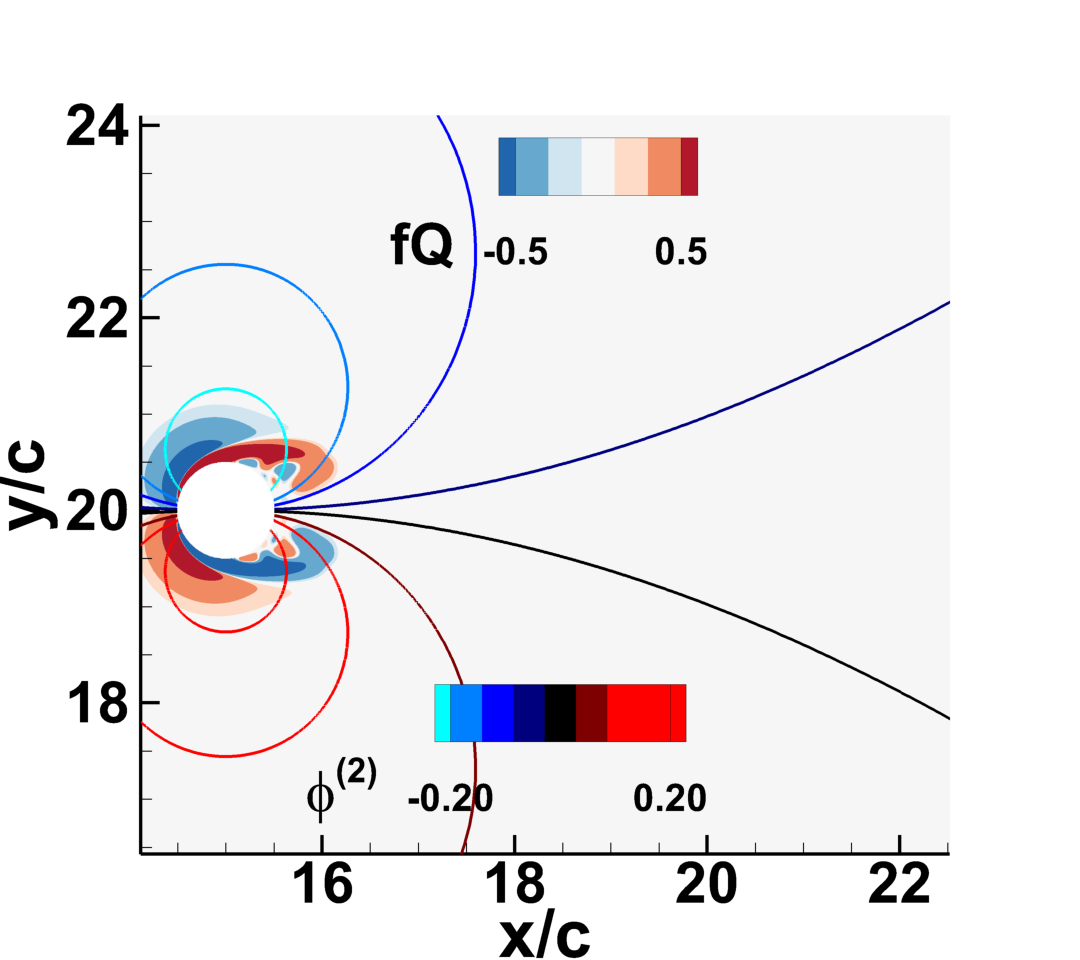}
         \caption{$f_{\hat{Q}_{00}}^{(2)}$}
         \label{fig:cc:ReD:QandfQ:g}
     \end{subfigure}
   \hfill
     \begin{subfigure}[b]{0.28\textwidth}
         \centering
         \includegraphics[width=\textwidth]{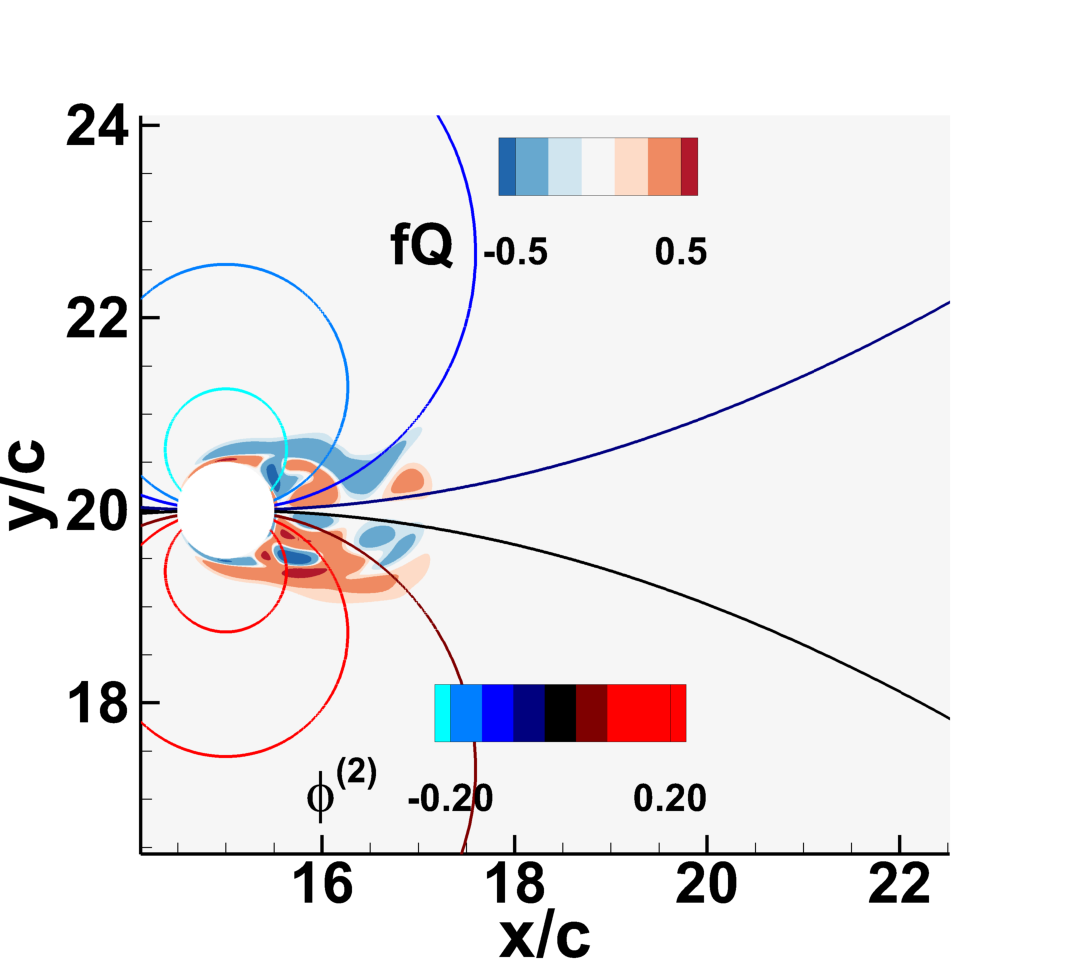}
         \caption{$f_{\hat{Q}_{11}}^{(2)}$}
         \label{fig:cc:ReD:QandfQ:h}
     \end{subfigure}
         \hfill
     \begin{subfigure}[b]{0.28\textwidth}
         \centering
         \includegraphics[width=\textwidth]{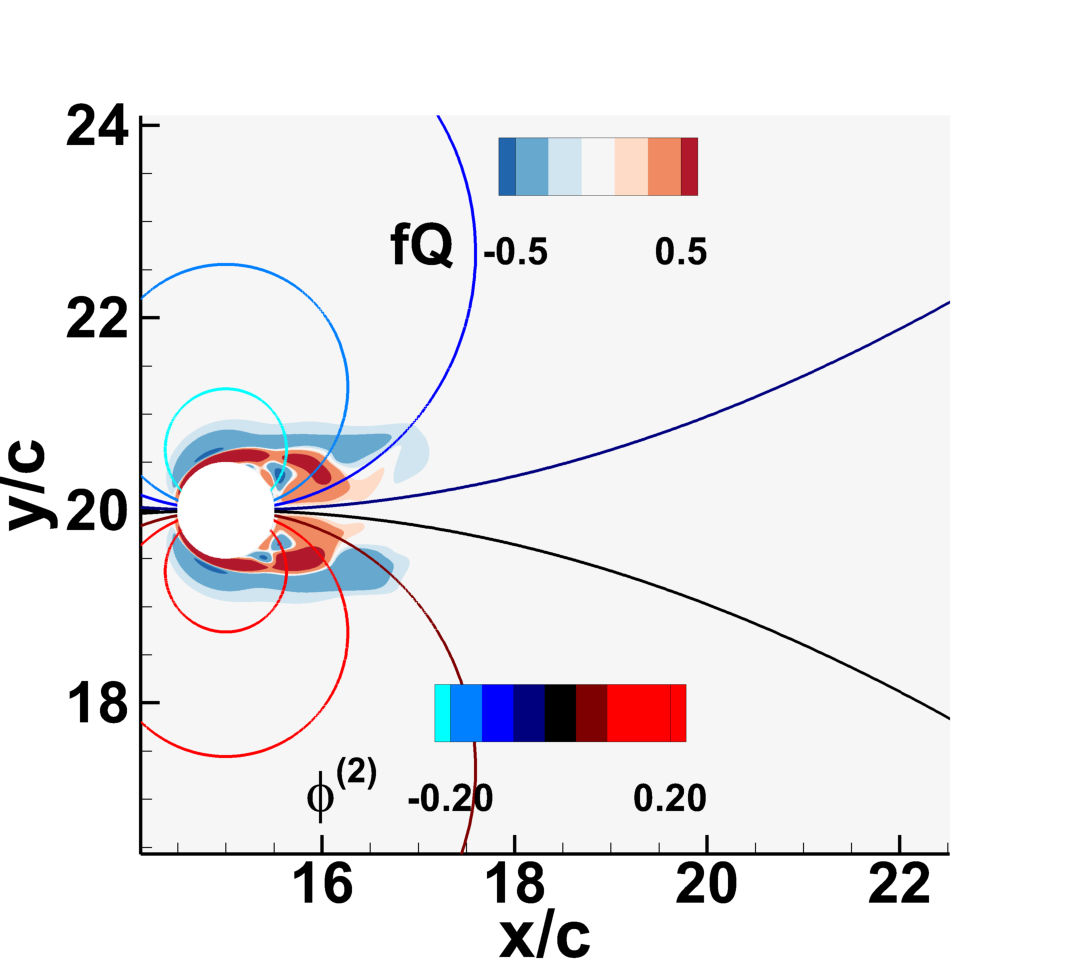}
         \caption{$f_{\hat{Q}_{01}}^{(2)}$}
         \label{fig:cc:ReD:QandfQ:i}
     \end{subfigure}
        \caption{Force partitioning based on Reynolds decomposition of the velocity field for the cylinder flow. Contours of $Q$ for the (a) mean flow, (b) the fluctuating component and (c) the interaction of the mean flow with the fluctuating component. Contours of the vortex-induced drag force density ${f^{(1)}_Q}$ and $\phi^{(1)}$ corresponding to (d) the mean flow, (\oc{e}) the fluctuating component and (\oc{f}) the interaction of the mean flow with the fluctuating component. Contours of the vortex-induced lift force density ${f^{(2)}_Q}$ and $\phi^{(2)}$ corresponding to (\oc{g}) the mean flow, (\oc{h}) the fluctuating component and (\oc{i}) the interaction of the mean flow with the fluctuating component.}
\label{fig:cc:ReD:QandfQ}
\end{figure}

The vortex shedding corresponding to a circular cylinder with \oc{a} Reynolds number of 300 based on the cylinder diameter ($d$) and the incoming flow velocity ($U_\infty$) is shown in figure \ref{fig:cc:rd:TotFlow}. The segment of the flow field we select for modal decomposition contains 5 cycles of the lift force with each cycle consisting of 48 snapshots. We begin by using the Reynolds decomposition to partition the total flow into a mean and a fluctuating component. Figure \ref{fig:cc:ReD:QandfQ}(a-c) shows the $Q$ fields corresponding to the mean mode $\hat{Q}_{00}$, the fluctuation mode $\hat{Q}_{11}$ and the interaction mode $\hat{Q}_{01}$ and we see that the mean mode $\bf{u}_0$ is symmetric about the wake centerline, and it captures vorticity near the cylinder surface, whereas the fluctuation mode $\bf{u}_1$ captures the shedding of the vortices in the wake. Since all modes other than the mean modes are a function of time, we always show these modes at an arbitrarily chosen time-instance.  

Figure \ref{fig:cc:ReD:QandfQ}(d-f) shows the contours of vortex-induced drag force density $f_{\hat{Q}_{mn}}^{(1)}$ for these modes along with the line contours of $\phi^{(1)}$, and figure \ref{fig:cc:ReD:QandfQ}(g-i) show the corresponding plots for lift. We note that $\phi^{(1)}$ is symmetric about the wake centerline and has high values in the front and rear of the cylinder. $\hat{Q}_{00}$ is also symmetric about the wake centerline, and therefore $f_{\hat{Q}_{00}}^{(1)}$, which is the product of these two symmetric functions, is also symmetric and the integral of this function will result in mean drag. Conversely, since $\phi^{(2)}$ is antisymmetric about the wake centerline, $f_{\hat{Q}_{00}}^{(2)}$ will be asymmetric about the centerline and will make no net contribution to the mean lift. 

\begin{figure}
 \begin{subfigure}[b]{0.5\textwidth}
      \centering
         \includegraphics[height=0.24\textheight]{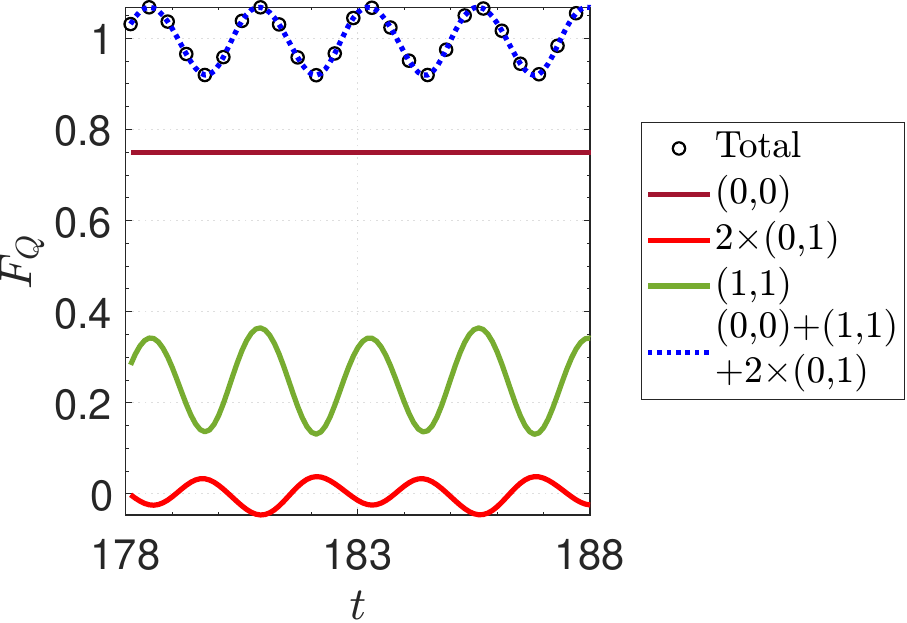}
         \caption{Drag}
         \label{fig:cc:red:fqvst:a}
     \end{subfigure}
      \hfill
         \begin{subfigure}[b]{0.5\textwidth}
        \centering
        \includegraphics[height=0.24\textheight]{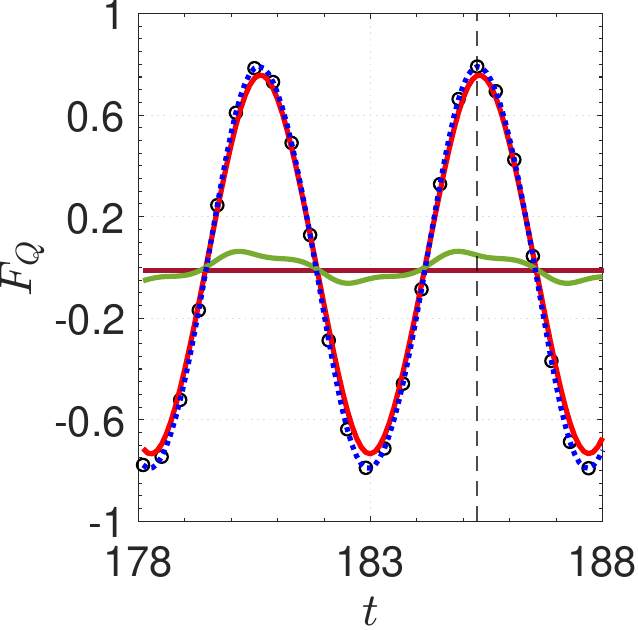}
        \caption{Lift}
        \label{fig:cc:red:fqvst:b}
    \end{subfigure}
\caption{Temporal variation of the non-dimensional vortex-induced (a) drag force ($F_Q^{(1)}$) and (b) lift force ($F_Q^{(2)}$) obtained from the Reynolds decomposition of the velocity field of the circular cylinder. \oc{The force for all the circular cylinder cases is normalized using the force coefficient ($0.5\rho U_\infty^2 d$) and the time is normalized using the flow time-scale ($d/U_\infty$).} The dashed vertical line on the right plot shows the time instance where all contour plots for the circular cylinder are shown.}
\label{fig:cc:rd:fqvst}
\end{figure}

The integrated drag and lift forces generated due to each of these modes are computed according to the modal force partitioning method, and the time variation of these components is shown in \oc{figure} \ref{fig:cc:rd:fqvst}. As indicated above, the $(0,0)$ mode is found to contribute to the mean drag force. Both the $(1,1)$ and $(0,1)$ modes contribute to the fluctuation in the drag. The total of these three modes is slightly less than the total pressure drag calculated directly from the integration of the pressure on the surface, and this is because beyond $F_Q$, the $F_\mu$ component (see Eq. \ref{fpmEqn}) also makes a small but non-negligible  contribution to the mean drag for this low Reynolds number flow. 
\begin{figure}
      \centering
         \includegraphics[height=0.3\textheight]{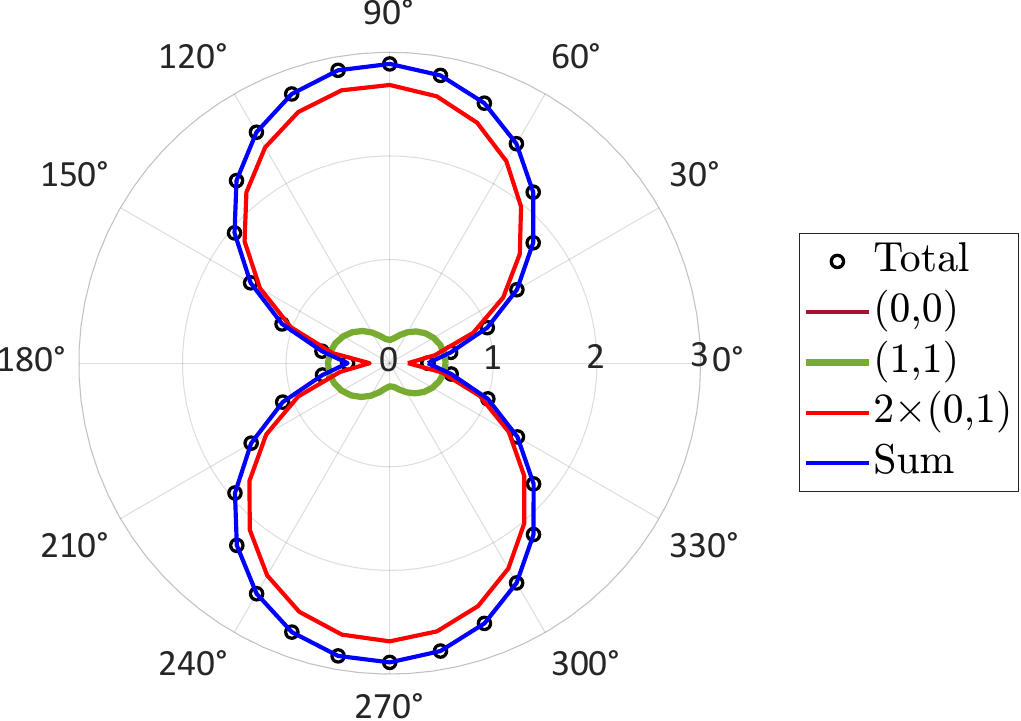}
\caption{Sound directivity plot based on modal force partitioning applied to the Reynolds decomposition of the velocity field for the circular cylinder flow showing directivity. The directivity shows $p'_{rms}[\times 10^{-5}]$ corresponding to a surface Mach number of 0.1 and is computed at a distance of 50$d$.}
\label{fig:cc:rd:dir}
\end{figure}

The lift is of particular importance here since it experiences large fluctuations that serve as the dominant source of aeroacoustic noise. As expected, the symmetric mean mode $(0,0)$ does not generate any lift. Interestingly, however, we observe that while the fluctuation mode $\bf{u}_1$ represents all of the velocity and vorticity fluctuations in the flow, this mode by itself contributes very little to the fluctuation in the lift force. Indeed, the variance of the $\hat{F}^{(2)}_{11}$ force component is only about 0.53\% of the variance in the total lift.  The only remaining modal contribution is the inter-modal component $2\hat{F}^{(2)}_{01}$, which is the interaction of the fluctuation mode with the mean mode. As shown in figure \ref{fig:cc:rd:fqvst}, this component actually provides the overwhelming majority (86.44 \%) of the variance in the lift force. Thus, the velocity mode that corresponds to the largest variation in the velocity field, i.e. $\bf{u}_1$, generates very little effect on the forces solely by itself. However, combined with the mean mode, it generates almost all the variation in the lift force. Thus, with regard to the lift force induced on the body, the mean mode acts as an ``amplifier'' for the higher $\bf{u}_1$ mode. This general observation will carry through for other modal decompositions as well. Finally we note that the sum of all these modal components very slightly under-predicts the total pressure induced lift force obtained directly from the integration of the pressure on the body. This difference is attributable to the viscous diffusion-induced pressure force ${F_\mu}$ and highlights the fact that even at these relatively low Reynolds numbers, ${F_\mu}$ is very small and therefore, the vortex-induced force (${F_\mu}$) provides an accurate estimate of the total pressure-induced force.  

Figure \ref{fig:cc:rd:dir} shows the directivity field of the sound associated with these modes. As expected from the modal contributions to drag and lift, we find that mode (1,1), which is the dominant mode in the velocity variation, only makes a small contribution to the sound field. On the other hand, the dominant contribution to the sound field comes from the mode (0,1). Thus, the results from even a simple Reynolds decomposition of a relatively low Reynolds number flow are counter-intuitive in that the mode most responsible for the variation in the flow field  by itself does not account for the dominant variation in the surface pressure and the generation of the noise. This is because the quantity $Q$ to which the surface force is directly dependent on, is a non-linear ``observable'' and it therefore encodes non-linear interactions in the modes. We explore this next via a POD of this flow.

\subsubsection{Proper Orthogonal Decomposition}
We now apply POD to the flow field of the circular cylinder and while the mean mode remains the same as the one for Reynolds decomposition, the POD procedure partitions the fluctuation component into multiple POD modes (in this case, there are 120 total POD modes). We further note that 7 modes (in addition to the mean mode) are required to reconstruct 98\% of the flow field.

\begin{figure}
    \centering
 \begin{subfigure}[b]{0.23\textwidth}
      \centering
         \includegraphics[width=\textwidth]{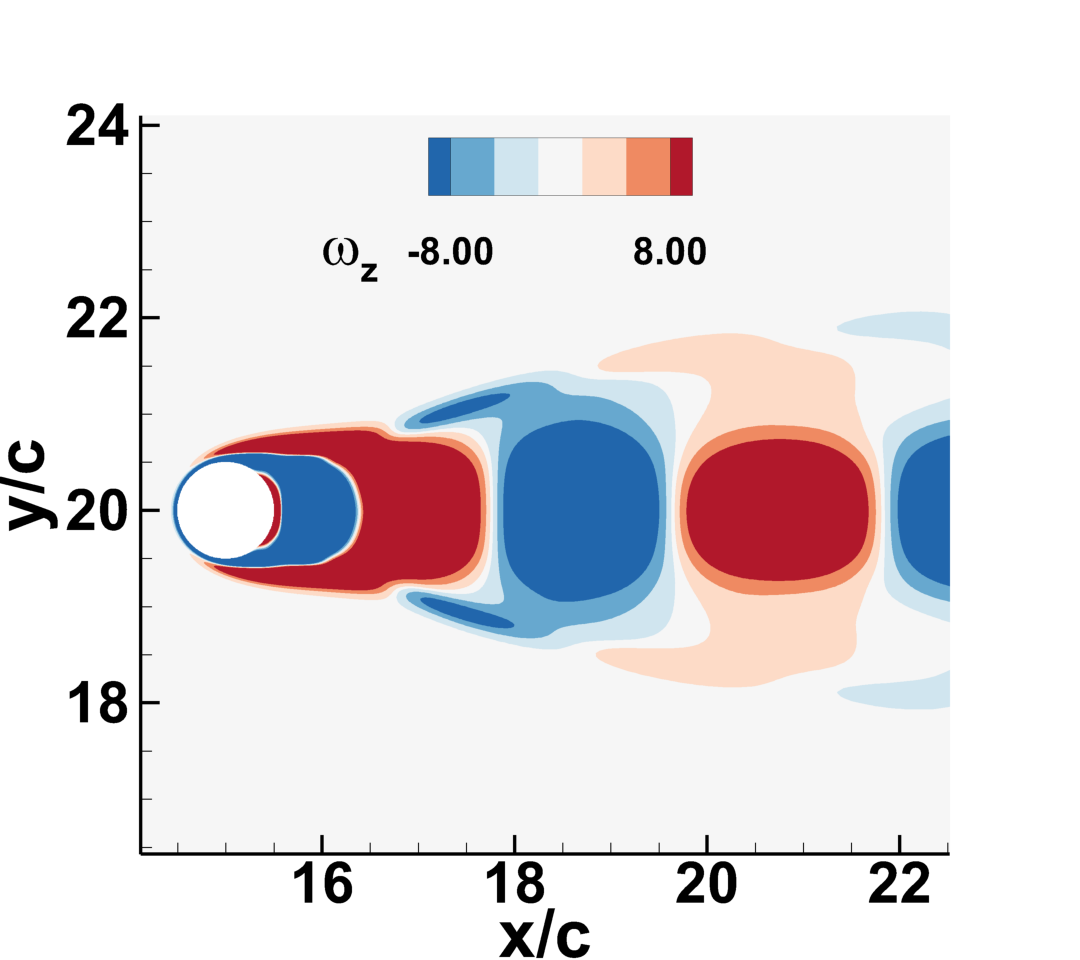}
         \caption{$\omega_z(\bf{u_1})$ }
         \label{fig:cc:pod:USigma:a}
     \end{subfigure}
     \begin{subfigure}[b]{0.23\textwidth}
         \centering
         \includegraphics[width=\textwidth]{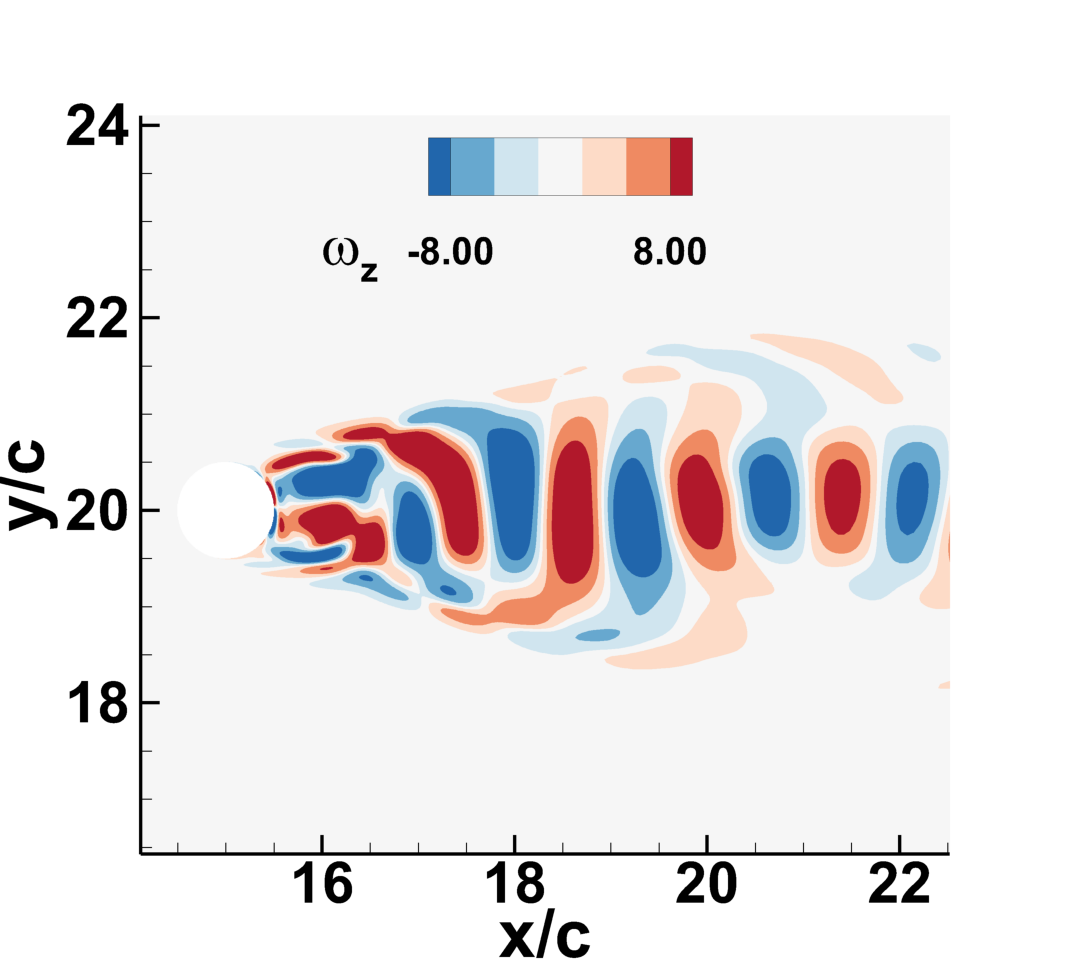}
         \caption{$\omega_z(\bf{u_2})$}
         \label{fig:cc:pod:USigma:b}
     \end{subfigure}
     \begin{subfigure}[b]{0.23\textwidth}
         \centering
         \includegraphics[width=\textwidth]{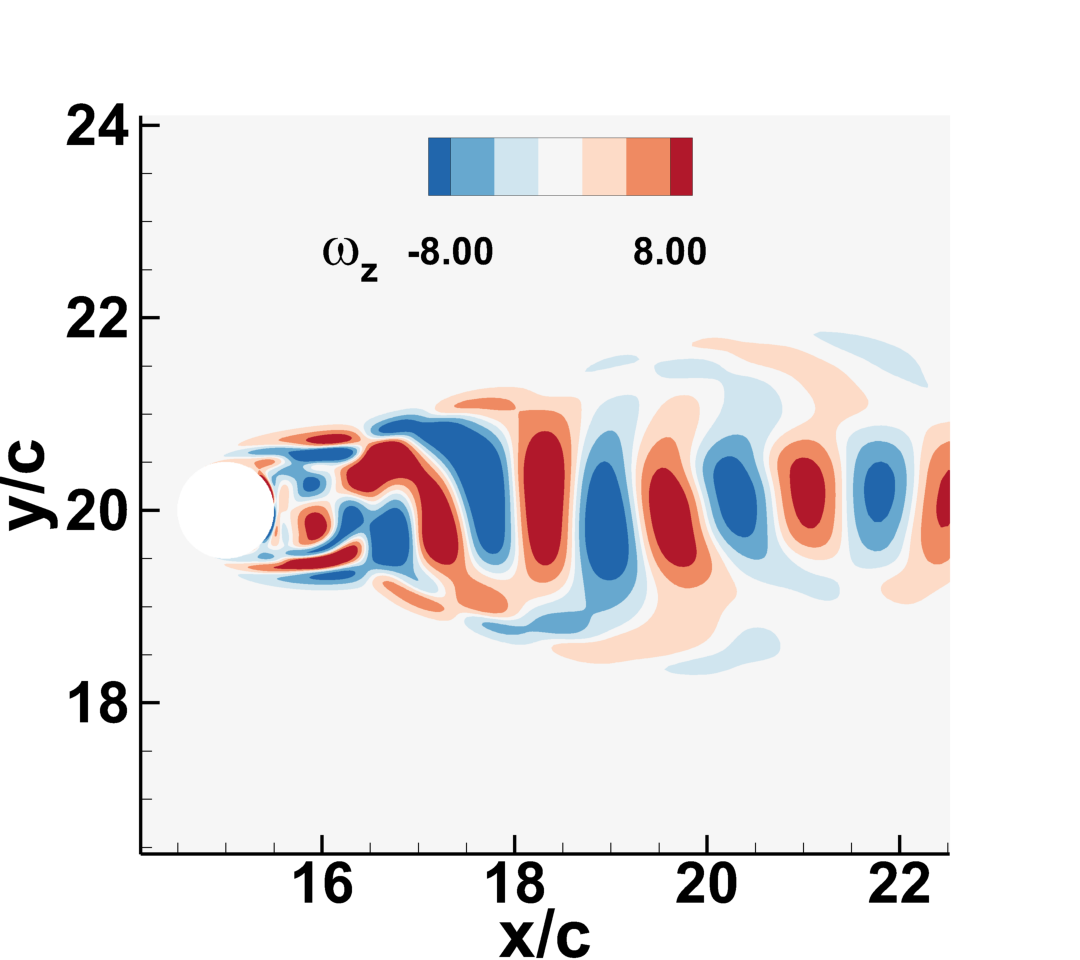}
         \caption{$\omega_z(\bf{u_3})$}
         \label{fig:cc:pod:USigma:c}
     \end{subfigure}
     \begin{subfigure}[b]{0.23\textwidth}
         \centering
         \includegraphics[width=\textwidth]{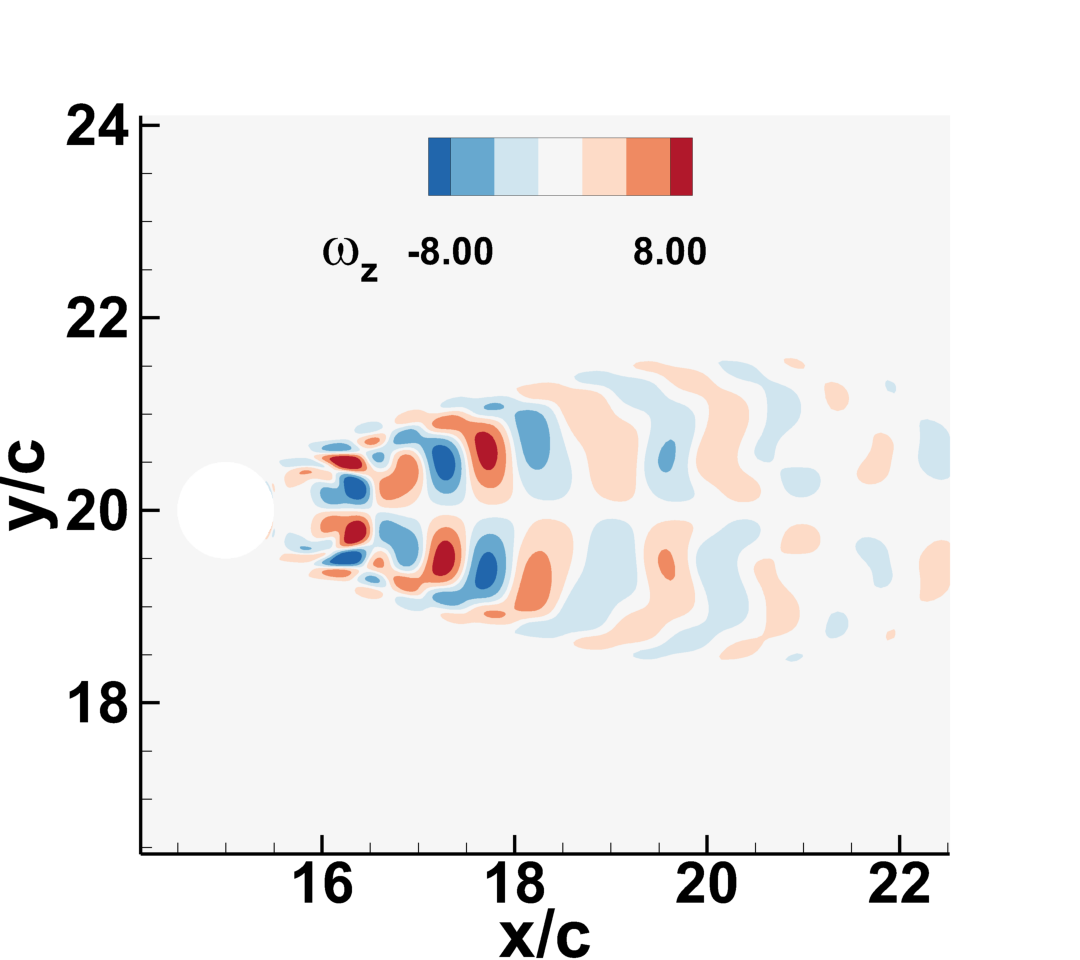}
         \caption{$\omega_z(\bf{u_4})$}
         \label{fig:cc:pod:USigma:d}
     \end{subfigure}

     \begin{subfigure}[b]{0.23\textwidth}
      \centering
         \includegraphics[width=\textwidth]{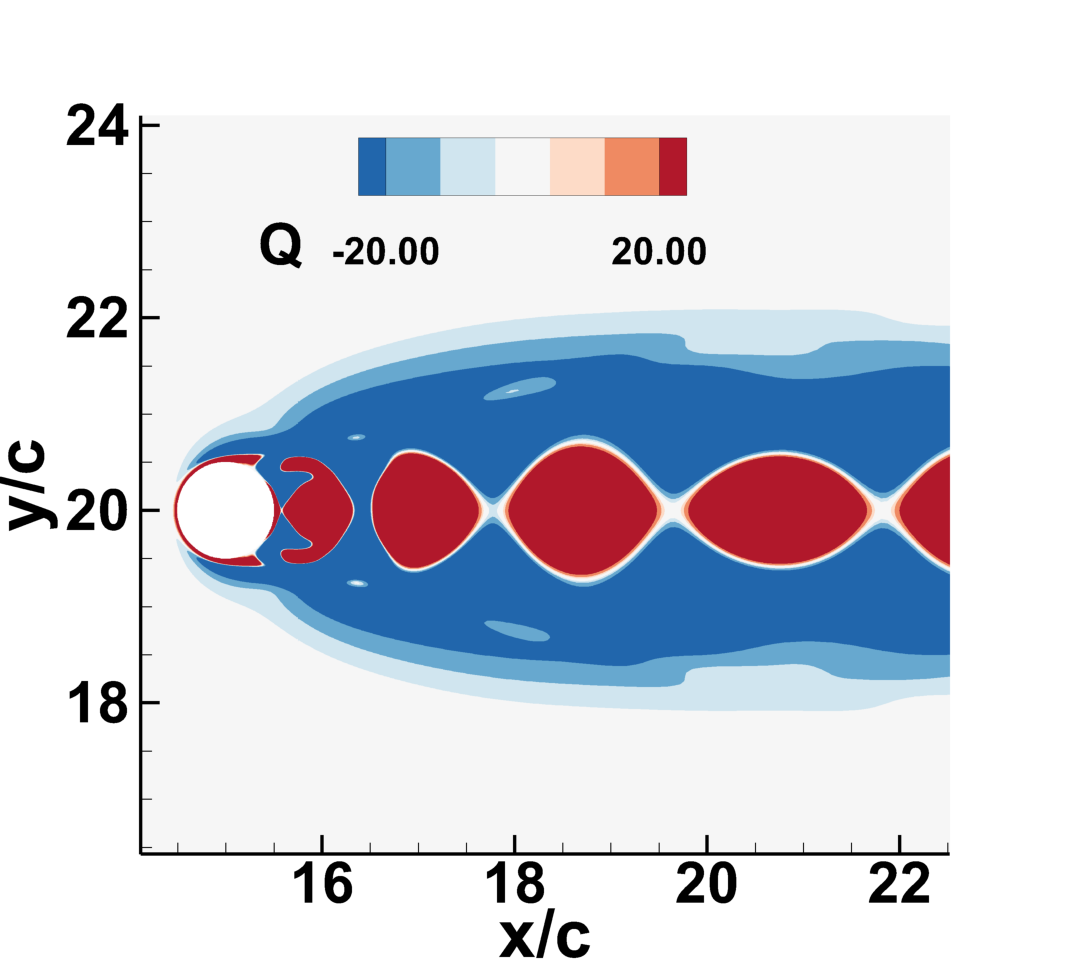}
         \caption{$\hat{Q}_{11}$ }
         \label{fig:cc:pod:USigma:e}
     \end{subfigure}
     \begin{subfigure}[b]{0.23\textwidth}
         \centering
         \includegraphics[width=\textwidth]{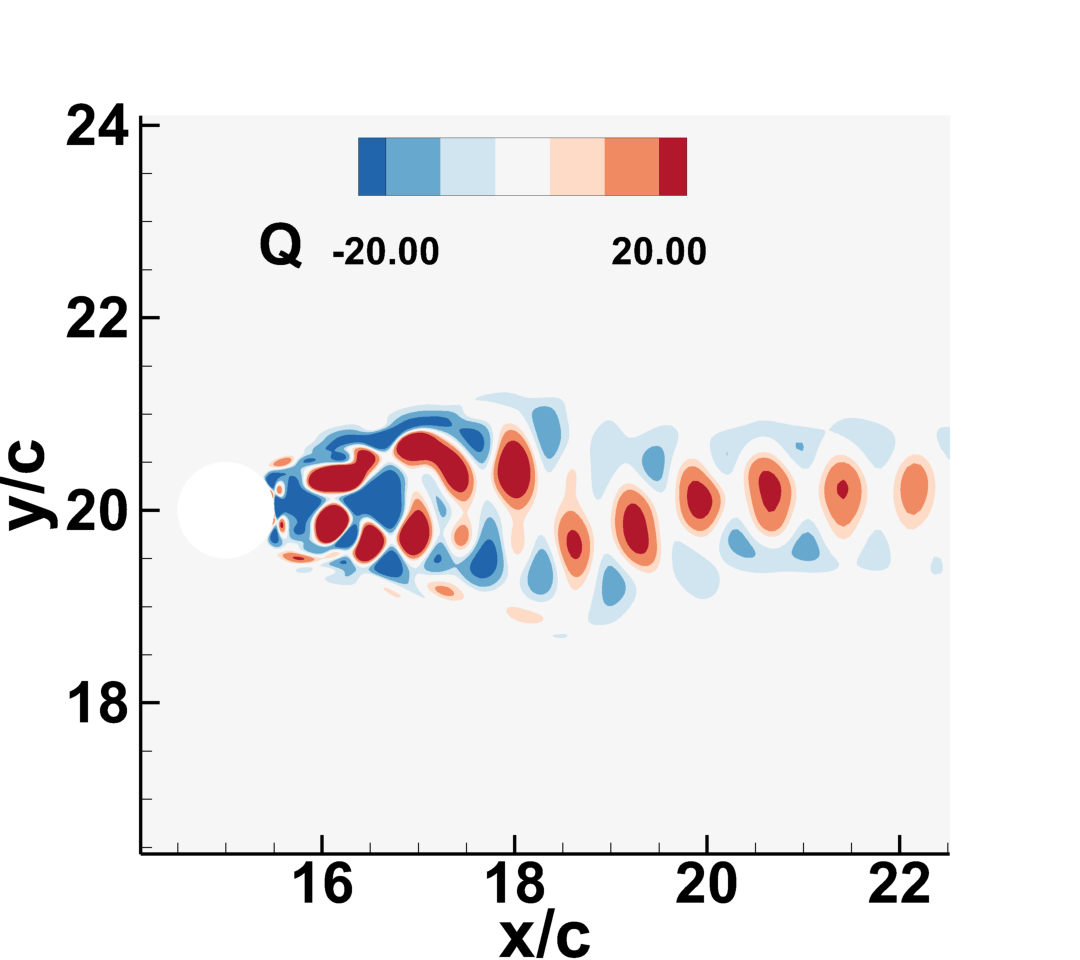}
         \caption{$\hat{Q}_{22}$}
         \label{fig:cc:pod:USigma:f}
     \end{subfigure}
     \begin{subfigure}[b]{0.23\textwidth}
         \centering
         \includegraphics[width=\textwidth]{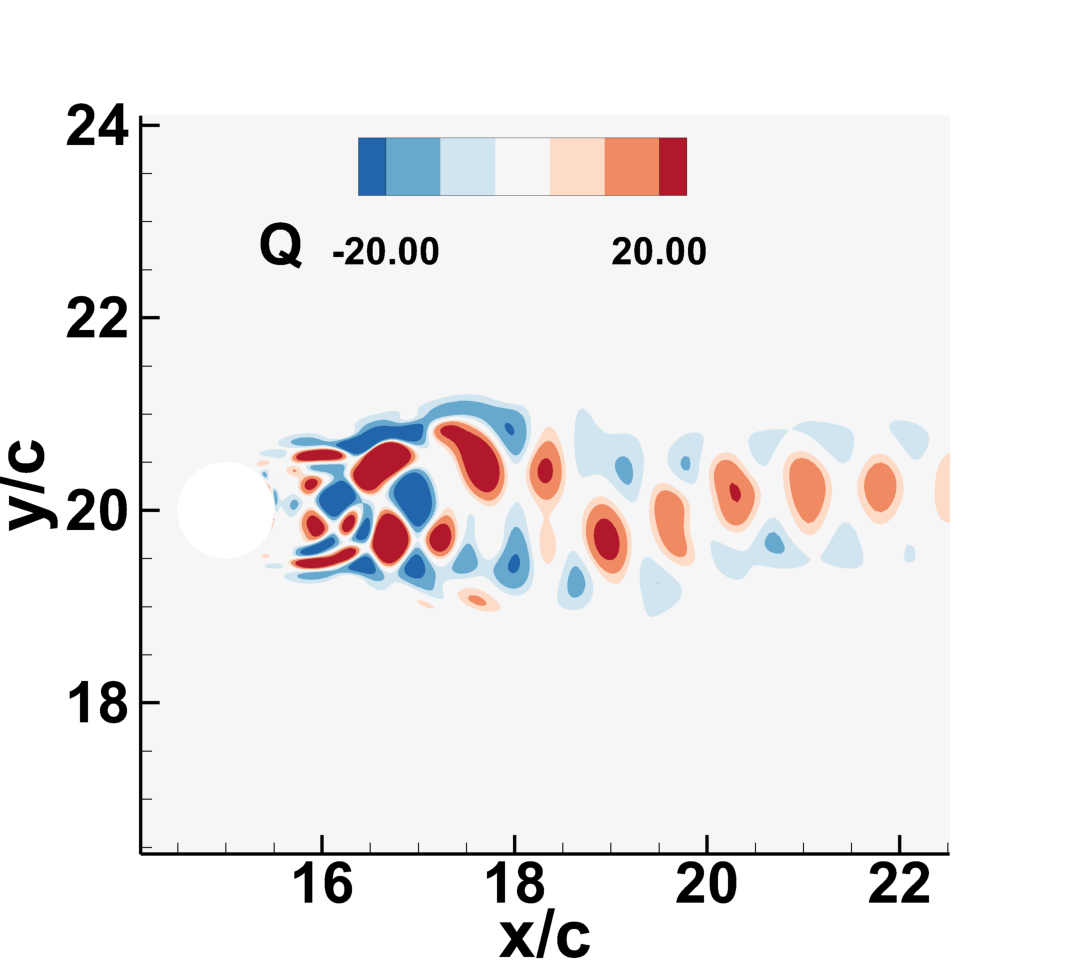}
         \caption{$\hat{Q}_{33}$}
         \label{fig:cc:pod:USigma:g}
     \end{subfigure}
     \begin{subfigure}[b]{0.23\textwidth}
         \centering
         \includegraphics[width=\textwidth]{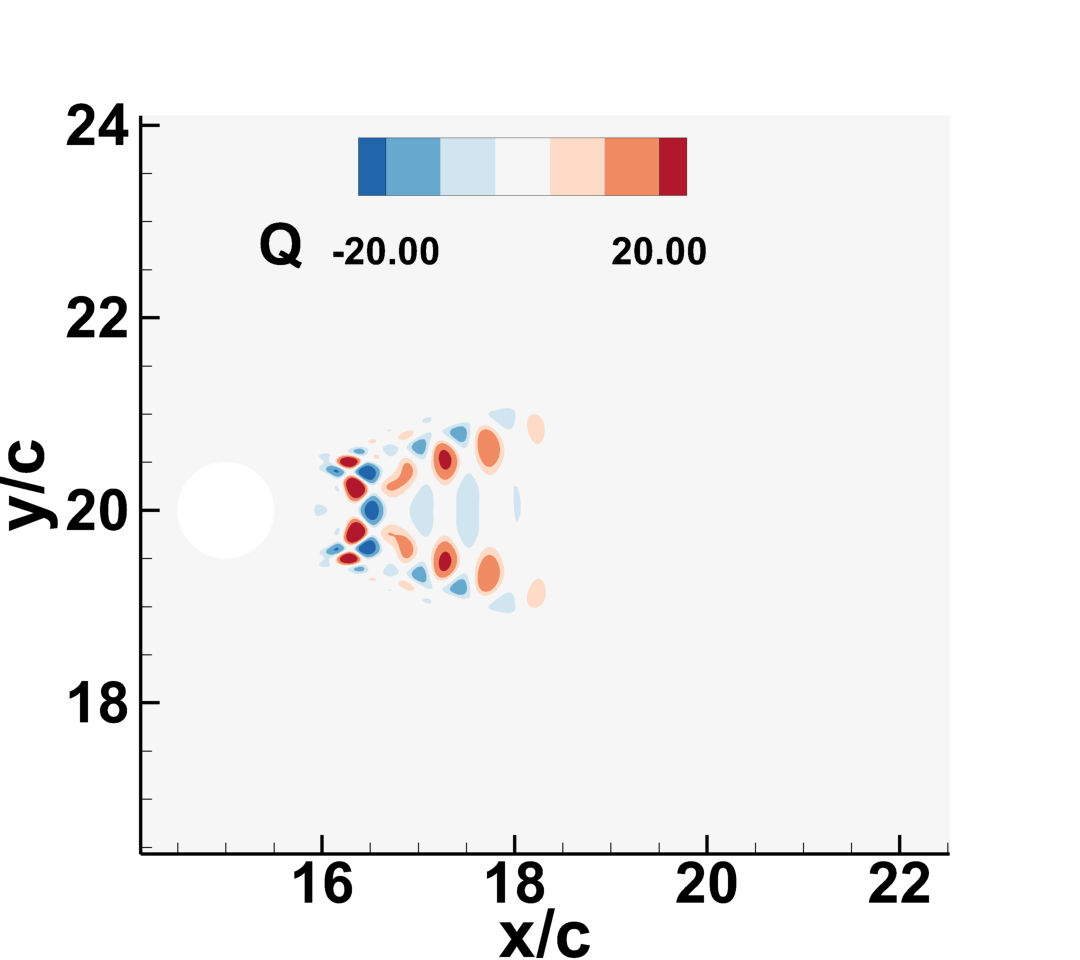}
         \caption{$\hat{Q}_{44}$}
         \label{fig:cc:pod:USigma:h}
     \end{subfigure}

        \begin{subfigure}[b]{0.23\textwidth}
         \centering
         \includegraphics[width=\textwidth]{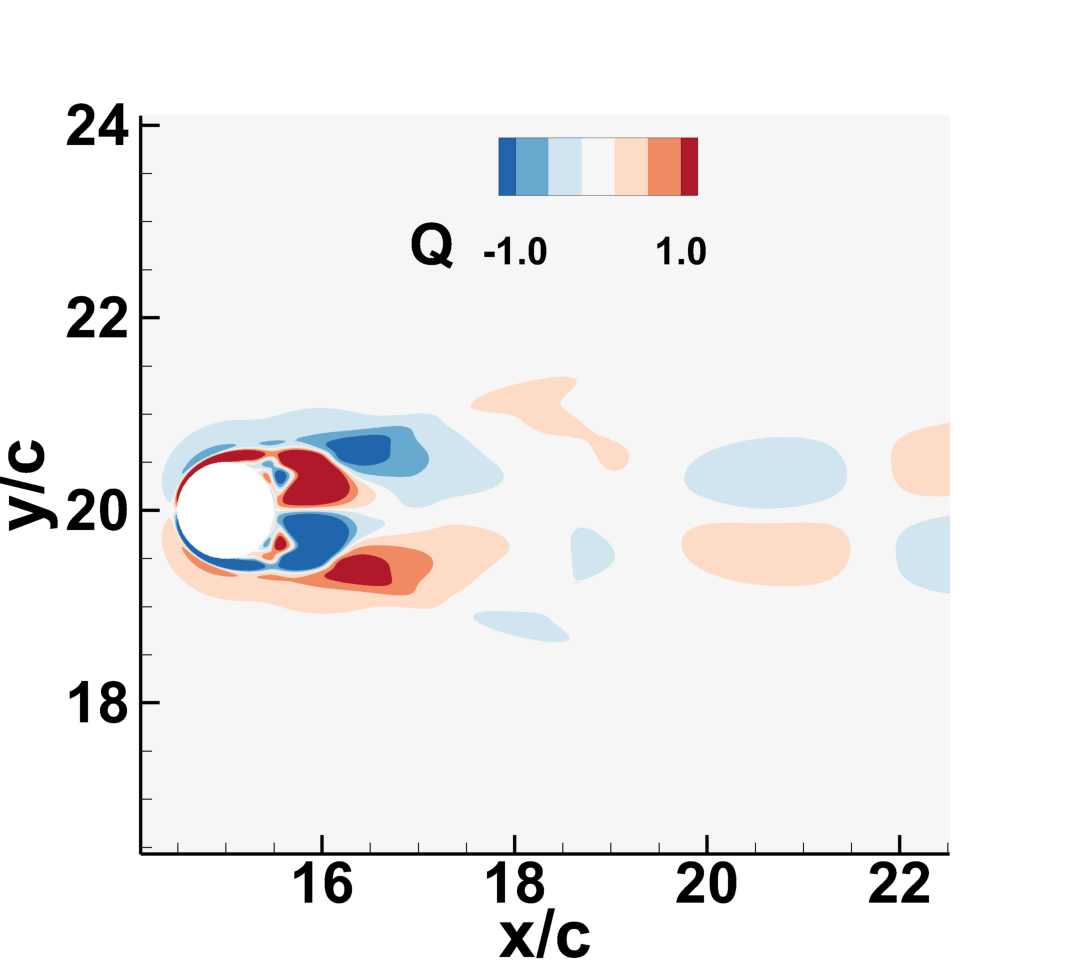}
         \caption{i.e. 2$\hat{Q}_{01}$}
         \label{fig:cc:pod:USigma:i}
     \end{subfigure} 
      \begin{subfigure}[b]{0.23\textwidth}
         \centering
         \includegraphics[width=\textwidth]{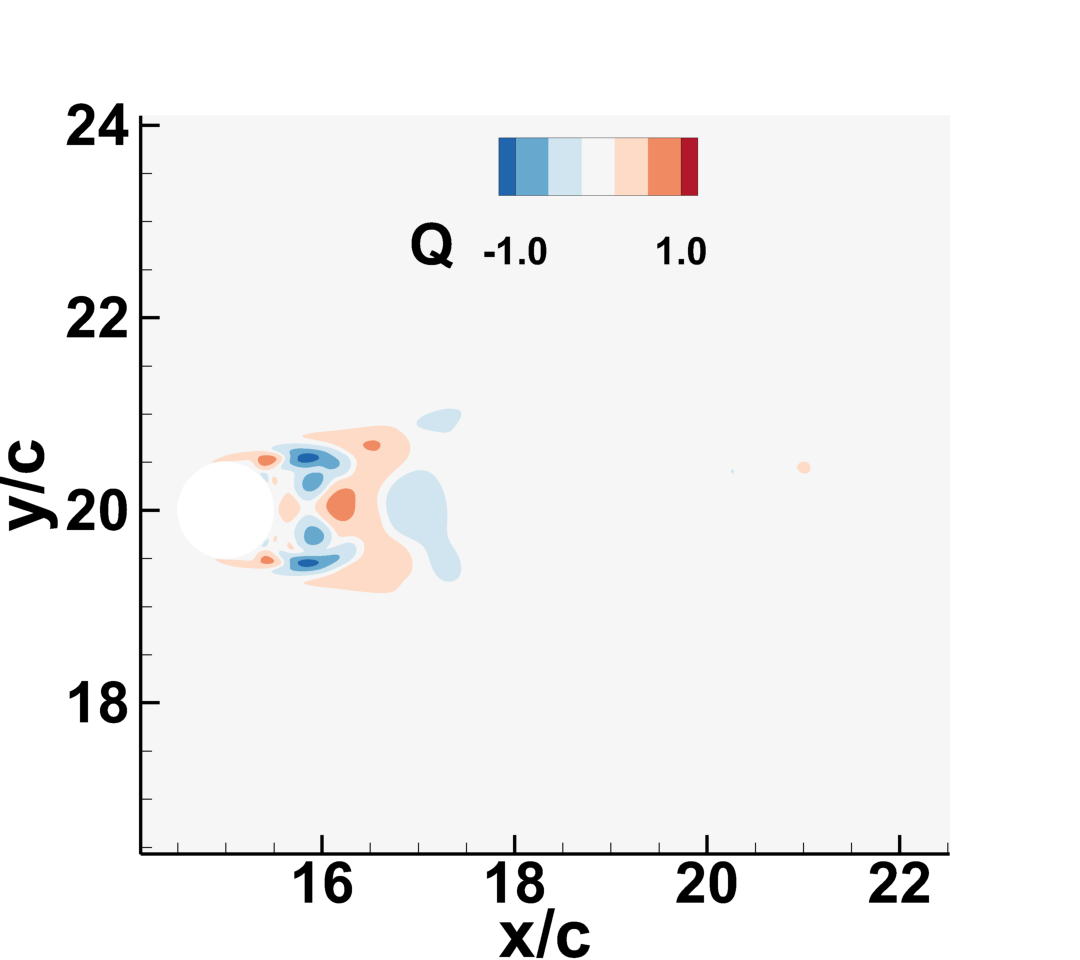}
         \caption{i.e. 2$\hat{Q}_{03}$}
         \label{fig:cc:pod:USigma:j}
     \end{subfigure} 
      \caption{POD applied to the velocity for circular cylinder flow. Spanwise vorticity shown for the spatial eigenvector ($U\Sigma$) corresponding to the (a) Mode-1, (b) Mode-2, (c) Mode-3 and (d) Mode-4. $Q$-field shown for the spatial eigenvectors ($U\Sigma$) corresponding to (e) Mode-1, (f) Mode-2, (g) Mode-3 and (h) Mode-4. $Q$-field corresponding to the interaction between the mean mode and POD modes (i) 0 and (j) 3.}
\label{fig:cc:pod:USigma}
\end{figure}

To investigate the spatial structures of the modal flow fields, we examined here the spatial eigenvector, $U\Sigma$ for each POD mode. The vorticity associated with the first 4 POD modes, which together constitute 99.9\% of the total variation in the flow are plotted in figure \ref{fig:cc:pod:USigma} a-d.  The symmetry properties of the flow about the wake centerline have important implications for the generation of forces on the body. A velocity field that is reflectionally symmetric about the wake centerline will generate drag but no lift, and a reflectionally anti-symmetric velocity field generates lift but no drag. These properties extend to the POD modes as well. In this regard , we note that Mode-1 is strictly symmetric across the wake centerline with alternating vortex structures aligned along the wake centerline. Mode-2 and 3 on the other hand, have a strong anti-symmetric topology in the near wake but transition to a more symmetric configuration in the downstream region. Mode-4 is different from the other modes in that it has a bifurcating topology in the wake and is anti-symmetric in the near as well as the downstream wake regions. However, force partitioning connects the pressure forces to the $Q$ field and it is therefore the symmetry properties of the $Q$ field that ultimately matter. Figure \ref{fig:cc:pod:USigma} (e-h) shows the $Q$-fields corresponding to these modes and we note that $Q_{11}$ is extremely symmetric and other modes ($Q_{22}$, $Q_{33}$ and $Q_{44}$ are also mostly symmetric in the near wake). Thus, we expect that none of these modes will make any significant contributions to the lift, but could make contributions to the drag force. Inter-modal interactions are also important, and we show $Q$ fields for two such modes (0,1) and (0,3). We note that $\hat{Q}_{01}$ is precisely antisymmetric. Inter-modal interactions between symmetric modes will also generate symmetric $Q$ fields, but beyond this, it is not readily apparent how to predict the symmetry properties of the $Q$ field corresponding to other complex inter-modal interactions such as for (0,3).

\begin{figure}
    \centering
     \begin{subfigure}[b]{0.8\textwidth}
         \centering
         \includegraphics[width=\textwidth]{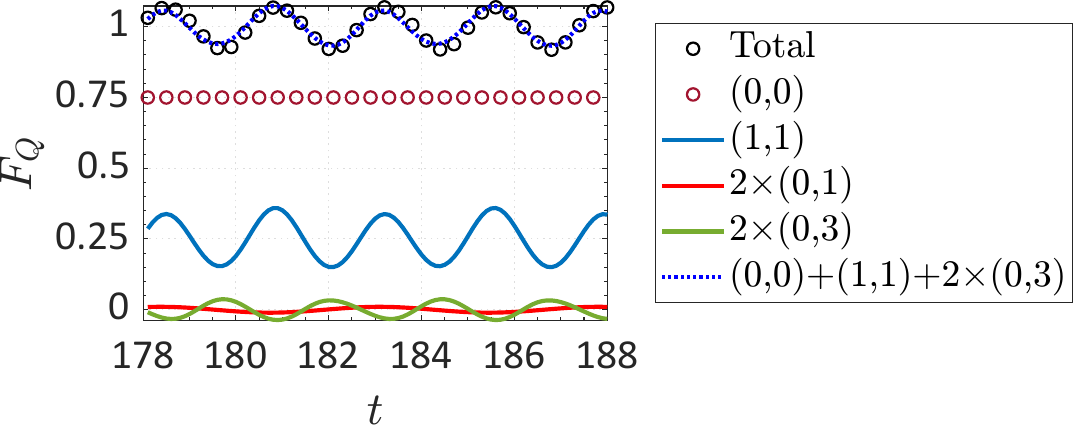}
         \caption{}
         \label{fig:cc:pod:fqvst:a}
     \end{subfigure}
   \hfill
     \begin{subfigure}[b]{0.8\textwidth}
         \centering
         \includegraphics[width=\textwidth]{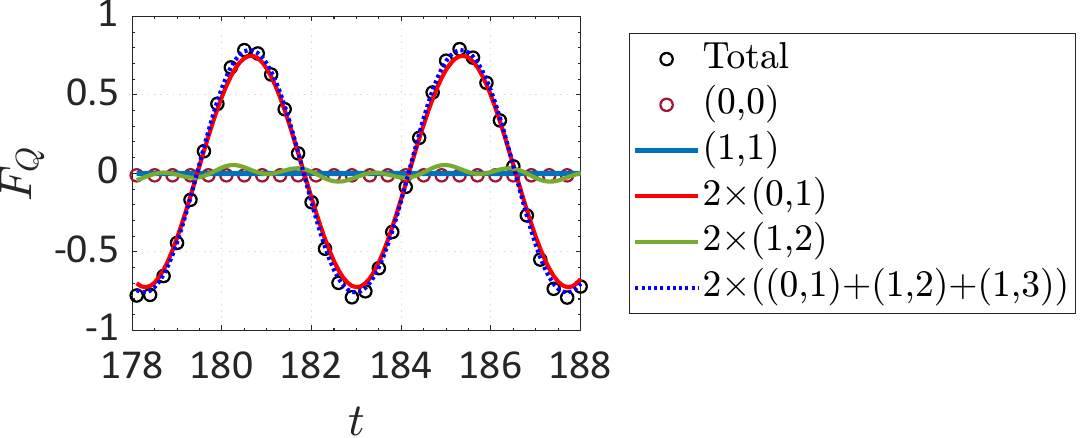}
         \caption{}
         \label{fig:cc:pod:fqvst:b}
     \end{subfigure}
       \hfill
        \caption{ Non-dimensional vortex-induced forces obtained for modes resulting from POD applied to the velocity field (a)   drag force ($F_Q^{(1)}$) and (b) lift force ($F_Q^{(2)}$). The plot shows intra-modal and inter-modal interactions.}
\label{fig:cc:pod:fqvst}
\end{figure}

Figure \ref{fig:cc:pod:fqvst} shows the contribution of the various POD modes, including the contributions of the inter-modal interactions to the pressure drag and lift. We only show those components that make a noticeable contribution to the force in question. The mean drag is connected with the (0,0) mode as expected. The time variation in the drag force is dominated by the symmetric (1,1) mode, but the inter-modal interactions (0,2) and (0,3) also make a small but noticeable contribution to the drag variation. For the lift, the symmetric modes including (0,0), (1,1), (2,2), (3,3) and (4,4) do not make any contribution to the lift. However, ,mode (0,1), which as shown earlier, is purely antisymmetric in $Q$ makes a dominant contribution to the lift force variation. The mode (1,2) makes a very small (barely noticeable) contribution to the lift variation, and other than that, all other contributions are negligible.

\begin{figure}
    \centering
 \begin{subfigure}[b]{0.45\textwidth}
      \centering
         \includegraphics[width=\textwidth]{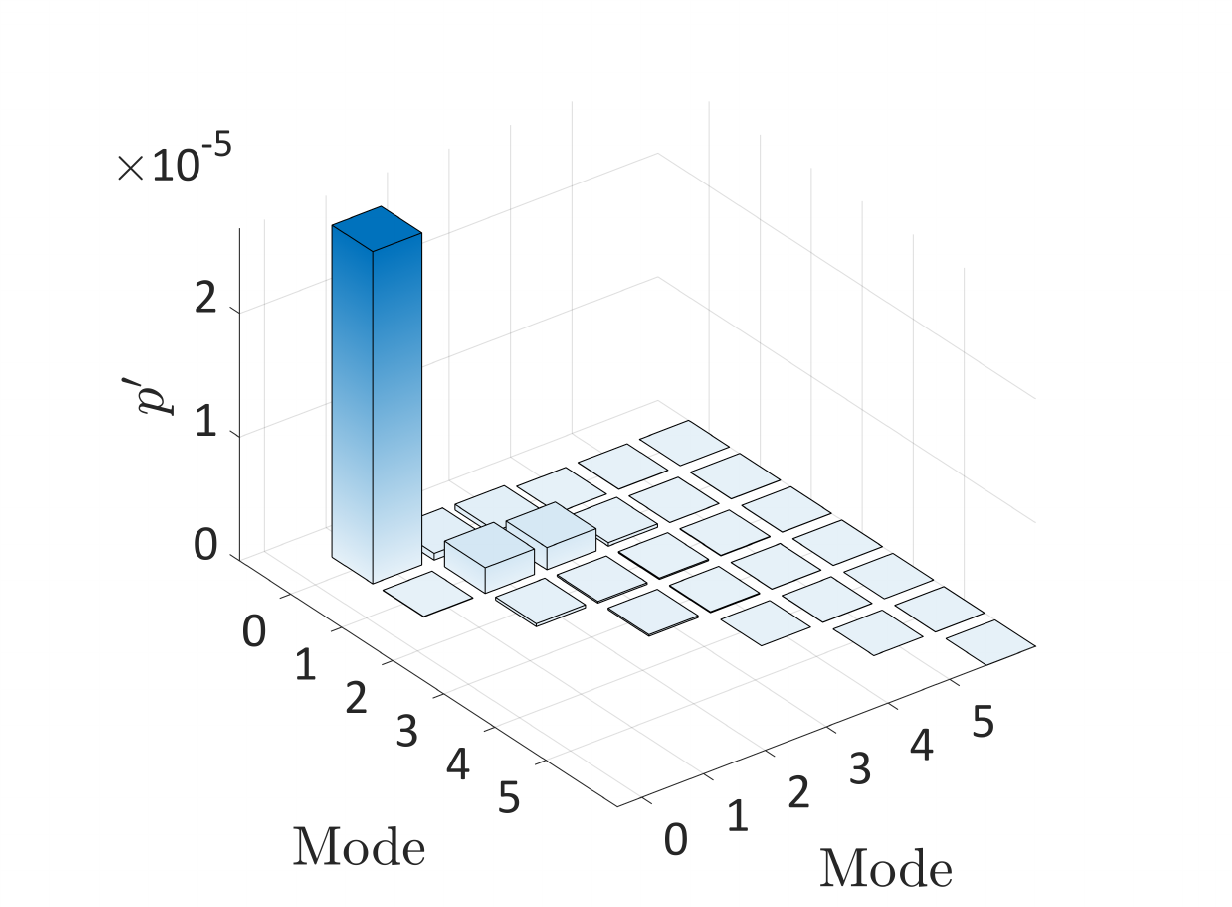}
         \caption{}
         \label{fig:cc:pod:spl:a}
     \end{subfigure}
  \hfill
     \begin{subfigure}[b]{0.5\textwidth}
         \centering
         \includegraphics[width=\textwidth]{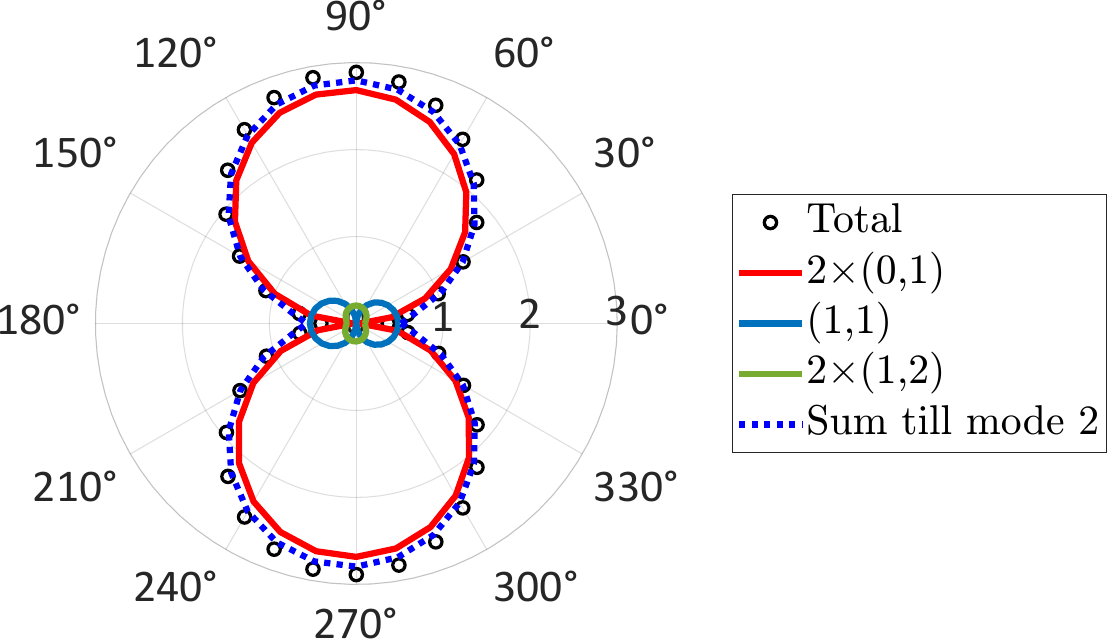}
         \caption{}
         \label{fig:cc:pod:spl:b}
     \end{subfigure}
        \caption{The aeroacoustic noise, calculated at a distance of $50d$ relative to the center of the cylinder and corresponding to a Mach number of 0.1 for the POD of the velocity field of the circular cylinder flow, show (a) the RMS of sound pressure level \oc{at location ($x$=0.0, $y$=-50$d$)} for the first 6 modal interactions and (b) the directivity ($p'_{rms}[\times 10^{-5}]$) shown for the dominant modes and their interactions.}
\label{fig:cc:pod:spl}
\end{figure}

The aerodynamic sound generated by these fluctuating forces is computed by the method described in section \ref{sec:fapm}. The flow Mach number is assumed to be $M=0.1$ and the RMS sound pressure shown in figure \ref{fig:cc:pod:spl:a} is computed at a distance of 50 diameters below the cylinder. The bars show the integrated sound pressure resulting from the intra-modal ((1,1), (2,2), ...), and  inter-modal ((0,1), (1,2), ...) interactions. The largest component of the pressure force fluctuation is that generated in the lift by the interaction of Mode-0 (mean) and Mode-1 and we see that this interaction captures most of the aeroacoustic noise in the form of a dipole oriented in the vertical direction. Mode-1 which generates the vast majority of the drag oscillations, also generates a small contribution as a dipole oriented in the horizontal direction. Very small contributions to noise are also generated by the interaction of Mode-1 with Modes-2 and 3.

We make two observations from these plots. First, Mode-1, the POD mode that captures most of the fluctuation in the velocity associated with the Karman vortex shedding, does not generate by itself the dominant component of the aeroacoustic noise. Rather, it is the inter-modal interaction between Mode-0 and Mode-1 that is responsible for most of the noise. Secondly, as shown from the bar chart of the sound pressure contribution, the appearance of (0,1) and other inter-modal interactions makes it difficult to pinpoint individual modes that are particularly important for noise generation. As will be shown later in the paper, this complexity increases very rapidly with increasing Reynolds numbers, since these flows have a wide range of modes with substantial energy.

\subsubsection{Modal Decomposition of $Q$}
Motivated by the above complication, we propose to examine modal decompositions of the $Q$-field as a way to directly access the features/modes in the flow that are responsible for the generation of pressure forces on the body. It is expected that this will circumvent the need to consider inter-modal interactions and result in a modal decomposition approach that more directly targets pressure forces.
\begin{figure}
    \centering
    \begin{subfigure}[b]{0.24\textwidth}
         \centering
         \includegraphics[width=\textwidth]{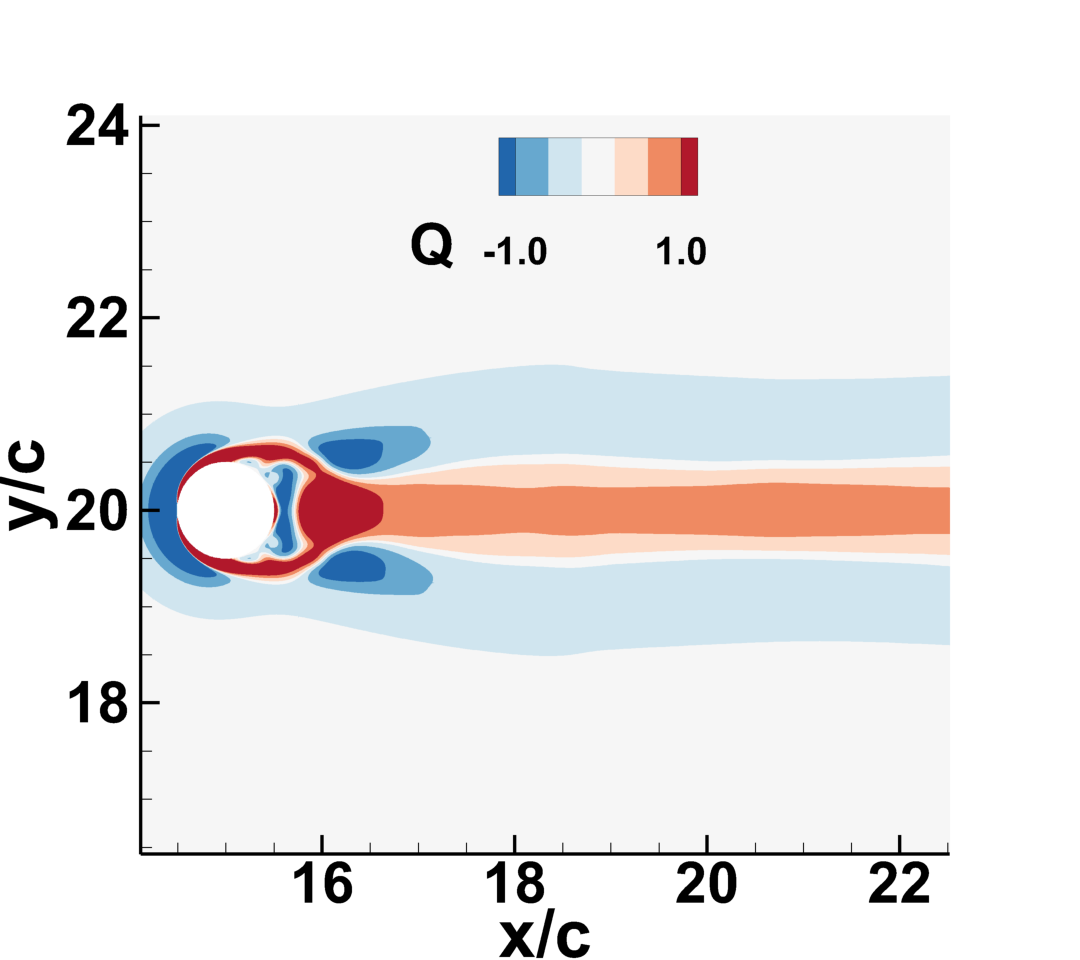}
         \caption{$\tilde{Q}_{0}$}
         \label{fig:cc:podQ:QandfQ:a}
     \end{subfigure}
     \hfill
      \begin{subfigure}[b]{0.24\textwidth}
         \centering
         \includegraphics[width=\textwidth]{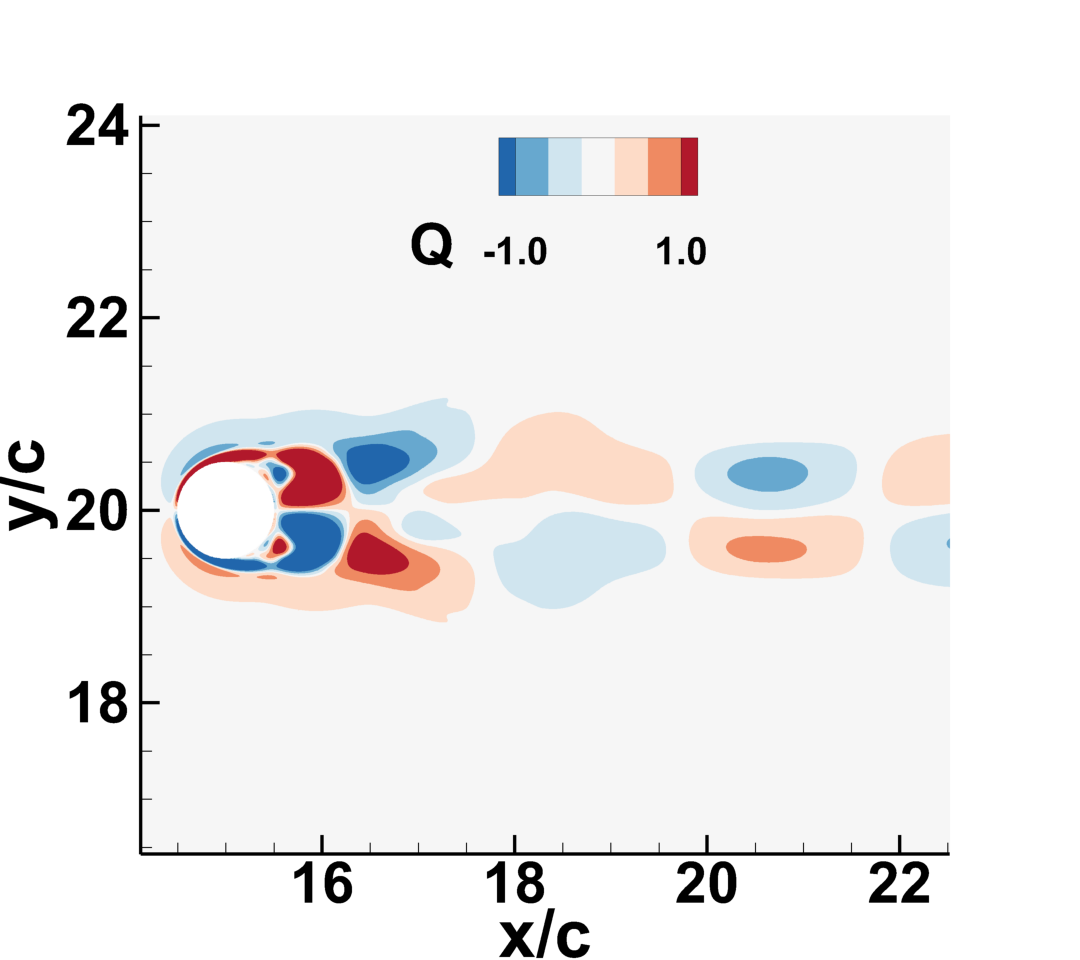}
         \caption{$\tilde{Q}_{1}$}
         \label{fig:cc:podQ:QandfQ:b}
     \end{subfigure}
     \hfill
     \begin{subfigure}[b]{0.24\textwidth}
         \centering
         \includegraphics[width=\textwidth]{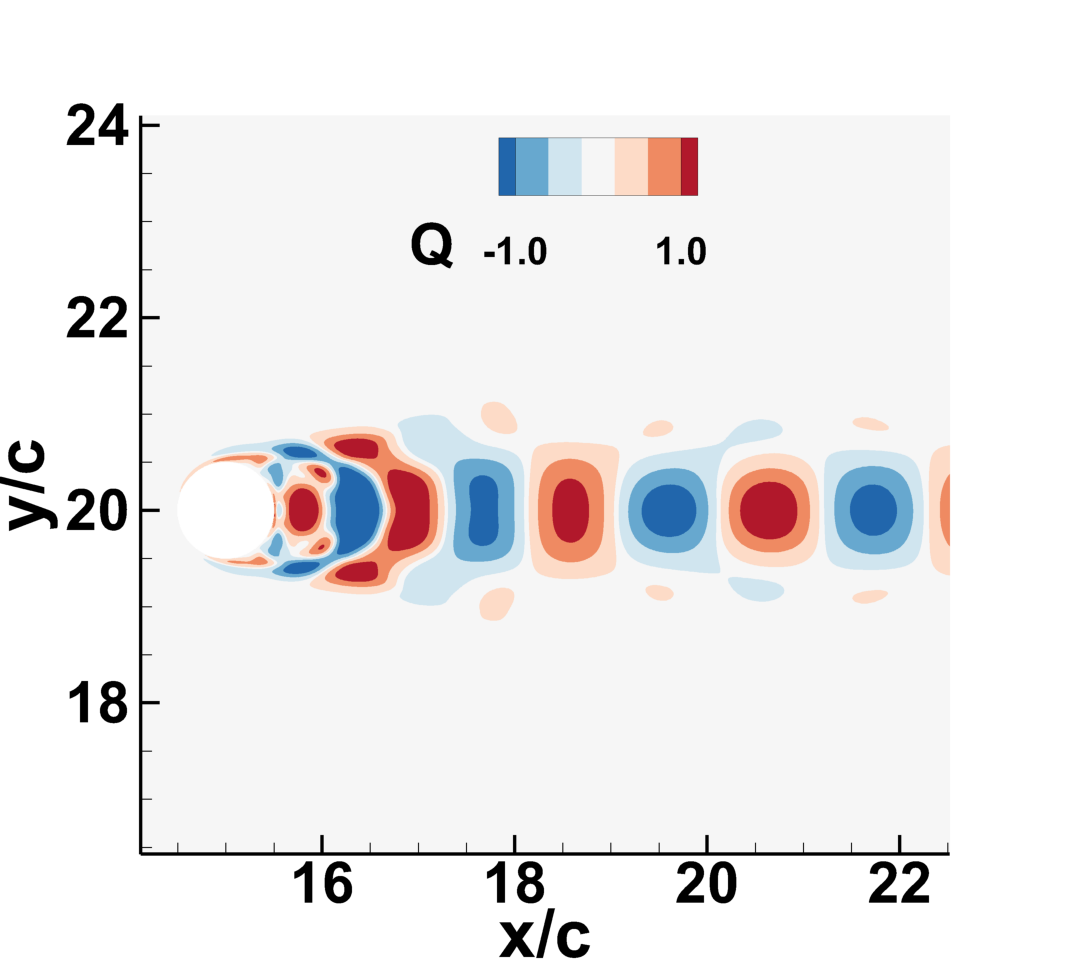}
         \caption{ $\tilde{Q}_{2}$}
         \label{fig:cc:podQ:QandfQ:c}
     \end{subfigure}
     \hfill
     \begin{subfigure}[b]{0.24\textwidth}
         \centering
         \includegraphics[width=\textwidth]{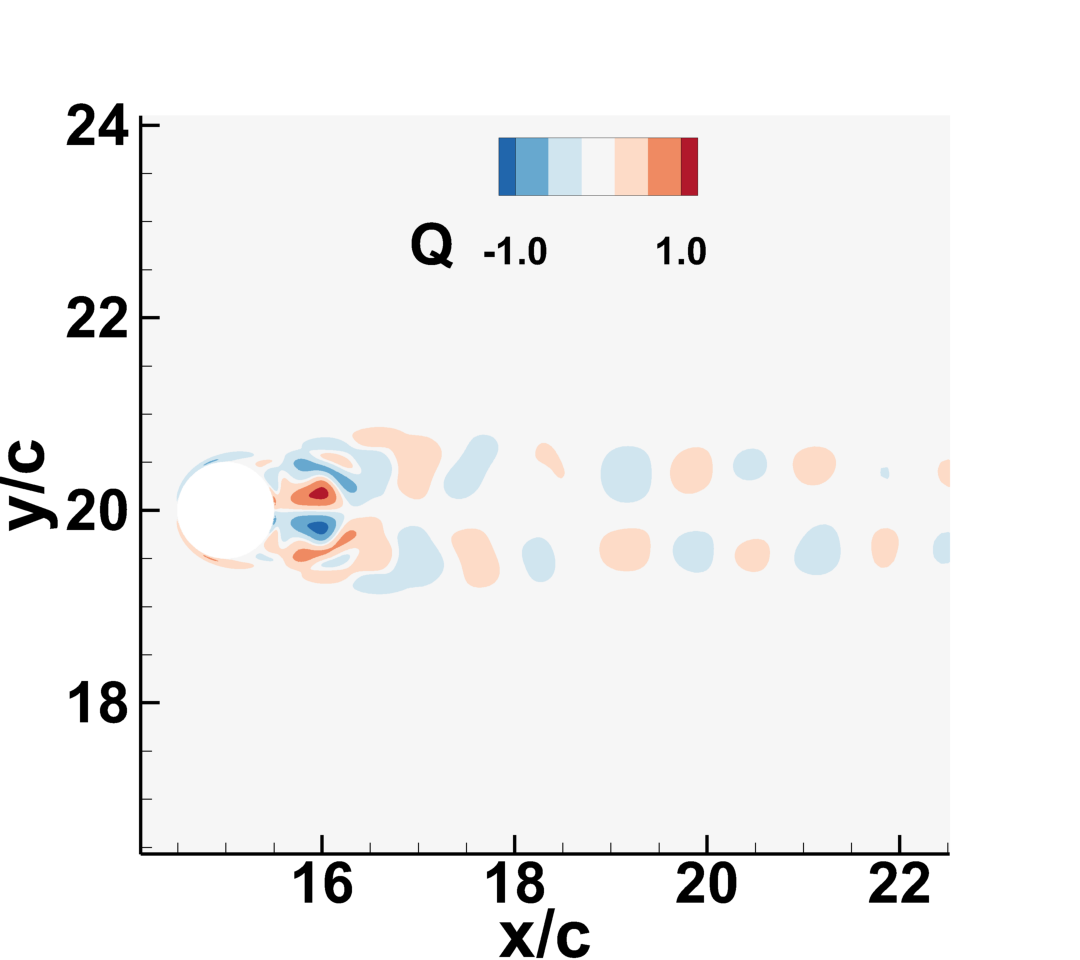}
         \caption{ $\tilde{Q}_{3}$}
         \label{fig:cc:podQ:QandfQ:d}
     \end{subfigure}
\hfill
     \begin{subfigure}[b]{0.24\textwidth}
         \centering
         \includegraphics[width=\textwidth]{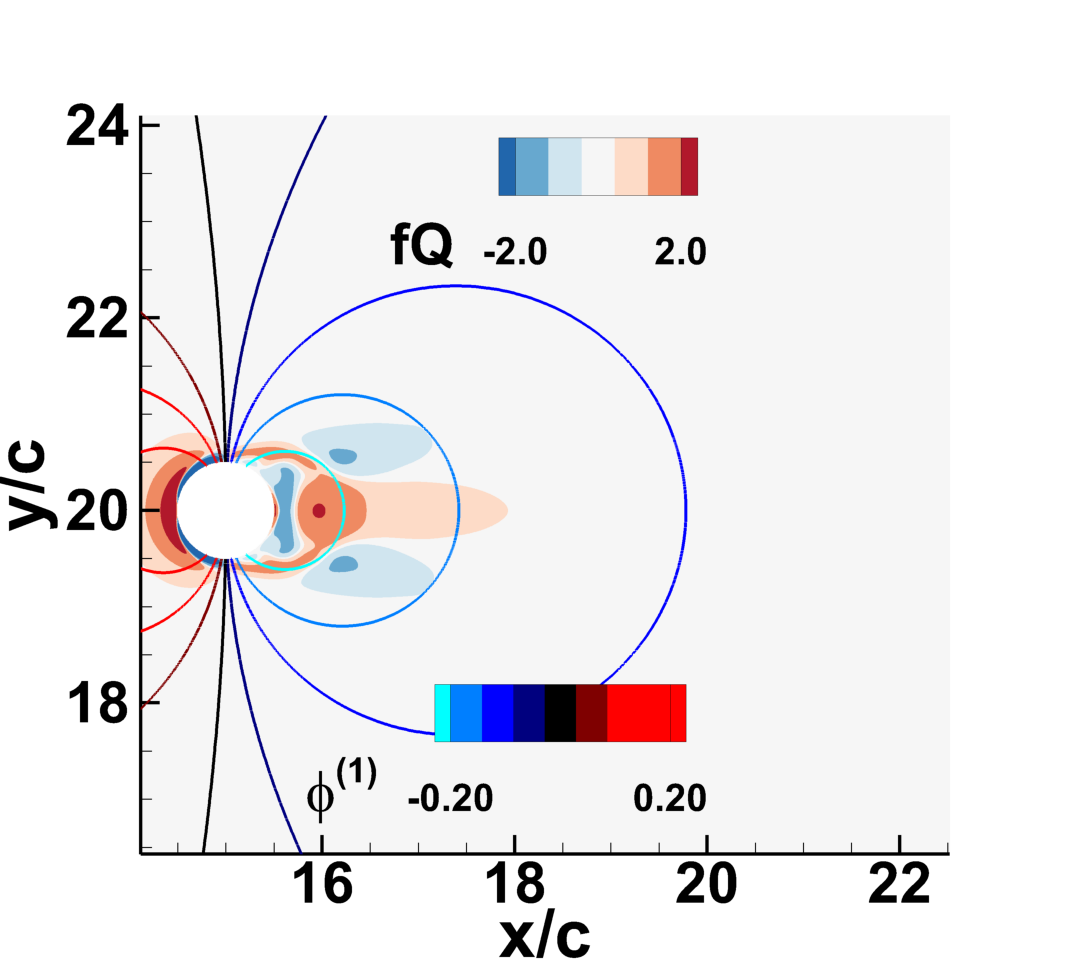}
         \caption{ $f_{\tilde{Q}_0}^{(1)}$}
         \label{fig:cc:podQ:QandfQ:e}
     \end{subfigure}
     \hfill
     \begin{subfigure}[b]{0.24\textwidth}
         \centering
         \includegraphics[width=\textwidth]{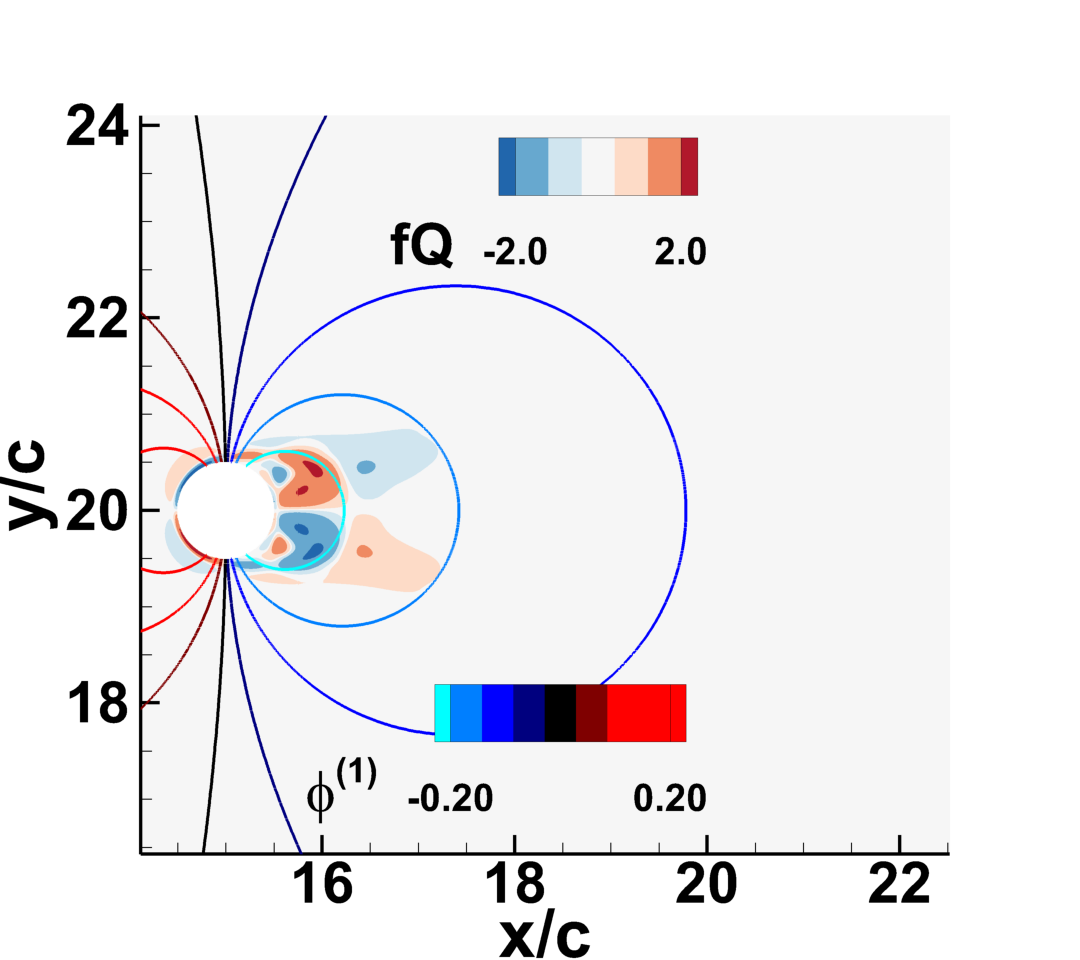}
         \caption{$f_{\tilde{Q}_1}^{(1)}$}
         \label{fig:cc:podQ:QandfQ:f}
     \end{subfigure}
     \hfill
     \begin{subfigure}[b]{0.24\textwidth}
         \centering
         \includegraphics[width=\textwidth]{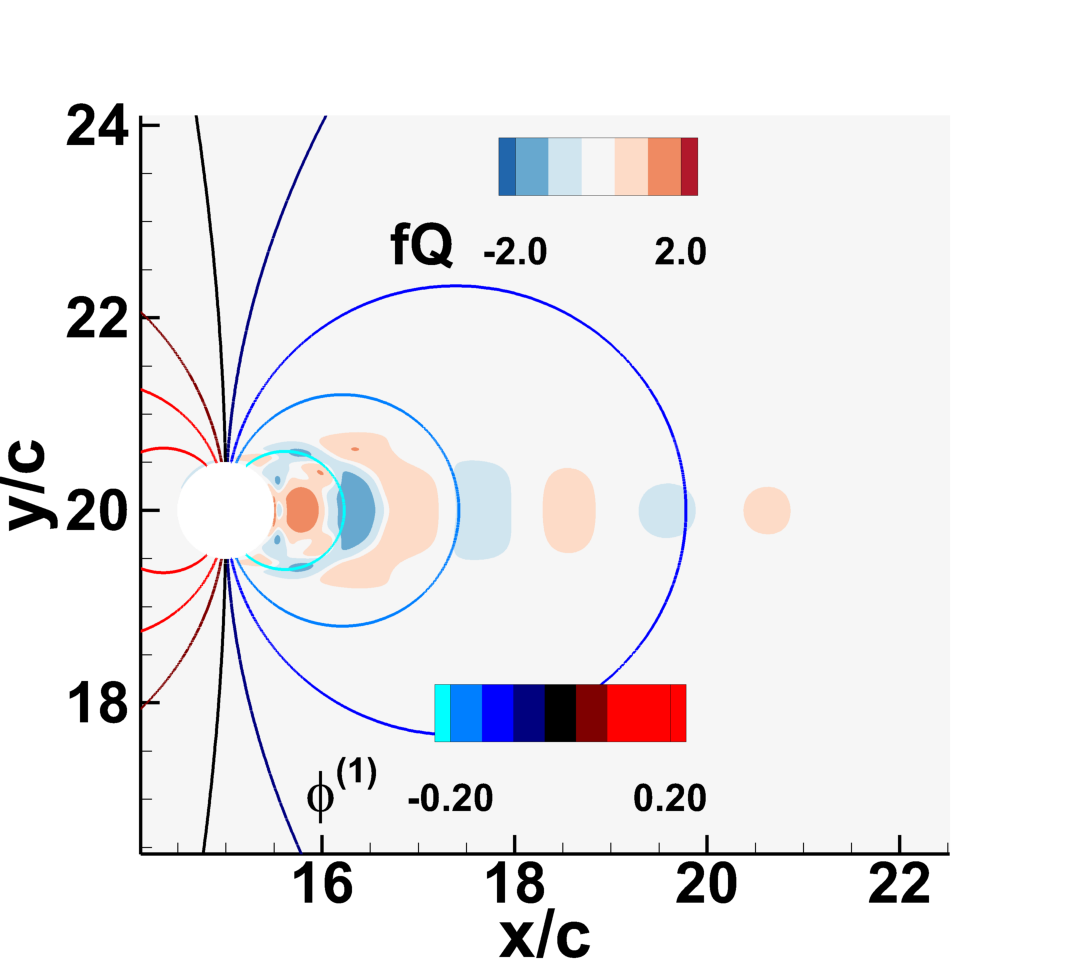}
         \caption{$f_{\tilde{Q}_2}^{(1)}$}
         \label{fig:cc:podQ:QandfQ:g}
     \end{subfigure}
\hfill
     \begin{subfigure}[b]{0.24\textwidth}
         \centering
         \includegraphics[width=\textwidth]{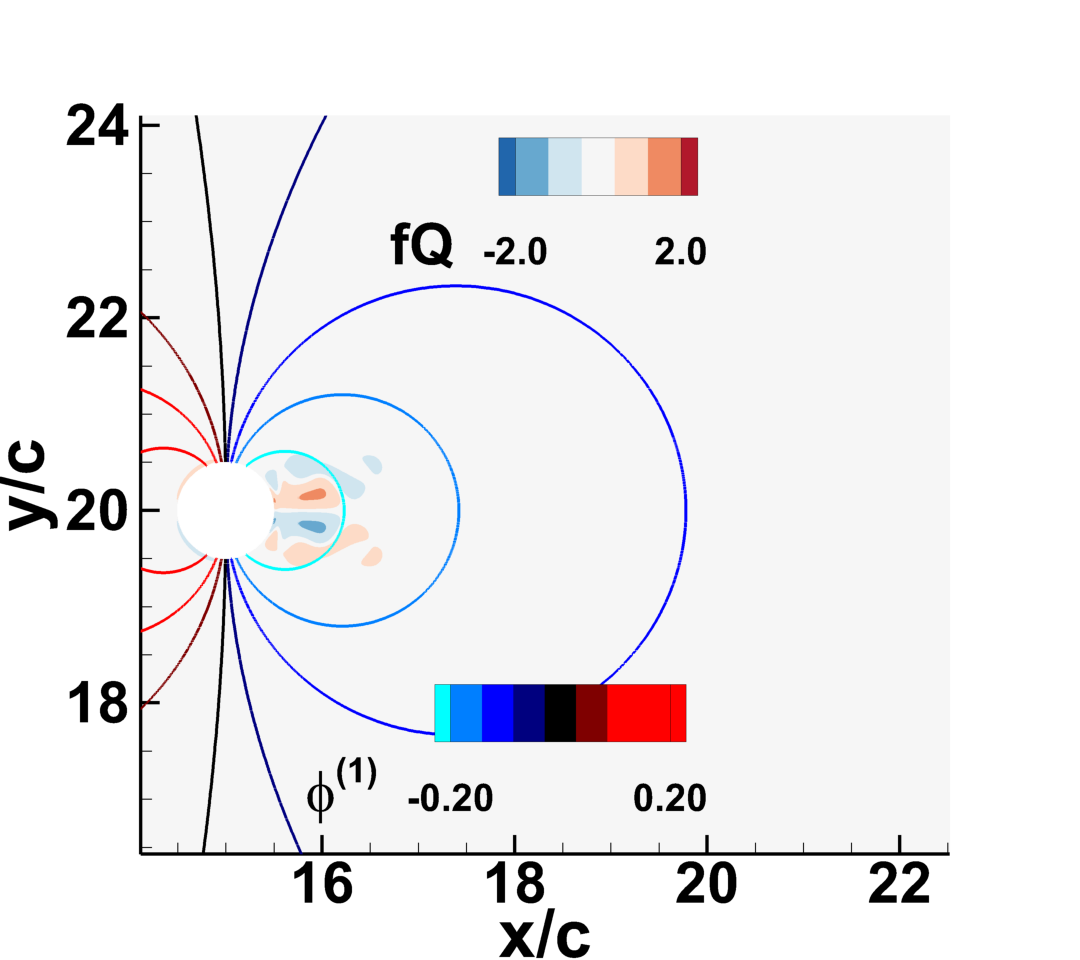}
         \caption{$f_{\tilde{Q}_3}^{(1)}$}
         \label{fig:cc:podQ:QandfQ:h}
     \end{subfigure}
     \begin{subfigure}[b]{0.24\textwidth}
      \centering
         \includegraphics[width=\textwidth]{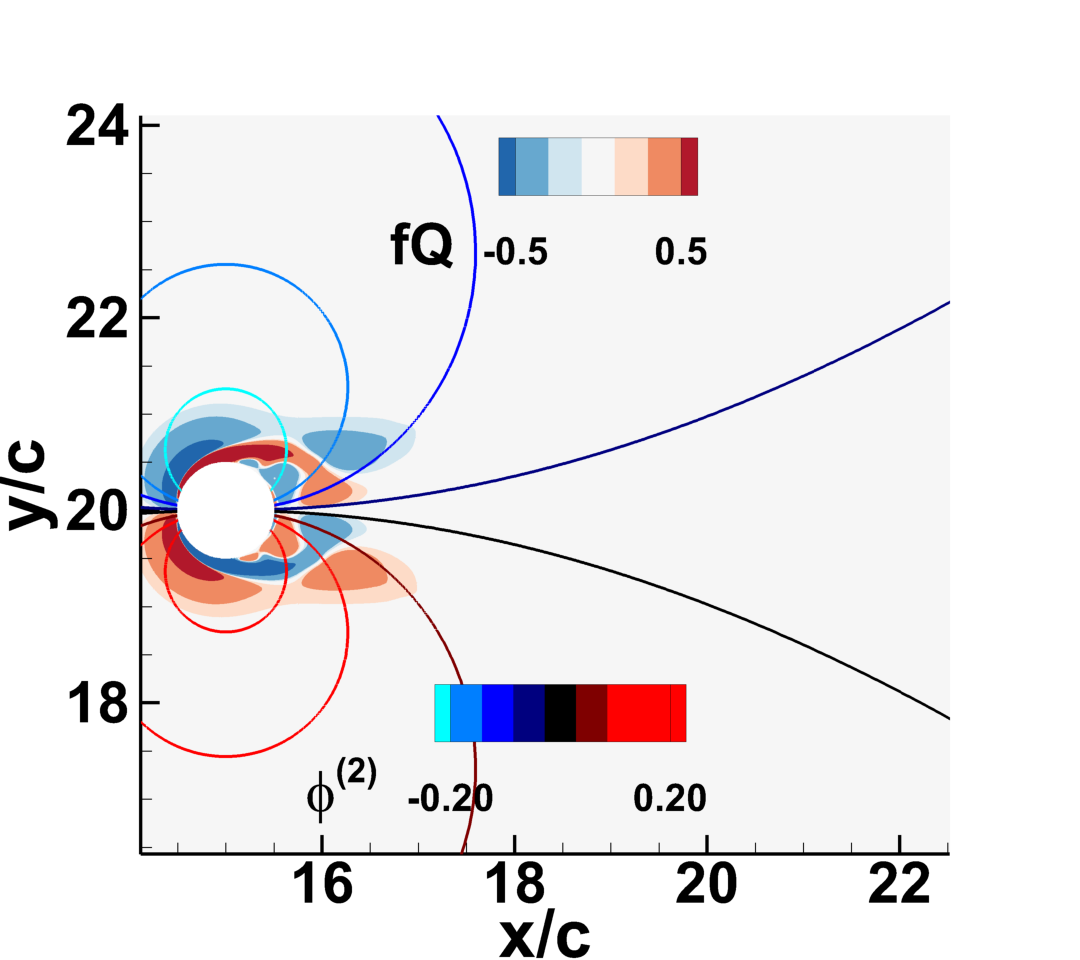}
         \caption{$f_{\tilde{Q}_0}^{(2)}$}
         \label{fig:cc:podQ:QandfQ:i}
     \end{subfigure}
     \hfill
 \begin{subfigure}[b]{0.24\textwidth}
      \centering
         \includegraphics[width=\textwidth]{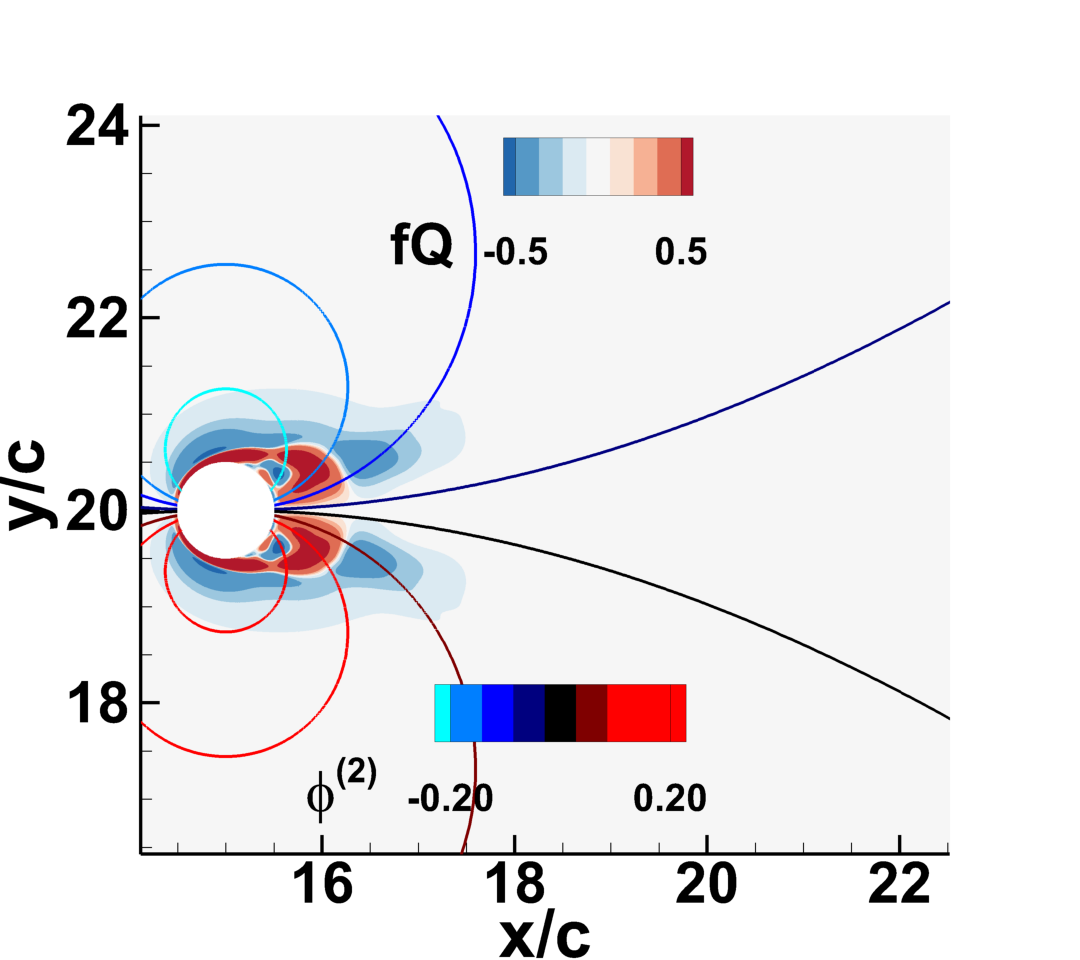}
         \caption{$f_{\tilde{Q}_1}^{(2)}$}
         \label{fig:cc:podQ:QandfQ:j}
     \end{subfigure}
  \hfill
     \begin{subfigure}[b]{0.24\textwidth}
         \centering
         \includegraphics[width=\textwidth]{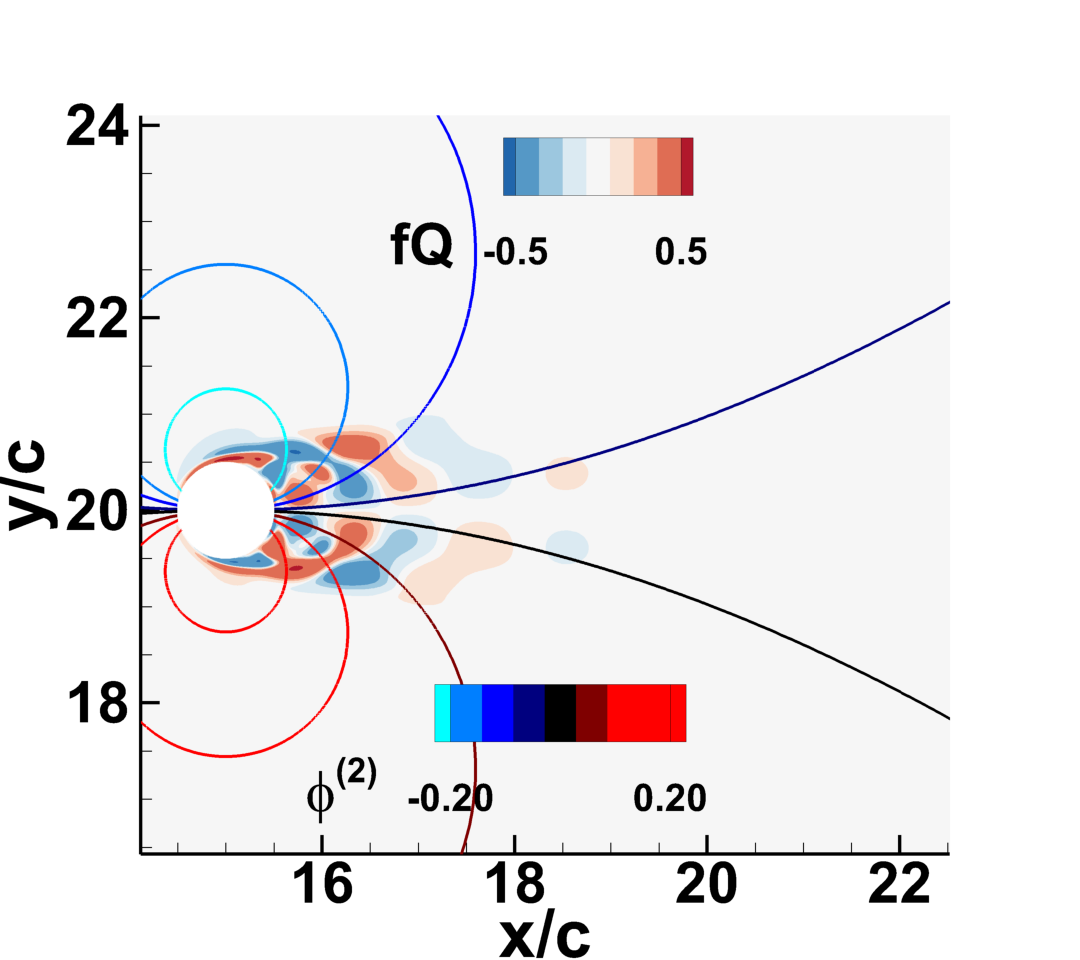}
         \caption{$f_{\tilde{Q}_2}^{(2)}$}
         \label{fig:cc:podQ:QandfQ:k}
     \end{subfigure}
   \hfill
     \begin{subfigure}[b]{0.24\textwidth}
         \centering
         \includegraphics[width=\textwidth]{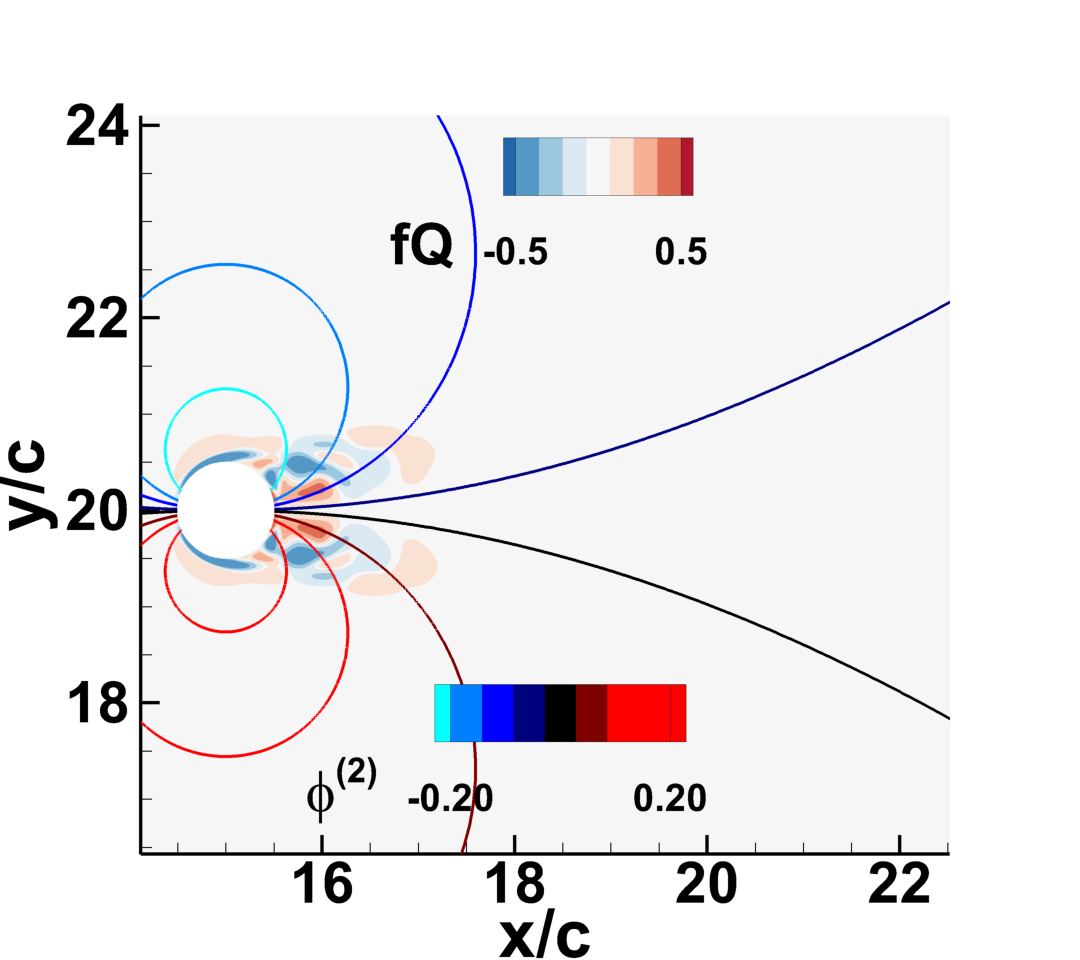}
         \caption{$f_{\tilde{Q}_3}^{(2)}$}
         \label{fig:cc:podQ:QandfQ:l}
     \end{subfigure}
     \hfill
     \caption{Modal force partitioning based on POD of $Q$-field. Contours of $Q$ for (a) the mean mode (i.e. $\tilde{Q}_{0}$), (b) Mode-1 (i.e. $\tilde{Q}_{1}$), (c) Mode-2 (i.e. $\tilde{Q}_{2}$) and (d) Mode-3 (i.e. $\tilde{Q}_{3}$). Contours of vortex-induced drag force density ${f^{(1)}_Q}$ and $\phi^{(1)}$ corresponding to (e) the mean Mode, (f) Mode-1, (g) Mode-2 and (h) Mode-3. Contours of vortex-induced lift force density ${f^{(2)}_Q}$ and $\phi^{(2)}$ corresponding to (i) the mean mode, (j) Mode-1, (k) Mode-2 and (l) Mode-3.}
\label{fig:cc:podQ:QandfQ}
\end{figure}

In this approach $Q$ fields are computed as a function of space and time from the total velocity and then subject to modal analysis. $Q$ would then be represented as a sum of $N$ modes as follows
\begin{equation}
    Q({\bf{x}},t) = \tilde{Q}_0 ({\bf{x}}) + \tilde{Q}_1({\bf{x}},t) +\tilde{Q}_2({\bf{x}},t) +\dots + \tilde{Q}_N({\bf{x}},t)
\end{equation}
where $\tilde{Q}_m$ is the m$^{th}$ mode of the $Q$ field. The force induced by this mode is given by:
\begin{equation}
    F_{\tilde{Q}_m}^{(i)}=-2\rho \int \tilde{Q}_{m}\phi^{(i)} dV \,\,\, ,
    \label{eqn:fq:Q}
\end{equation}
and the corresponding vortex-induced noise is given by,
\begin{equation}
p'_{\tilde{Q}_m}=\frac{{\bf r}}{4\pi r^2}\left[\left(\frac{1}{c} \frac{\partial}{\partial t}+\frac{1}{r}\right)\cdot {\bf{F}}_{\tilde{Q}_m} \right]_{(t-\frac{r}{c})} .
\label{speqn:Q}
\end{equation}

We apply POD to the $Q$-field and the first three POD modes are shown in figure \ref{fig:cc:podQ:QandfQ}. Mode-1 shown in the figure \ref{fig:cc:podQ:QandfQ:a} is antisymmetric about the wake centerline, and the topology of the modes is indicative of the alternative vortex shedding in the near wake and the large oscillation of the lateral velocity in the wake. Mode-2 and 3 are symmetric and anti-symmetric respectively. Mode-2 in particular is comparable in magnitude to Mode-1 and it is indicative of the symmetric fluctuations in streamwise velocity generated by the vortices shed in the wake. Due to these symmetry properties, we expect that while Mode-1 and 3 will contribute to lift only, Mode-2 will contribute only to drag. The corresponding contour plots of $f_Q$ for drag and lift are shown below these plots. The multiplication of $Q$ with these influence fields tends to diminish the influence of distant features. Furthermore, while $\phi^{(1)}$ due to its symmetric nature tends to emphasize the vortex structures in the stagnation and wake regions, anti-symmetric $\phi^{(2)}$ diminishes the influence of the structures in these regions and emphasizes the features above and cylinder.

\begin{figure}
    \centering
     \begin{subfigure}[b]{0.48\textwidth}
         \centering
         \includegraphics[height=0.15\textheight]{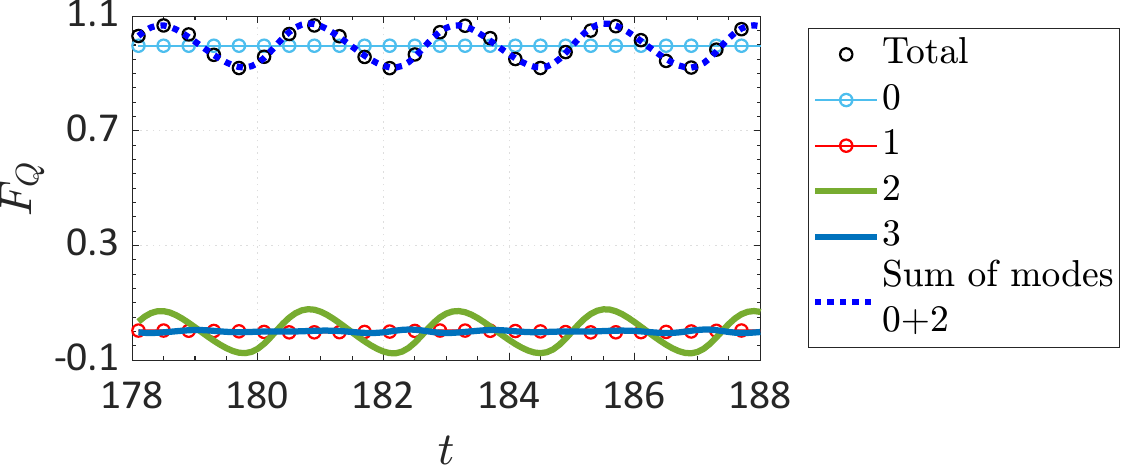}
         \caption{}
         \label{fig:cc:podQ:fqvst:a}
     \end{subfigure}
      \hfill
     \begin{subfigure}[b]{0.48\textwidth}
         \centering
         \includegraphics[height=0.15\textheight]{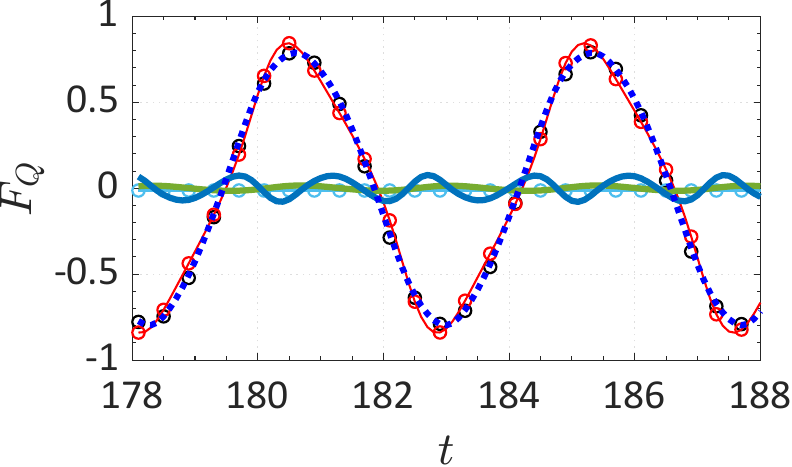}
         \caption{}
         \label{fig:cc:podQ:fqvst:c}
     \end{subfigure}
        \caption{Temporal variation of the non-dimensional vortex-induced (a) drag force ($F_Q^{(1)}$) and (b) lift force ($F_Q^{(2)}$) for the dominant POD modes obtained from application of POD applied to the $Q$-field of the circular cylinder flow.}
\label{fig:cc:podQ:fqvst}
\end{figure}
Figures \ref{fig:cc:podQ:fqvst:a} and \ref{fig:cc:podQ:fqvst:c} show plots of vortex-induced drag and lift force vs. time for these modes, respectively. With respect to drag, we note that Mode-0 provides the mean drag and Mode-2 provides most of the time variation in this quantity. Modes 1 and 3, which are antisymmetric modes, do not provide any contribution to drag. With regard to lift, Mode-1 and 3 provide almost all of the time variation in this quantity, with Mode-1 having a high amplitude and Mode-3 having a much lower amplitude and higher frequency (by a factor of 3) compared to Mode-1. The other modes do not provide any significant contribution to drag. 

\begin{figure}
    \centering
     \begin{subfigure}[b]{0.4\textwidth}
         \centering
         \includegraphics[width=\textwidth]{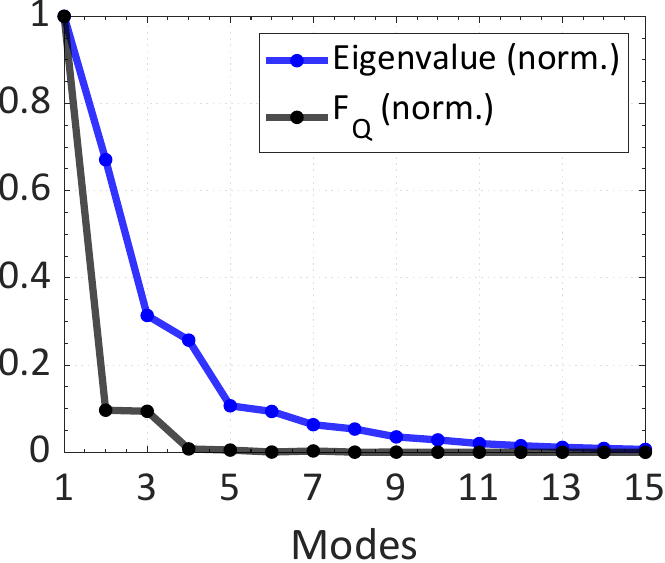}
         \caption{total force}
         \label{fig:cc:podQ:dir:a}
     \end{subfigure}
      \begin{subfigure}[b]{0.5\textwidth}
         \centering
         \includegraphics[width=\textwidth]{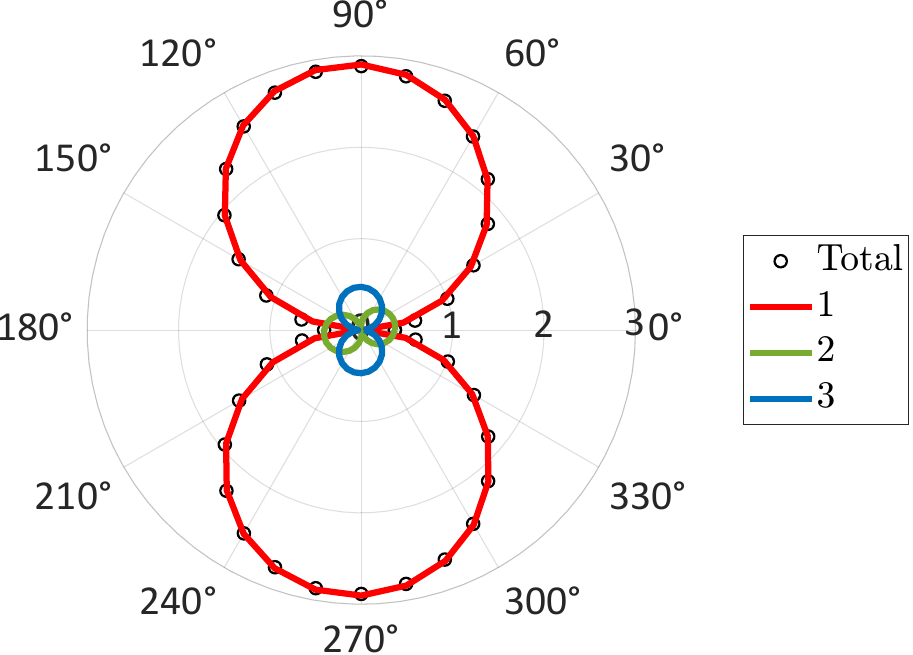}
         \caption{directivity}
         \label{fig:cc:podQ:dir:b}
     \end{subfigure}
        \caption{Application of POD applied to the $Q$-field for the circular cylinder showing (a) normalized eigenvalues (with 12 modes required to reconstruct 98\% of the Q-field) and vortex-induced total force ($\sqrt{(F_Q^{(1)})^2+(F_Q^{(2)})^2}$) and (b) the sound directivity ($p'_{rms}[\times 10^{-5}]$). The values were calculated at a distance of 50$d$ away and correspond to surface Mach of 0.1. }
\label{fig:cc:podQ:dir}
\end{figure}

Figure \ref{fig:cc:podQ:dir:a} shows the normalized and cumulative values (also normalized by the sum) of the eigenvalues for the POD modes of $Q$. The vortex-induced force is calculated for each of the POD modes of the $Q$-field and the corresponding values normalized by the vortex-induced lift force value of the first mode are shown using the solid black line. We see from the peaks that the important modes related to the lift force are Modes-1 and 3. The dashed black line shows the cumulative value of the vortex-induced lift force as more modes are added, and this value is normalized by the total vortex-induced lift force. 

Figure \ref{fig:cc:podQ:dir:b} shows the directivity plots of the aeroacoustic sound associated with these modes. We find that Mode-1 generates the vast majority of the sound, and this in the form of a vertically oriented dipole. Mode-2 and 3 provide much lower but similar levels of overall sound intensities, but while Mode-2 sound is a dipole directed in the horizontal direction, Mode-3 sound is a dipole in the vertical direction.

The above discussion shows that a direct decomposition of $Q$ generates a simpler description of the influence of the decomposed modes on the pressure forces and the induced sounds. This is primarily due to the elimination of complex inter-modal interactions that are generated in the $Q$-field when the modal decomposition is based on the velocity field. We also note that for this simple circular cylinder case, the modes obtained from the decomposition of $Q$ exhibit useful symmetries (as in figure \ref{fig:cc:podQ:QandfQ}) that are connected with their influence on the vector pressure forces induced on the body. Based on this, it is clear that the application of modal force partitioning is particularly useful when paired with a direct decomposition of $Q$ and we focus primarily on this approach for the remaining cases in this paper.

\oc{We also simulated and decomposed the $Q$-field of the circular cylinder at Re=150 keeping other parameters consistent with the Re=300 case. We observed that the modes of the resulting decomposition at this lower Reynolds number exhibit a noise directivity pattern similar to the Re=300 case shown in the figure \ref{fig:cc:podQ:dir:b}. Specifically, Modes-1 and 3 are vertically oriented dipoles while the Mode-2 represents a horizontally oriented dipole. Thus, the behavior is similar despite a factor of two difference in the Reynolds number} 


\subsection{Flow past an airfoil at Re=2500}
\label{sec:airfoil}

Next, a NACA 0015 airfoil at a high angle-of-attack (20$^\circ$) and higher Reynolds number (Re$_c=2500$, based on the chord ($c$) and the incoming flow velocity ($U_\infty$)) is considered, and this provides a greater level of complexity than the low Reynolds number circular cylinder flow. The instantaneous vorticity field for this case is plotted in figure \ref{fig:af:td:vel:vor:a} which shows shedding of vortices from the leading as well as trailing-edges. As will be shown shortly, this flow not only has a dominant cyclic component but also exhibits significant cycle-to-cycle variations. Thus, this flow is a good candidate to demonstrate the application of modal force partitioning to a triple-decomposition of a flow. 

\subsubsection{Triple decomposition of the velocity field}
\begin{figure}
    \centering
 \begin{subfigure}[b]{0.24\textwidth}
      \centering
         \includegraphics[width=\textwidth]{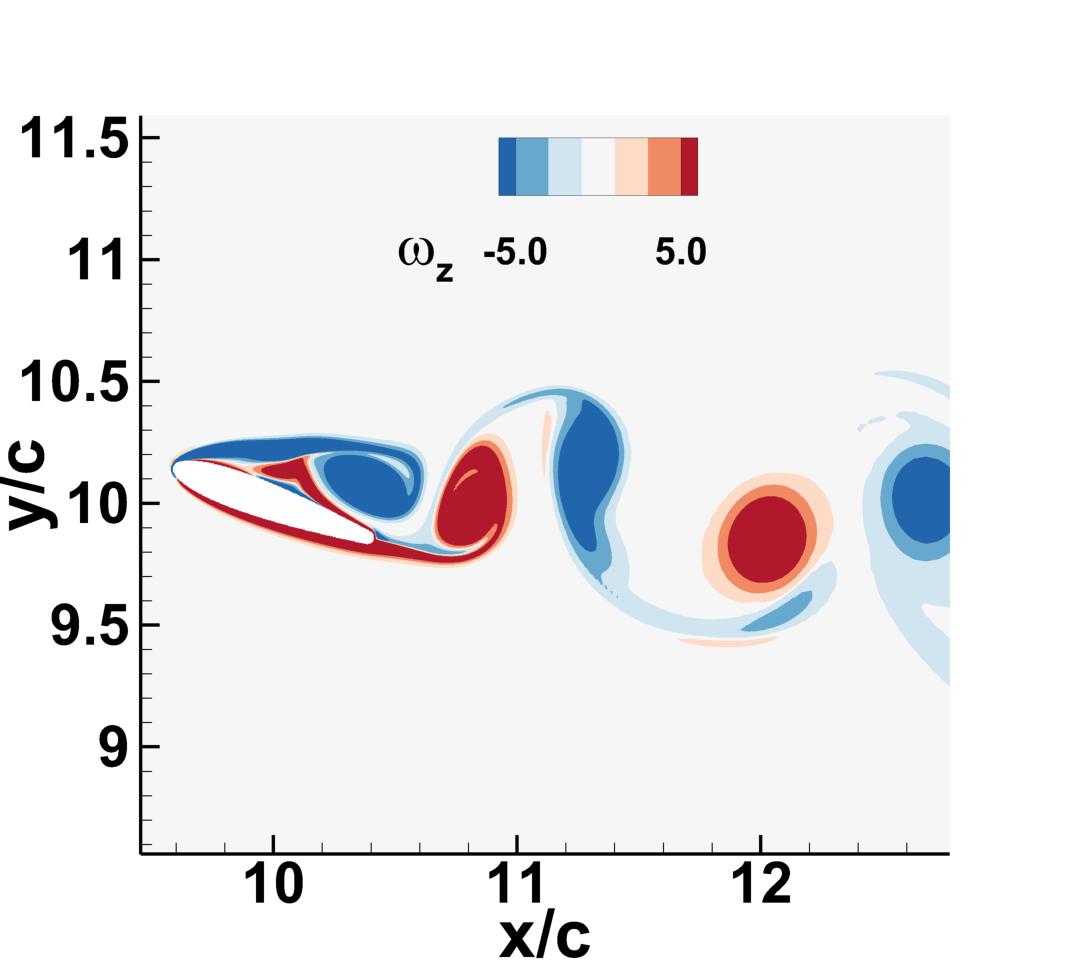}
         \caption{$\omega_z(\bf{u})$}
         \label{fig:af:td:vel:vor:a}
     \end{subfigure}
  \hfill
     \begin{subfigure}[b]{0.24\textwidth}
         \centering
         \includegraphics[width=\textwidth]{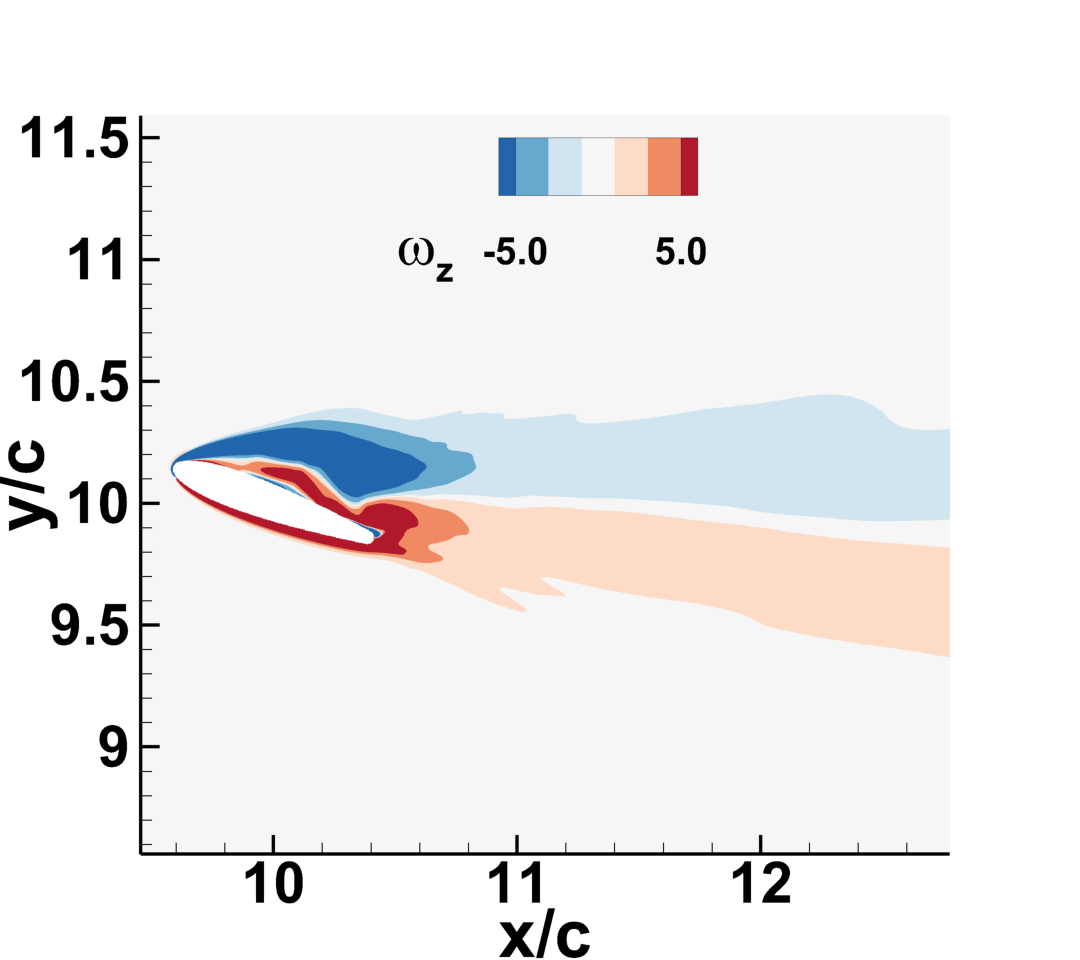}
         \caption{$\omega_z(\bf{u_0})$}
         \label{fig:af:td:vel:vor:b}
     \end{subfigure}
   \hfill
     \begin{subfigure}[b]{0.24\textwidth}
         \centering
         \includegraphics[width=\textwidth]{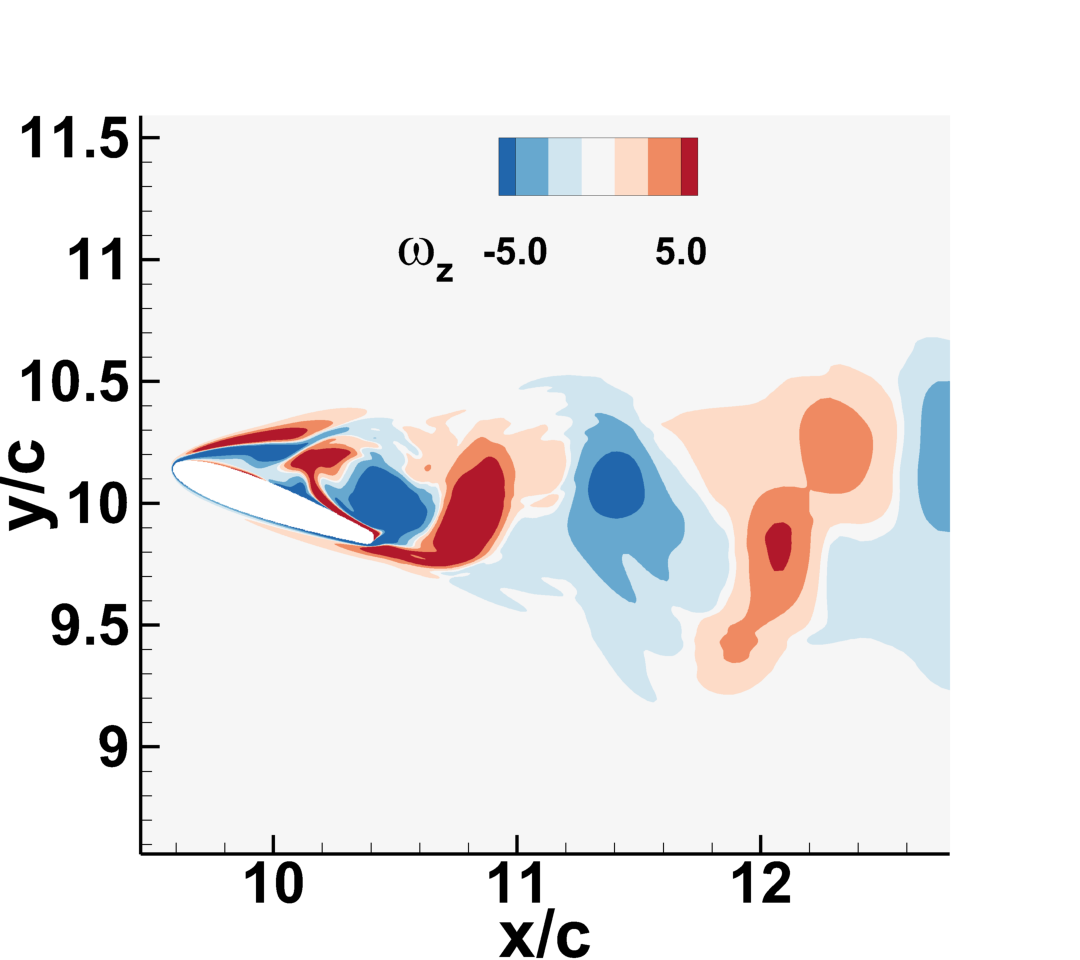}
         \caption{$\omega_z(\bf{u_1})$}
         \label{fig:af:td:vel:vor:c}
     \end{subfigure}
     \hfill
     \begin{subfigure}[b]{0.24\textwidth}
         \centering
         \includegraphics[width=\textwidth]{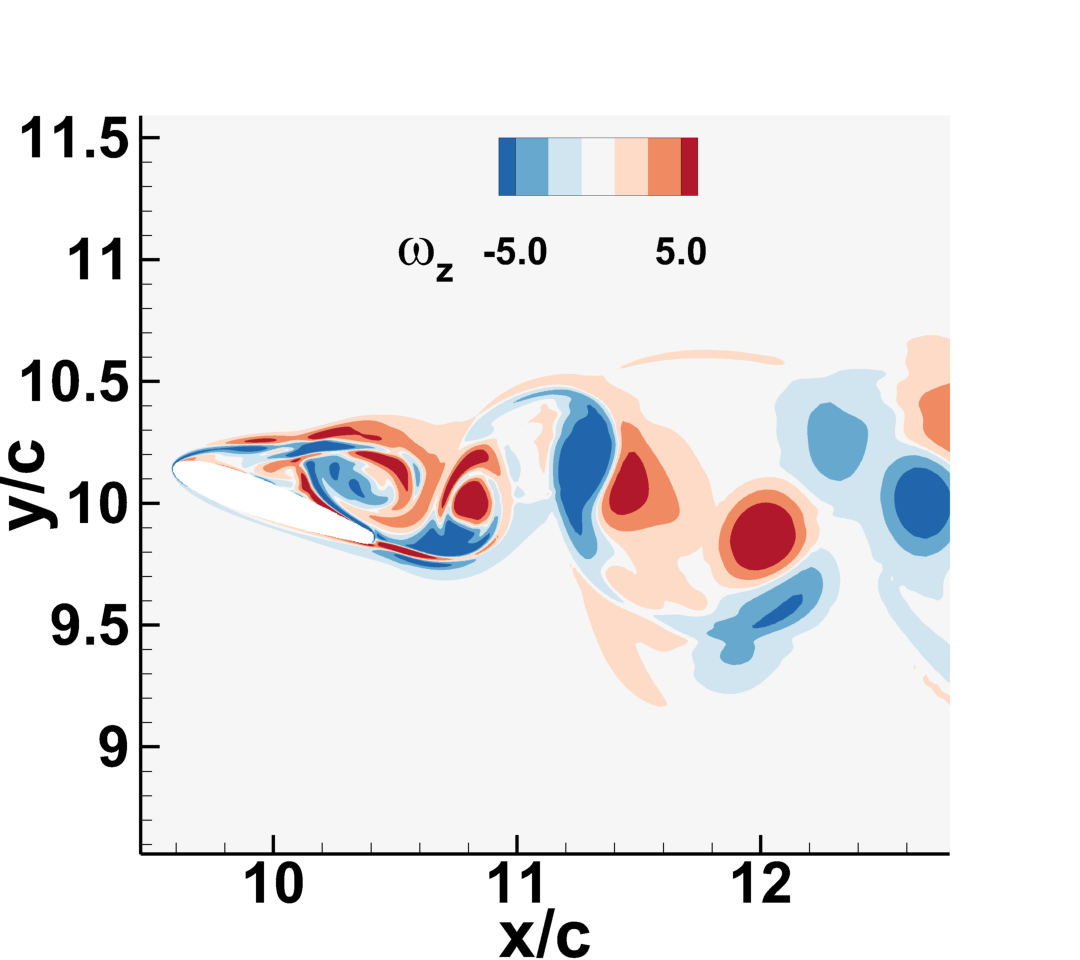}
         \caption{$\omega_z(\bf{u_2})$}
         \label{fig:af:td:vel:vor:d}
     \end{subfigure}
        \caption{Spanwise vorticity associated with the modes associated with the triple decomposition of the velocity field for the airfoil flow showing (a) a snapshot of the flow field before decomposition, (b) the mean mode ($\bf{u_0}$), (c) the coherent mode ($\bf{u_1}$) and (d) the incoherent mode ($\bf{u_2}$).}
\label{fig:af:td:vel:vor}
\end{figure}

Triple decomposition \cite{hussain_mechanics_1970} is used to partition the fluctuating component into coherent ($\bf{u}_1$) and incoherent ($\bf{u}_2$) components with the flow field interval chosen such that it contains 9 cycles of the coherent wave with 32 snapshots in each cycle. The flow structures generated by these components via the decomposition of the velocity field are shown in Figure \ref{fig:af:td:vel:vor} and we observe that the coherent component captures the separation of the leading-edge vortex (LEV) and the Karman-like vortex shedding in the wake. The incoherent component captures variations along the periphery of the primary vortices (the LEV and the wake vortices), since these regions are more significantly affected by the chaotic nature of the flow than the cores of the vortices.  The vortices seen in the coherent mode are also arranged sequentially in series of counterclockwise vorticity originating from trailing-edge and clockwise vorticity originating from the leading-edge while the vortices shed in incoherent mode appear in thin sheet and in vortex pairs. 

\begin{figure}
    \centering
 \begin{subfigure}[b]{0.4\textwidth}
      \centering
         \includegraphics[width=\textwidth]{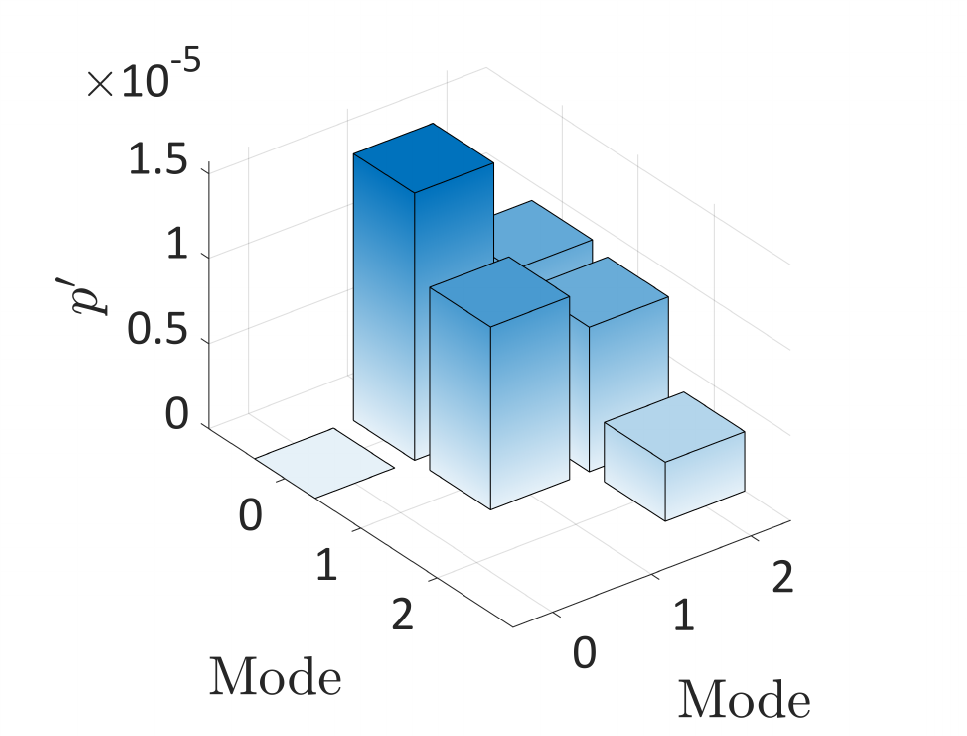}
         \caption{}
         \label{fig:af:tripD:vel:dir:a}
     \end{subfigure}
  \hfill
     \begin{subfigure}[b]{0.5\textwidth}
         \centering
         \includegraphics[width=\textwidth]{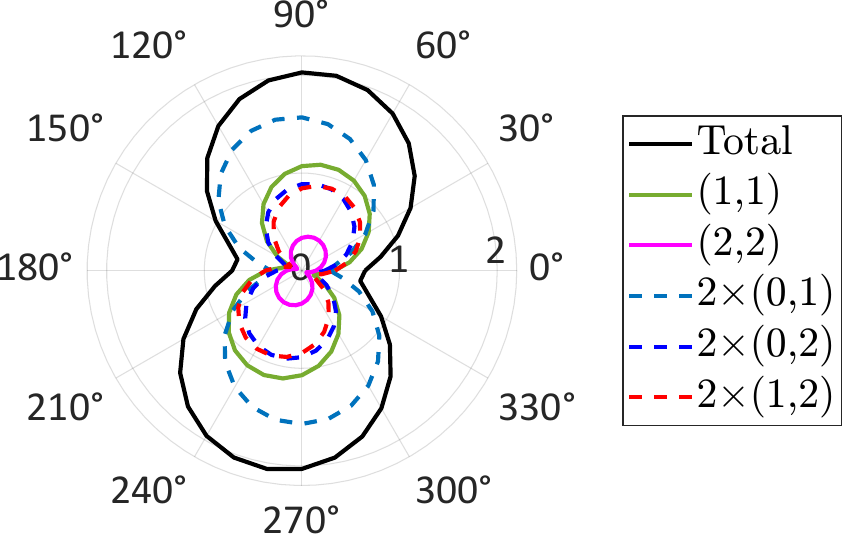}
         \caption{}
         \label{fig:af:tripD:vel:dir:b}
     \end{subfigure}
        \caption{(a) The RMS value of sound pressure calculated at 270$^\circ$ and associated with the intra-modal and inter-modal interactions and (b) the corresponding directivity ($p'_{rms}[\times 10^{-5}]$) pattern.}
\label{fig:af:tripD:vel:dir}
\end{figure}

The contributions of the decomposed flow modes to the aerodynamic sound are obtained by applying the FAPM as described above. The flow Mach number is set to $M=0.1$ and the sound pressure is evaluated at a distance of $57c$ from the airfoil. The RMS value of sound pressure at a location directly above the airfoil at a distance of $57c$ is shown in figure \ref{fig:af:tripD:vel:dir:a}. As we have seen for the circular cylinder case, the aerodynamic sound is characterized by the intra-modal as well as inter-modal interactions. We find that the interaction between the mean and coherent modes (0,1) generates the strongest dipole sound, but other inter-modal interactions also contribute substantially to the radiated sound. The directivity patterns for the sound generated by modal interactions at a distance of $57c$ are plotted in figure \ref{fig:af:tripD:vel:dir:b}, and these also show that the directivities of the inter-modal interactions can be quite distinct. 

This result reaffirms the observation made for the POD of the cylinder case that if the modal decomposition is applied to the velocity field, it is not easy to identify the dominant noise modes and the associated flow structures, since they are entangled by inter-modal interactions. This is even more apparent for this airfoil flow at a higher Reynolds number, since this flow generates substantial energy in the higher mode, resulting in more substantial inter-modal interactions.

\subsubsection{Triple decomposition of the $Q$-field}
\begin{figure}
    \centering
 \begin{subfigure}[b]{0.24\textwidth}
      \centering
         \includegraphics[width=\textwidth]{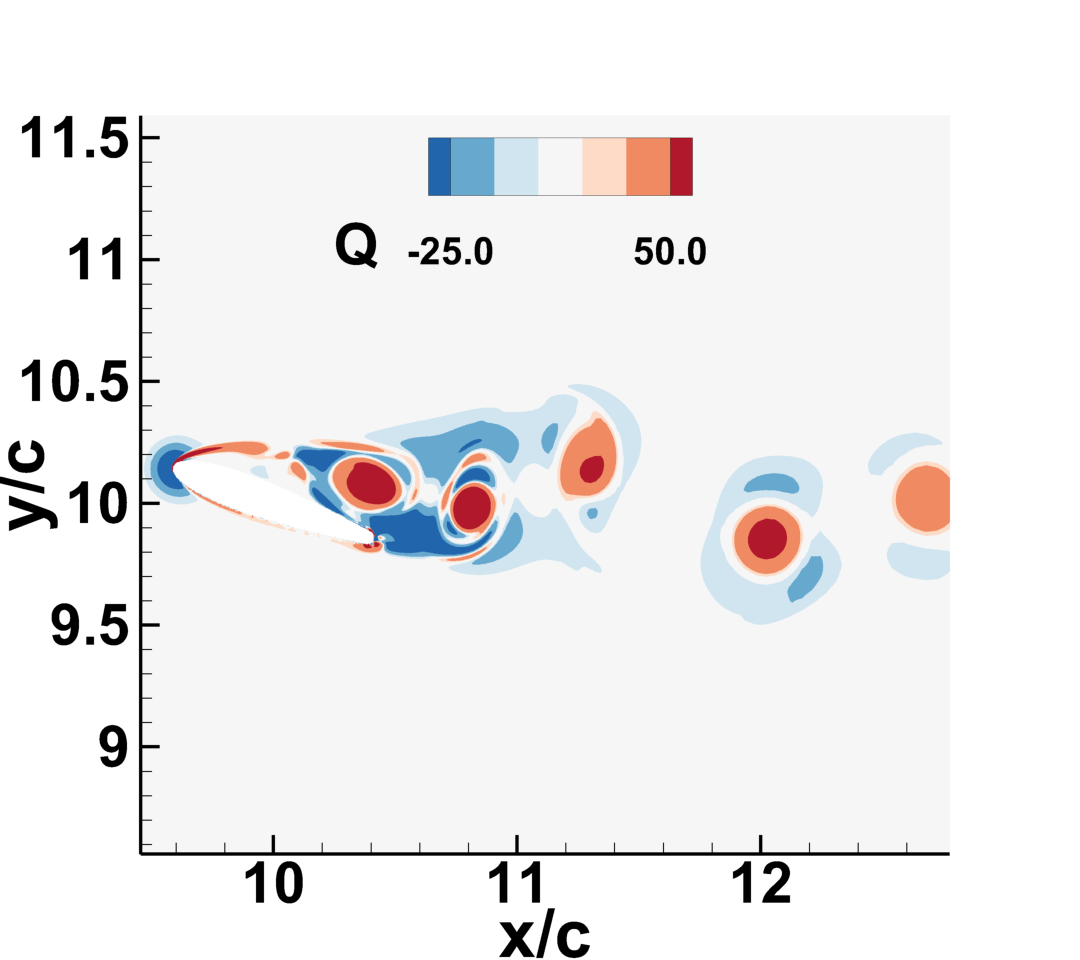}
         \caption{$Q$}
         \label{fig:af:td:flowfq:a}
     \end{subfigure}
  \hfill
  \begin{subfigure}[b]{0.24\textwidth}
      \centering
         \includegraphics[width=\textwidth]{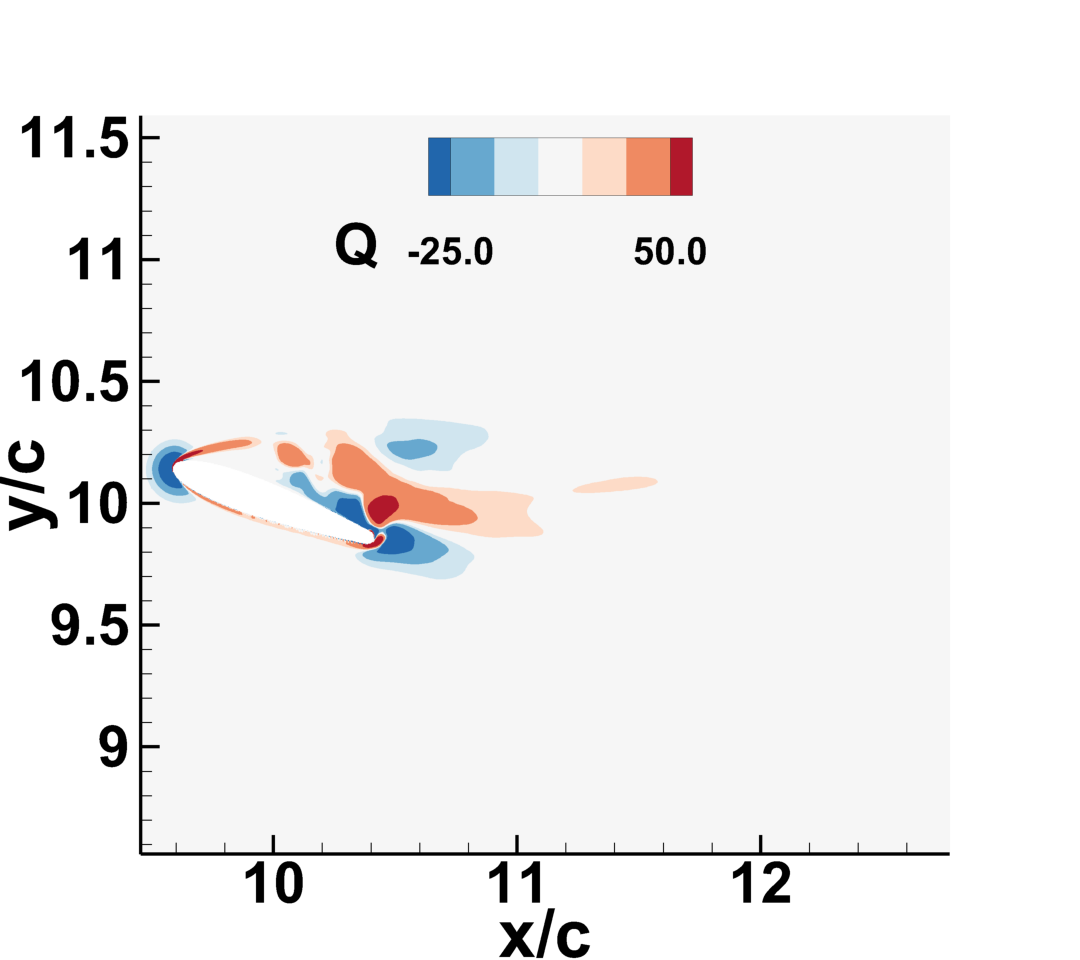}
         \caption{$\tilde{Q}_{0}$}
         \label{fig:af:td:flowfq:b}
     \end{subfigure}
  \hfill
     \begin{subfigure}[b]{0.24\textwidth}
         \centering
         \includegraphics[width=\textwidth]{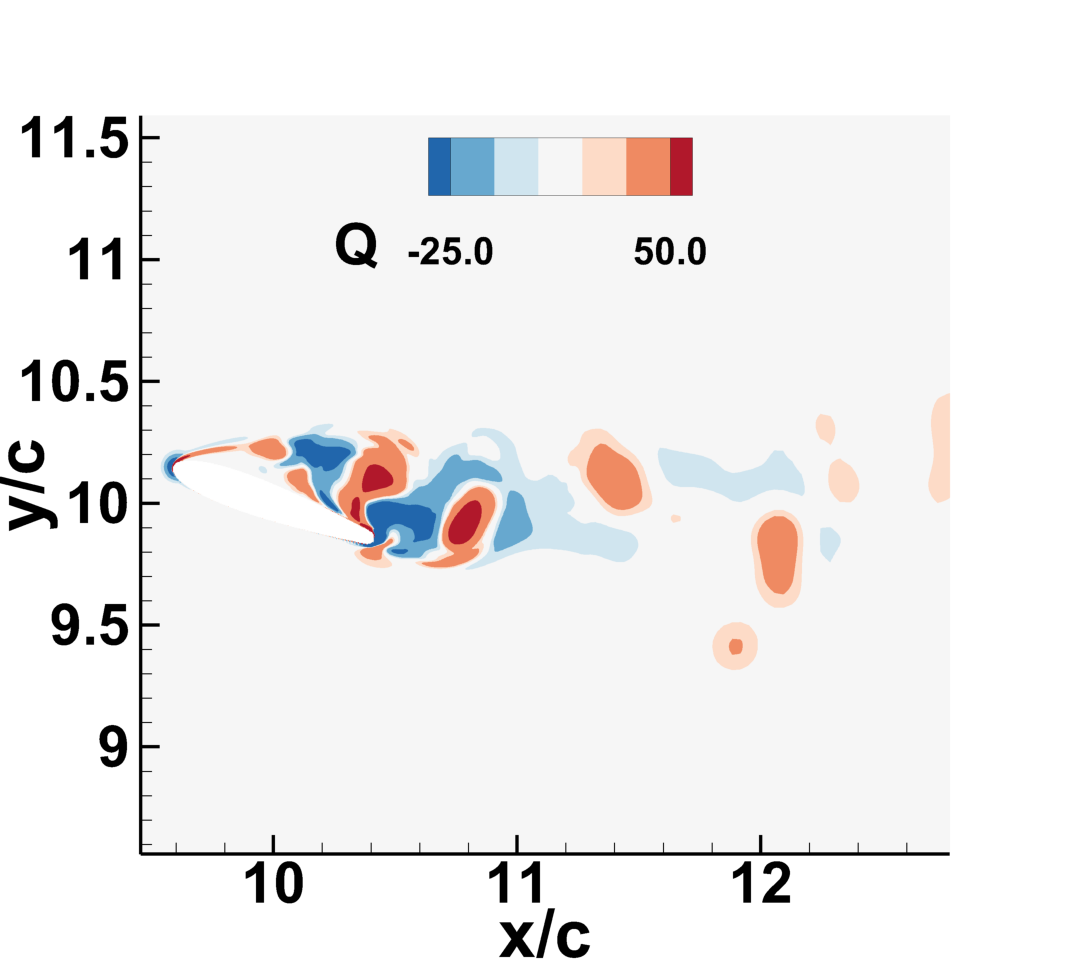}
         \caption{$\tilde{Q}_{1}$}
         \label{fig:af:td:flowfq:c}
     \end{subfigure}
   \hfill
     \begin{subfigure}[b]{0.24\textwidth}
         \centering
         \includegraphics[width=\textwidth]{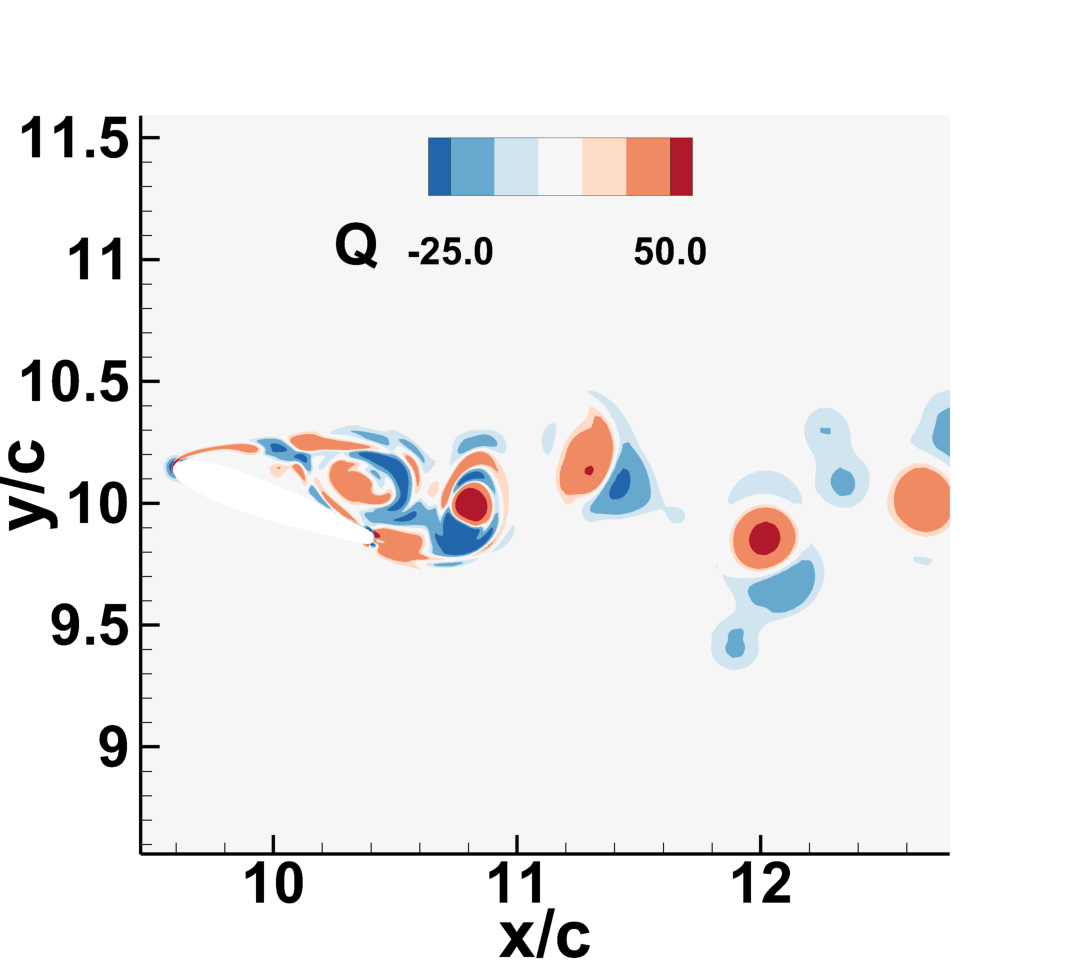}
         \caption{$\tilde{Q}_{2}$}
         \label{fig:af:td:flowfq:d}
     \end{subfigure}
     \hfill
     \begin{subfigure}[b]{0.24\textwidth}
      \centering
         \includegraphics[width=\textwidth]{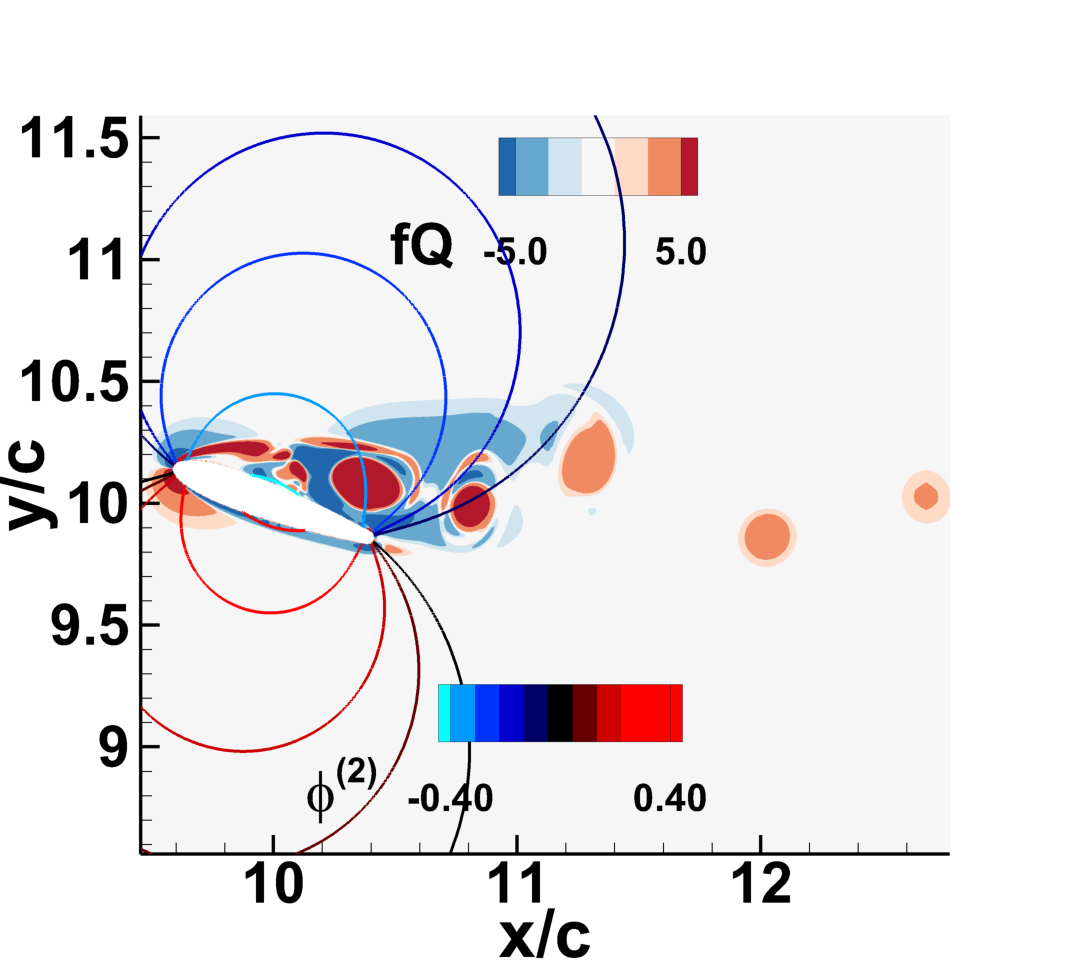}
         \caption{$f_{\tilde{Q}_{Total}}^{(2)}$}
         \label{fig:af:td:flowfq:e}
     \end{subfigure}
  \hfill
  \begin{subfigure}[b]{0.24\textwidth}
      \centering
         \includegraphics[width=\textwidth]{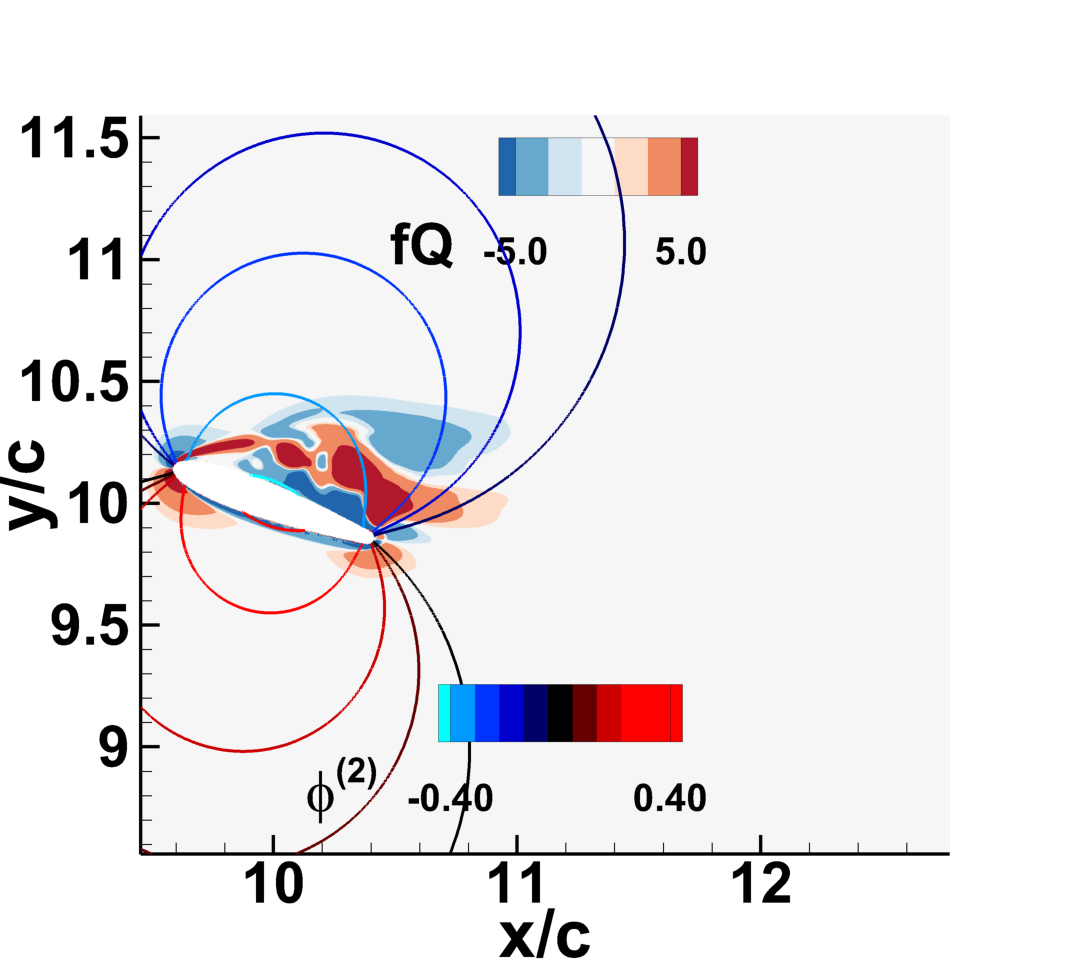}
         \caption{$f_{\tilde{Q}_0}^{(2)}$}
         \label{fig:af:td:flowfq:f}
     \end{subfigure}
  \hfill
     \begin{subfigure}[b]{0.24\textwidth}
         \centering
         \includegraphics[width=\textwidth]{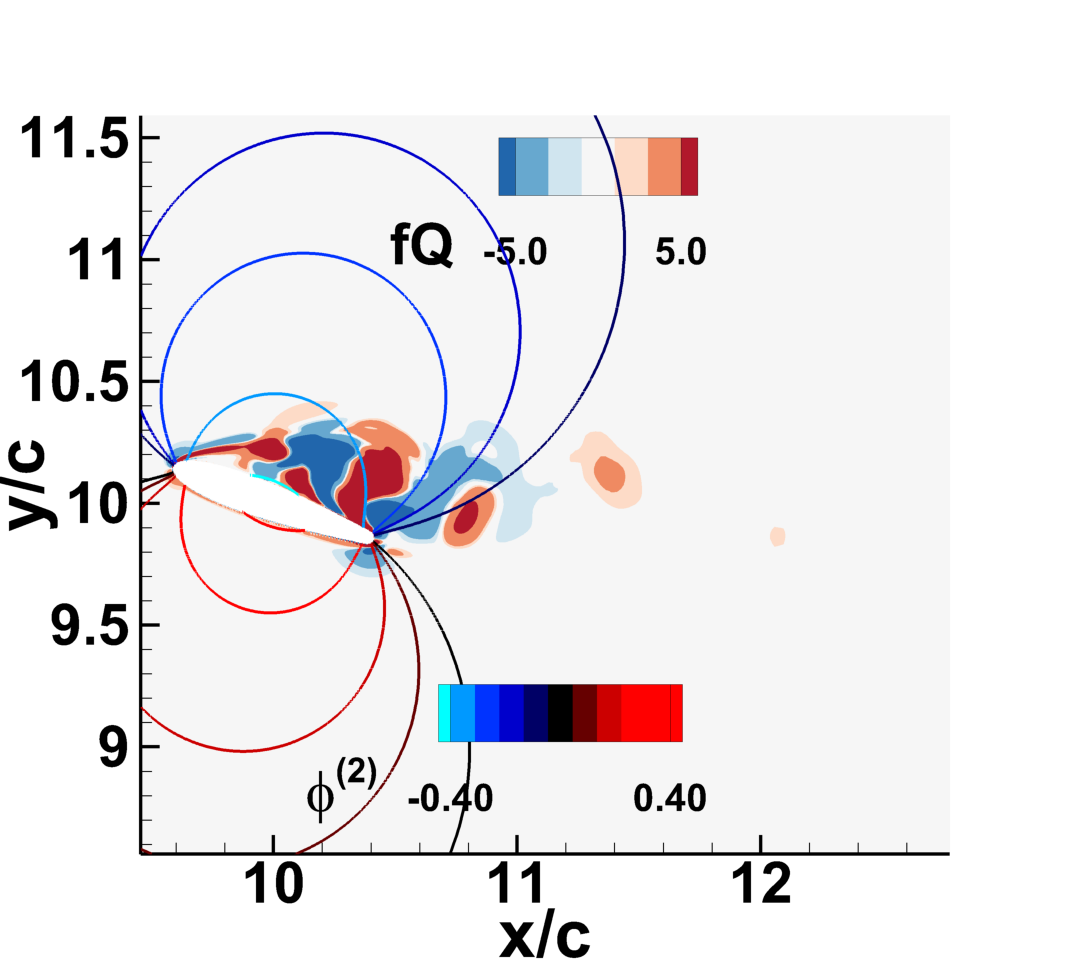}
         \caption{$f_{\tilde{Q}_1}^{(2)}$}
         \label{fig:af:td:flowfq:g}
     \end{subfigure}
   \hfill
     \begin{subfigure}[b]{0.24\textwidth}
         \centering
         \includegraphics[width=\textwidth]{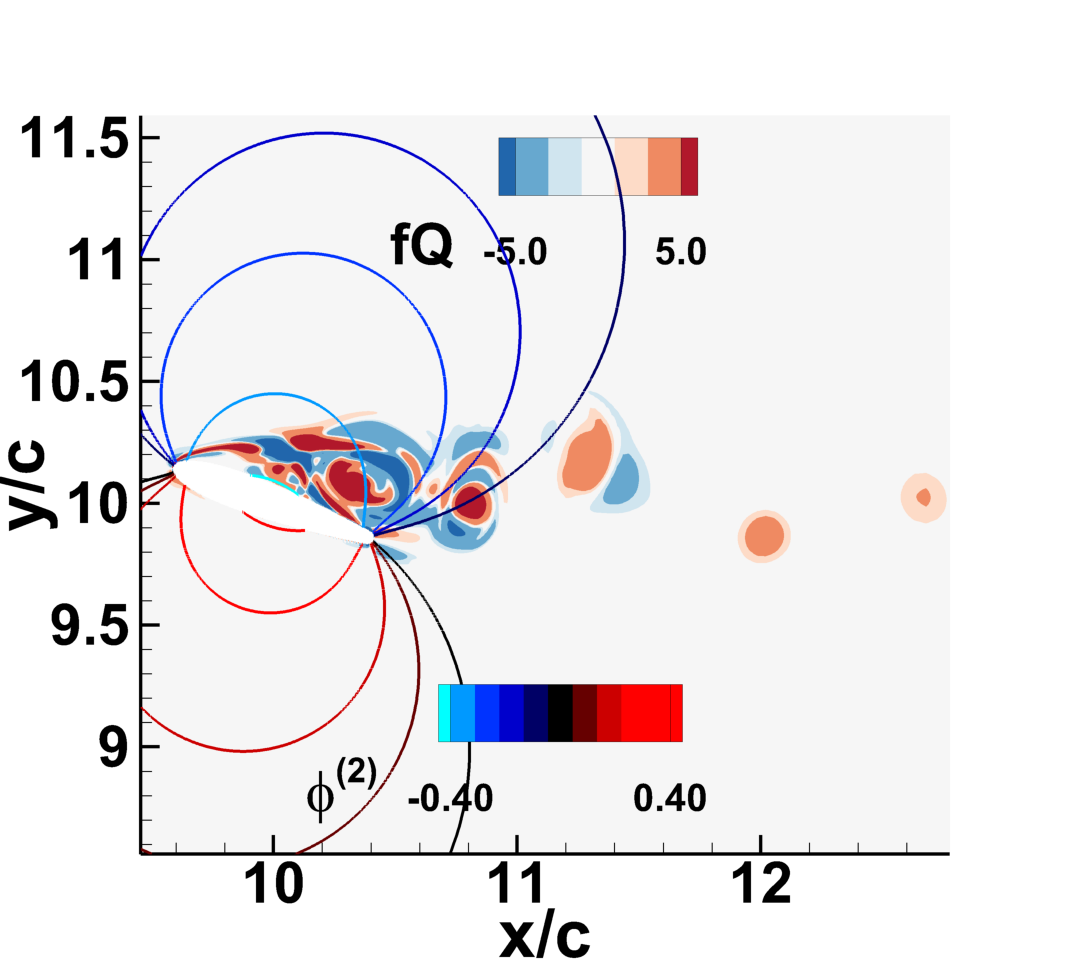}
         \caption{$f_{\tilde{Q}_2}^{(2)}$}
         \label{fig:af:td:flowfq:h}
     \end{subfigure}
       \hfill
     \begin{subfigure}[b]{0.24\textwidth}
      \centering
         \includegraphics[width=\textwidth]{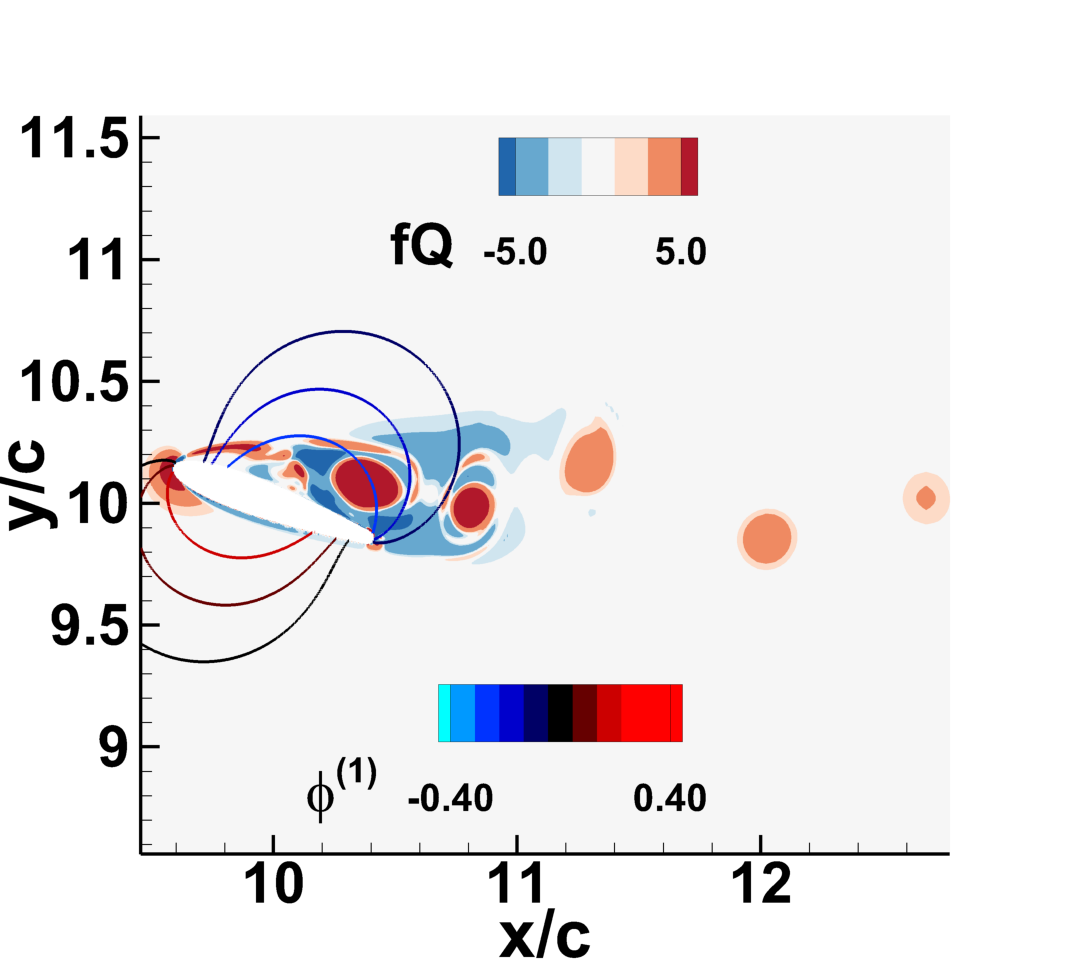}
         \caption{\oc{$f_{\tilde{Q}_{Total}}^{(1)}$}}
         \label{fig:af:td:flowfq:i}
     \end{subfigure}
  \hfill
  \begin{subfigure}[b]{0.24\textwidth}
      \centering
         \includegraphics[width=\textwidth]{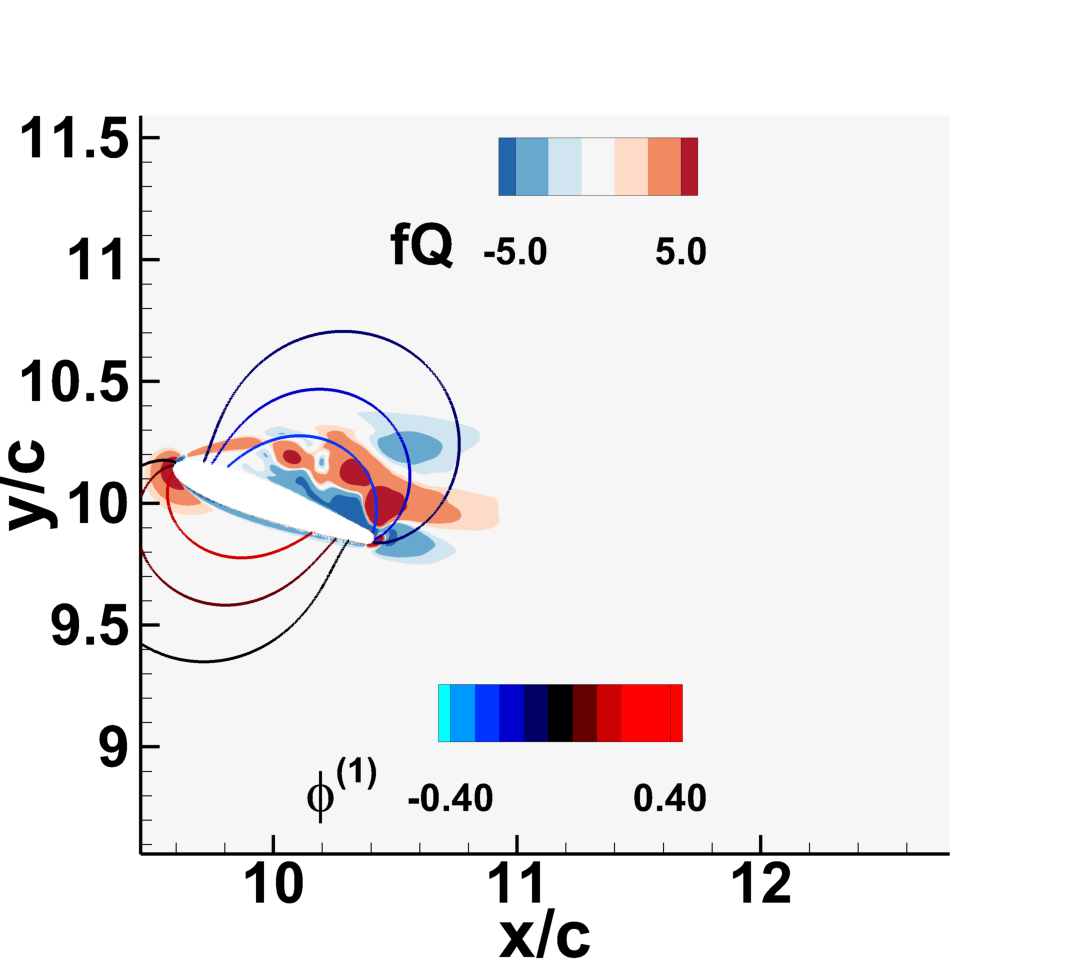}
         \caption{\oc{$f_{\tilde{Q}_0}^{(1)}$}}
         \label{fig:af:td:flowfq:j}
     \end{subfigure}
  \hfill
     \begin{subfigure}[b]{0.24\textwidth}
         \centering
         \includegraphics[width=\textwidth]{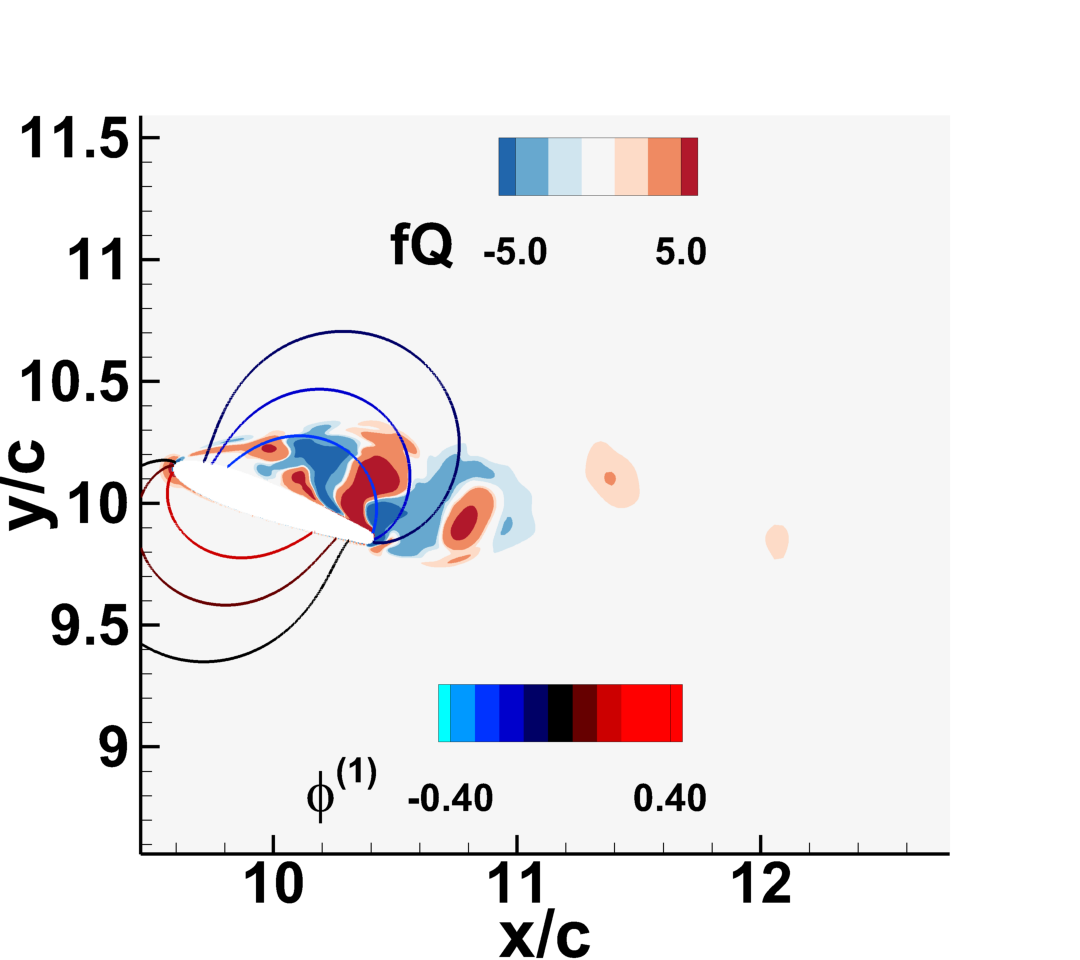}
         \caption{\oc{$f_{\tilde{Q}_1}^{(1)}$}}
         \label{fig:af:td:flowfq:k}
     \end{subfigure}
   \hfill
     \begin{subfigure}[b]{0.24\textwidth}
         \centering
         \includegraphics[width=\textwidth]{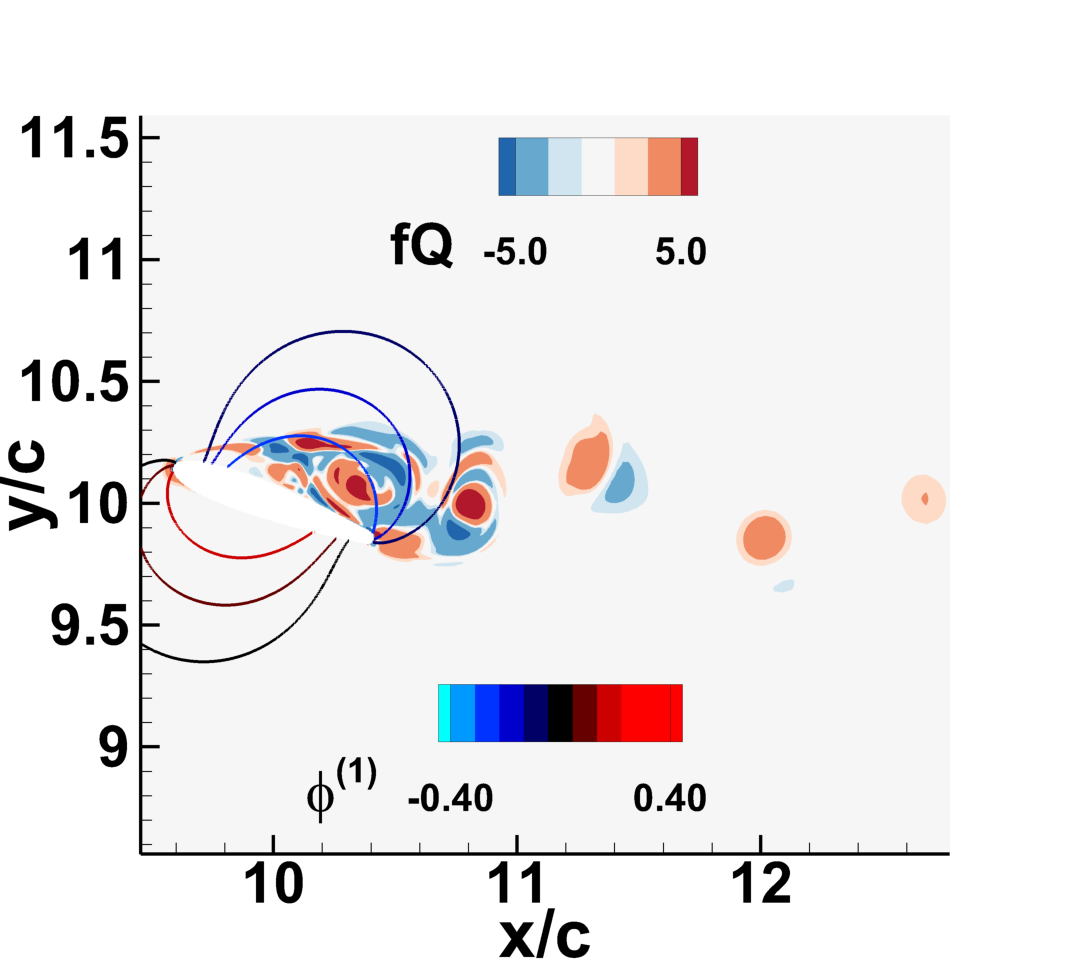}
         \caption{\oc{$f_{\tilde{Q}_2}^{(1)}$}}
         \label{fig:af:td:flowfq:l}
     \end{subfigure}
        \caption{Triple decomposition applied directly to the $Q$-field of the airfoil showing contours of $Q$ for (a) a snapshot of the flow, (b) the mean flow, (c) the coherent part of the flow and (d) the non-coherent part of the flow. Figures (e)-(h) are the corresponding contours plots of vortex-induced lift force density ($f_Q^{(2)}$) and $\phi^{(2)}$ \oc{and figures (i)-(l) are the corresponding contour plots of vortex-induced drag force density ($f_Q^{(1)}$) and $\phi^{(1)}$}.}
\label{fig:af:td:flowfq}
\end{figure}

As shown in the previous section, based on the FPM and FAPM formulations, if we decompose the $Q$ field directly, the issues associated with inter-modal interactions can be avoided. Thus, the triple-decomposition is now applied directly to the $Q$-field. The temporally and spatially varying $Q$ field is directly decomposed into a mean component (figure \ref{fig:af:td:flowfq:b}), a coherent part (figure \ref{fig:af:td:flowfq:c}), and the remaining incoherent part (figure \ref{fig:af:td:flowfq:d}). The mean flow depicts the vortical structure due to the flow separation over the airfoil, and the coherent part captures the shedding of leading and trailing-edge vortices. The incoherent part shows thin sheets of positive and negative $Q$ around the shed vortices, which represent the effect of cycle-to-cycle variations that occur mostly along the periphery of the large vortices as they interact with the other vortices in the flow. Thus, modes obtained from the direct decomposition of the $Q$-field are also amenable to interpretation, which is quite often the key in gaining some insight into these phenomena.

The influence field in the vertical direction ($\phi^{(2)}$) \oc{along with} the vortex-induced lift force density are 
plotted for each mode in figure \ref{fig:af:td:flowfq} \oc{(e-h) and the influence field in the horizontal direction ($\phi^{(1)}$) along with the vortex induced drag force density are plotted in figure \ref{fig:af:td:flowfq} (i-l)}. As noted earlier, the vortical structures near the airfoil contribute more significantly to the pressure force on the body. For the coherent part, the shedding of the leading-edge vortex is key to the generation of periodic force fluctuations. The cycle-to-cycle variation of the size and strength of this vortex-induced force is captured by the incoherent mode.

\begin{figure}
    \centering
 \begin{subfigure}[b]{0.47\textwidth}
      \centering
         \includegraphics[width=\textwidth]{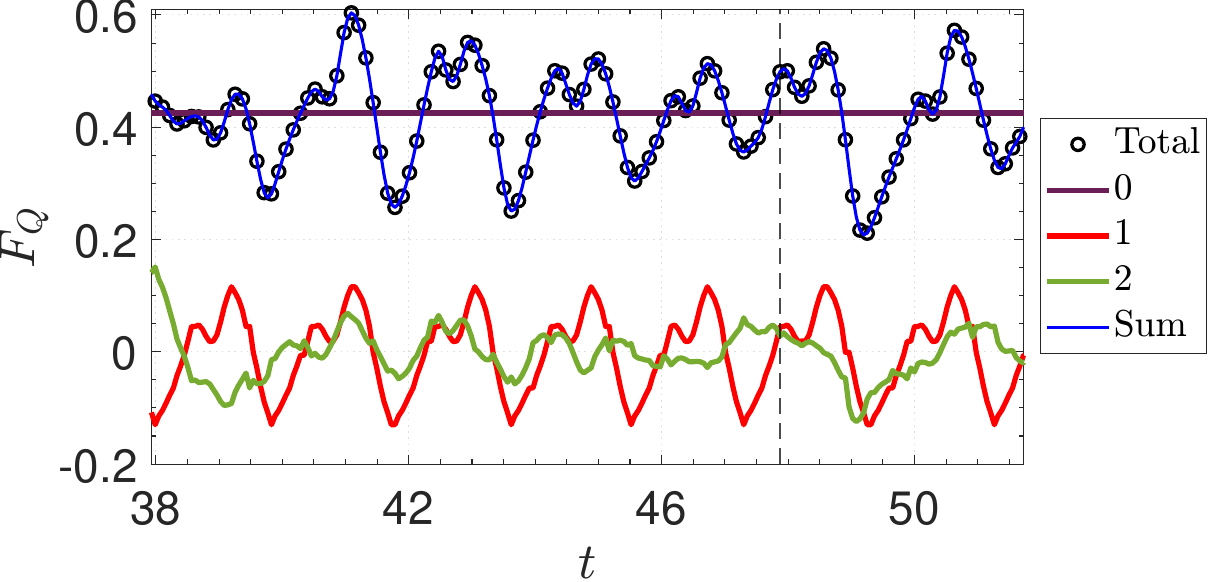}
         \caption{}
         \label{fig:af:td:fq:a}
     \end{subfigure}
  \hfill
   \begin{subfigure}[b]{0.47\textwidth}
      \centering
         \includegraphics[width=\textwidth]{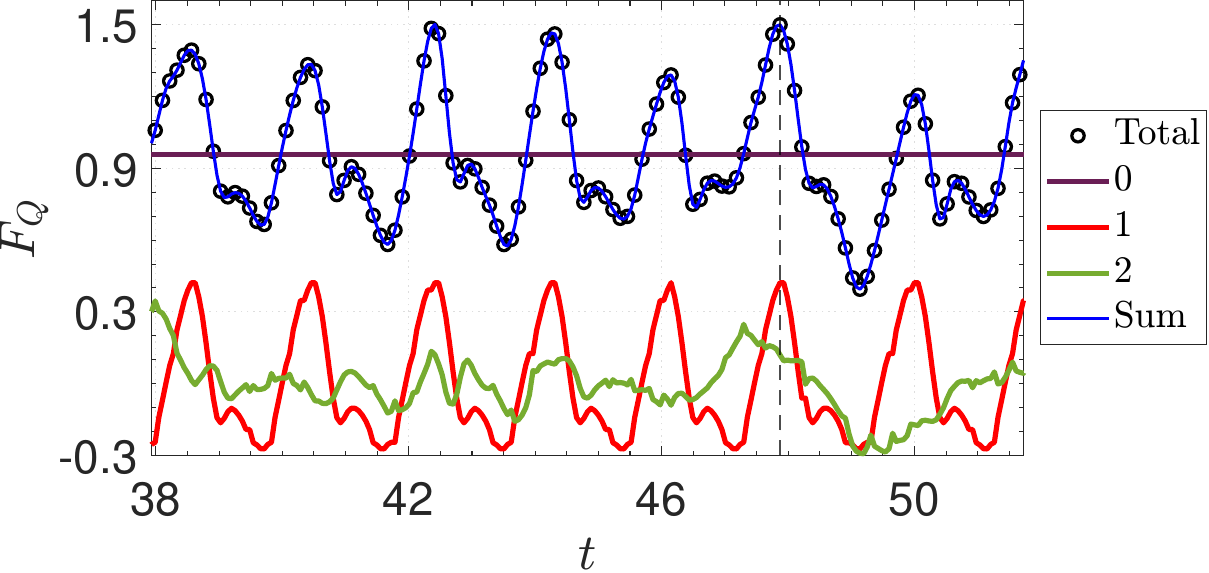}
         \caption{}
         \label{fig:af:td:fq:b}
     \end{subfigure}
  \hfill
  \begin{subfigure}[b]{0.40\textwidth}
      \centering
         \includegraphics[width=\textwidth]{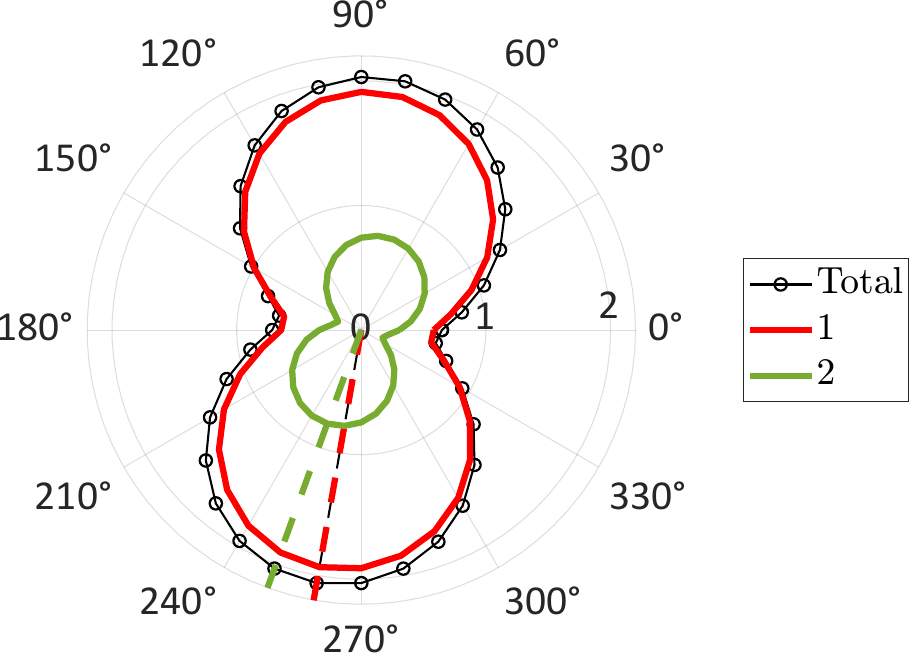}
         \caption{}
         \label{fig:af:td:fq:c}
     \end{subfigure}
        \caption{Results from triple decomposition applied directly to the $Q$-field of the airfoil. Temporal variation of the vortex-induced (a) drag force ($F_Q^{(1)}$) and (b) lift force ($F_Q^{(2)}$) corresponding to the modes of the triple decomposition.\oc{The vortex induced lift force is normalized using the force coefficient ($0.5\rho U_\infty^2 c$) and the time is normalized using the flow time-scale ($c/U_\infty$).} (c) The directivity ($p'_{rms}[\times 10^{-5}]$) calculated at a distance of \oc{57c} corresponding to a Mach number of 0.1 with dashed line showing directivity of highest sound intensity.}
\label{fig:af:td:fq}
\end{figure}

The time variations of the vortex-induced drag($F_Q^{(1)}$) and lift ($F_Q^{(2)}$) forces obtained by applying the FPM with the triple-decomposed $Q$-fields are shown in the figure \ref{fig:af:td:fq}, and we see that the total vortex-induced force has a dominant cyclical component, it deviates from a strictly periodic variation and provides the motivation for the use of a triple decomposition for this case. As expected, the mean mode alone produces the mean vortex-induced drag and lift forces, but it lacks temporal variance and hence, is  incapable of generating any aeroacoustic sound. The coherent part of the flow (Mode-1) consists of a periodic variation and produces a major portion of the force fluctuations while the incoherent part (Mode-2) is responsible for a smaller and nonperiodic portion of the fluctuation. Note that by directly decomposing the $Q$-field, the periodic and non-periodic components of the force are represented by individual modes and no inter-modal interactions exist to complicate the interpretation of the modal contributions. This decomposition also provides some useful insights into distinctive contributions of the two modes: for Mode-1, the ratio of the RMS lift fluctuation to RMS drag fluctuation is about 2, and thus Mode-1 is clearly lift-fluctuation dominant. For Mode-2 however, this ratio is reduced to 1.5. This is because the dominant feature for Mode-1 is the leading-edge vortex shedding on top of the airfoil, and this preferentially affects the lift force fluctuation. On the other hand, Mode-2 related structures are stronger in the aft portion of the airfoil (i.e. near the trailing-edge) since cycle-to-cycle variations grow as the vortices convect downstream. The airfoil surface near the trailing-edge has a more significant orientation in the drag direction and this results in a relatively larger drag fluctuation contribution from Mode-2.

The aerodynamic sound is again computed by applying the FAPM to the triple-decomposed $Q$-fields. The directivity patterns for the dipole sounds generated by the fluctuating aerodynamic forces are shown in figure  \ref{fig:af:td:fq:c}. The aerodynamic sound is evaluated at a distance of $57c$ away from the airfoil with a flow Mach number of 0.1 and. As one can expect, most of the noise produced by the airfoil comes from the high amplitude coherent mode(Mode-1), while the noise from the incoherent mode is noticeably weaker. It is interesting to note that the directivity peaks for the two modes are slightly different as marked in figure \ref{fig:af:td:fq:c}. Indeed, the directivity of the incoherent mode is tilted further away (by about $10^o$) from the vertical axis compared to the coherent mode and this is directly connected with the decreased lift-to-drag fluctuation ratio of the incoherent mode compared to the coherent mode as discussed above. Thus, modal force partitioning when applied to the $Q$-field results in interpretable insights that are difficult to obtain using other approaches. Furthermore, modal force partitioning provides a novel and useful method to connect the coherent and incoherent modes to distinct characteristics of the aerodynamic force as well as the amplitude and directivity characteristics of the aeroacoustic sound.

\subsection{Revolving Wing}
\label{sec:revolvingRotor}
In this section, we model a single rotor blade with the geometry based on the experimental study of \cite{45RotorExp}. The rotor blade is a rectangular flat plate with an aspect ratio of 5, the angle-of-attack is set to 45$^\circ$. The blade revolves with an angular velocity of $0.25 v_t/R_c$, where $v_t$ is the tip velocity and $R_c$ is the distance from the center of revolution to the tip of the blade. The blade is modeled as a zero thickness plate, and the simulation is performed in a non-inertial rotating reference frame with a Reynolds number of 3300 based on the tip velocity($v_t$) and the radius measured from center of revolution ($R_c$). The Reynolds number based on the tip velocity and the chord length is 500. The domain size used for this simulation is $8R_c \times 10R_c \times 10R_c$. The mesh and the location of the rotor (shown in red) is shown in figure \ref{fig:rw:setup:a} and we point out that a very fine mesh is employed over the region $5c \times 2.4R_c \times 2.4R_c$  to resolve the vortices generated over the blade and in its wake. 

\subsubsection{Flow characteristics and aerodynamic loads}
\begin{figure}
    \centering
 \begin{subfigure}[b]{0.4\textwidth}
      \centering
         \includegraphics[width=\textwidth]{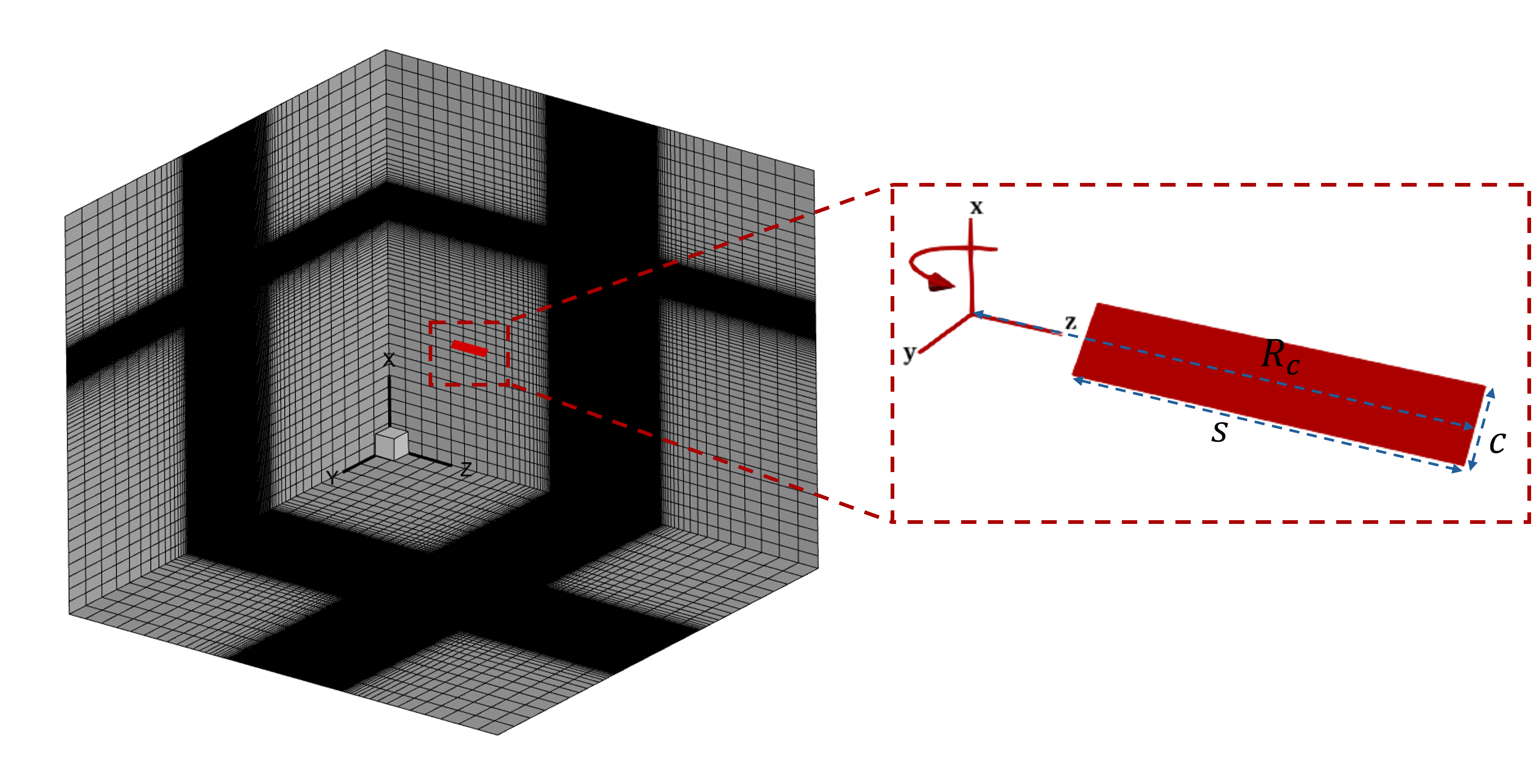}
         \caption{}
         \label{fig:rw:setup:a}
     \end{subfigure}
  \hfill
     \begin{subfigure}[b]{0.5\textwidth}
         \centering
         \includegraphics[width=\textwidth]{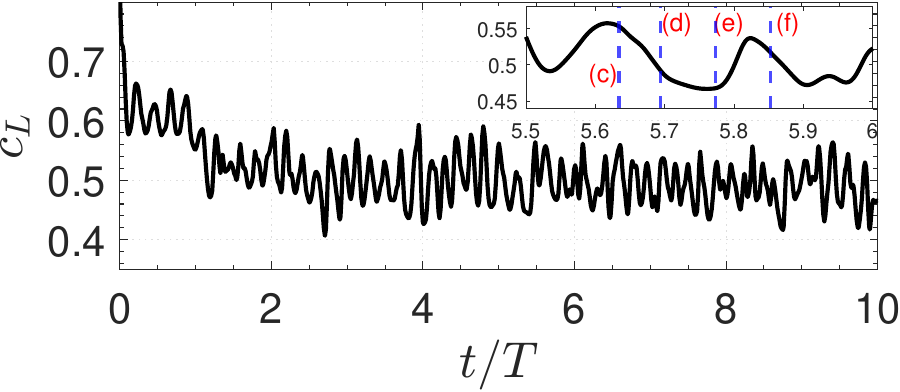}
         \caption{}
         \label{fig:rw:setup:b}
     \end{subfigure}
      \begin{subfigure}[b]{0.19\textwidth}
      \centering
         \includegraphics[width=\textwidth]{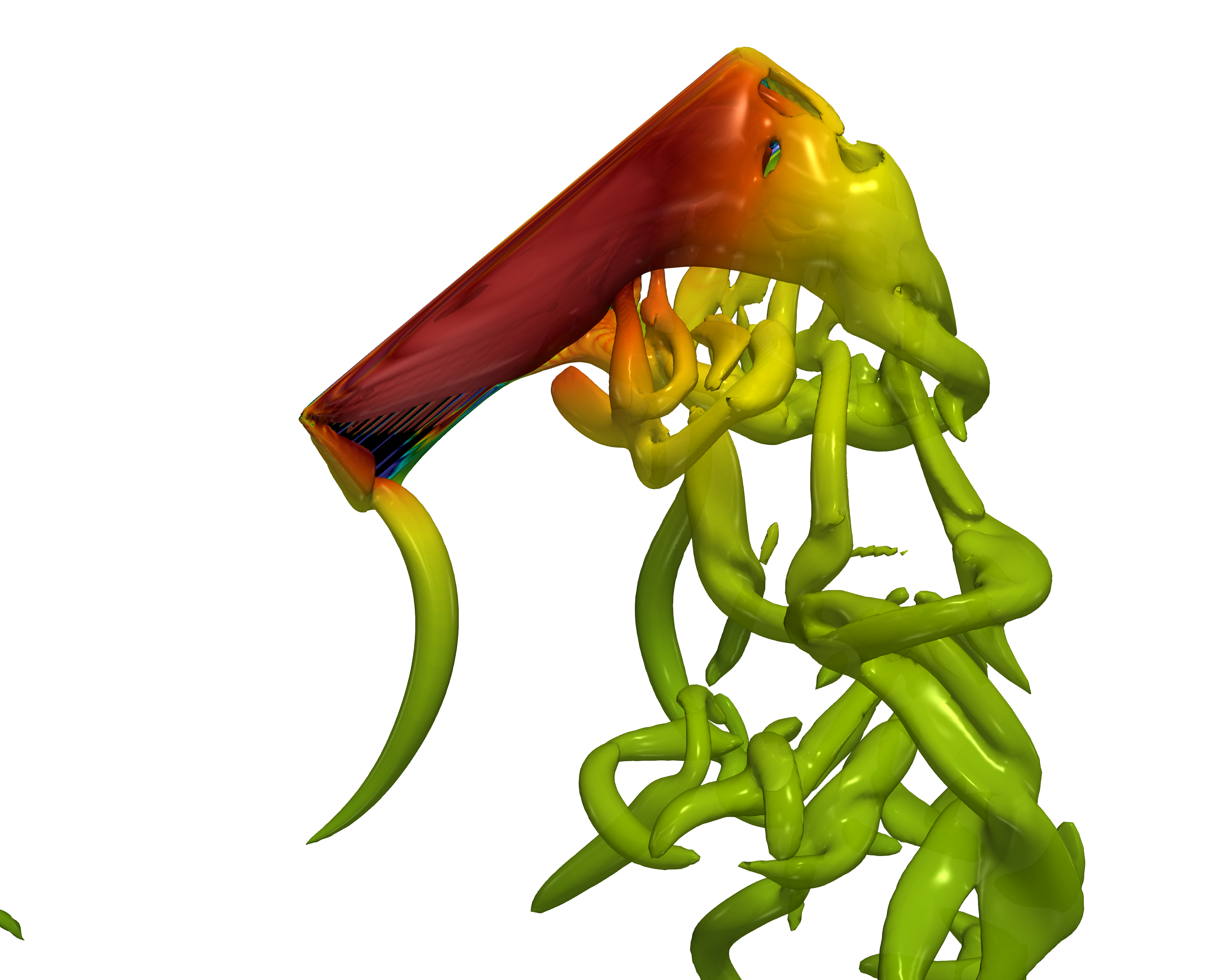}
         \caption{}
         \label{fig:rw:setup:c}
     \end{subfigure}
  \hfill
     \begin{subfigure}[b]{0.19\textwidth}
         \centering
         \includegraphics[width=\textwidth]{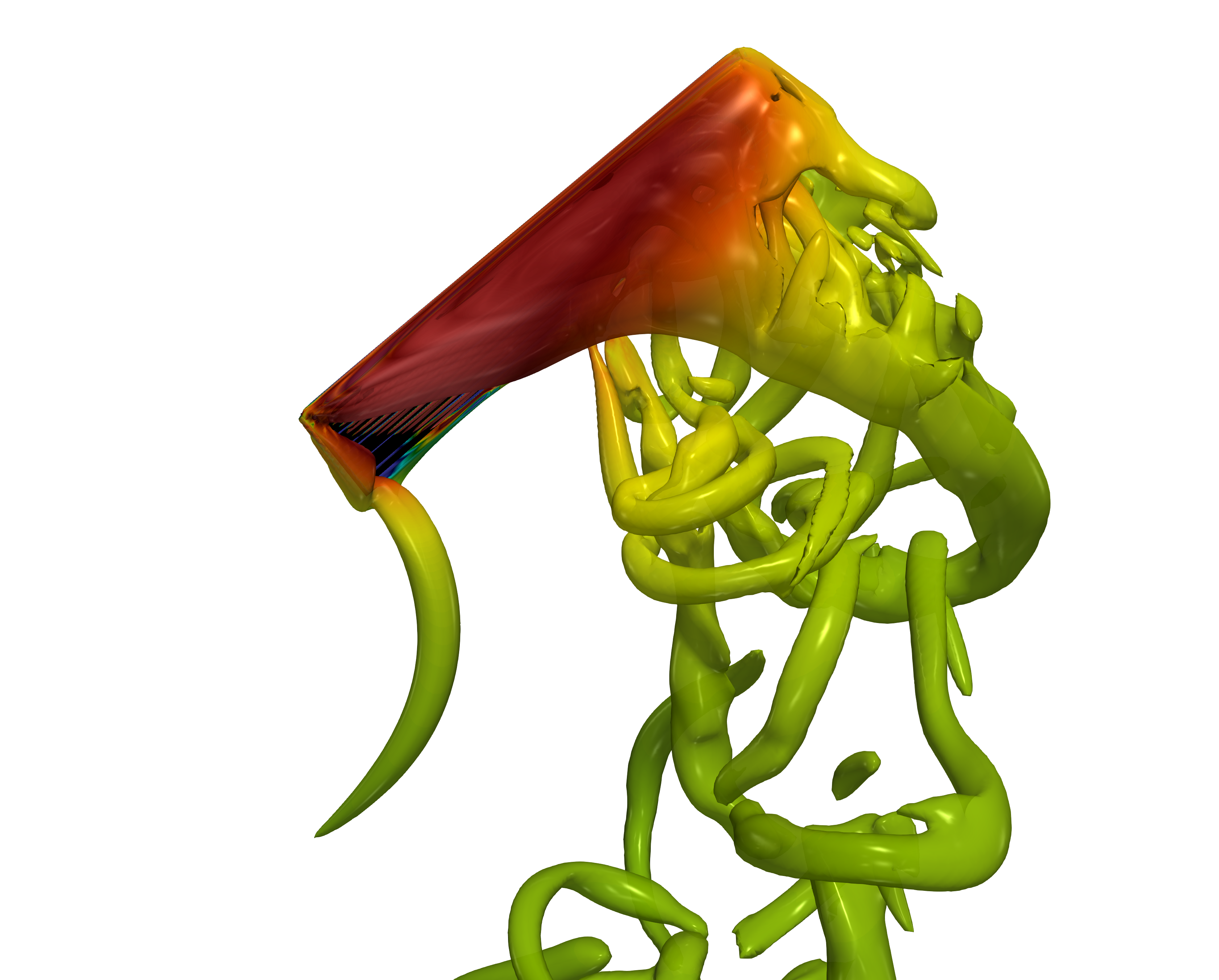}
         \caption{}
         \label{fig:rw:setup:d}
     \end{subfigure}
   \hfill
     \begin{subfigure}[b]{0.19\textwidth}
         \centering
         \includegraphics[width=\textwidth]{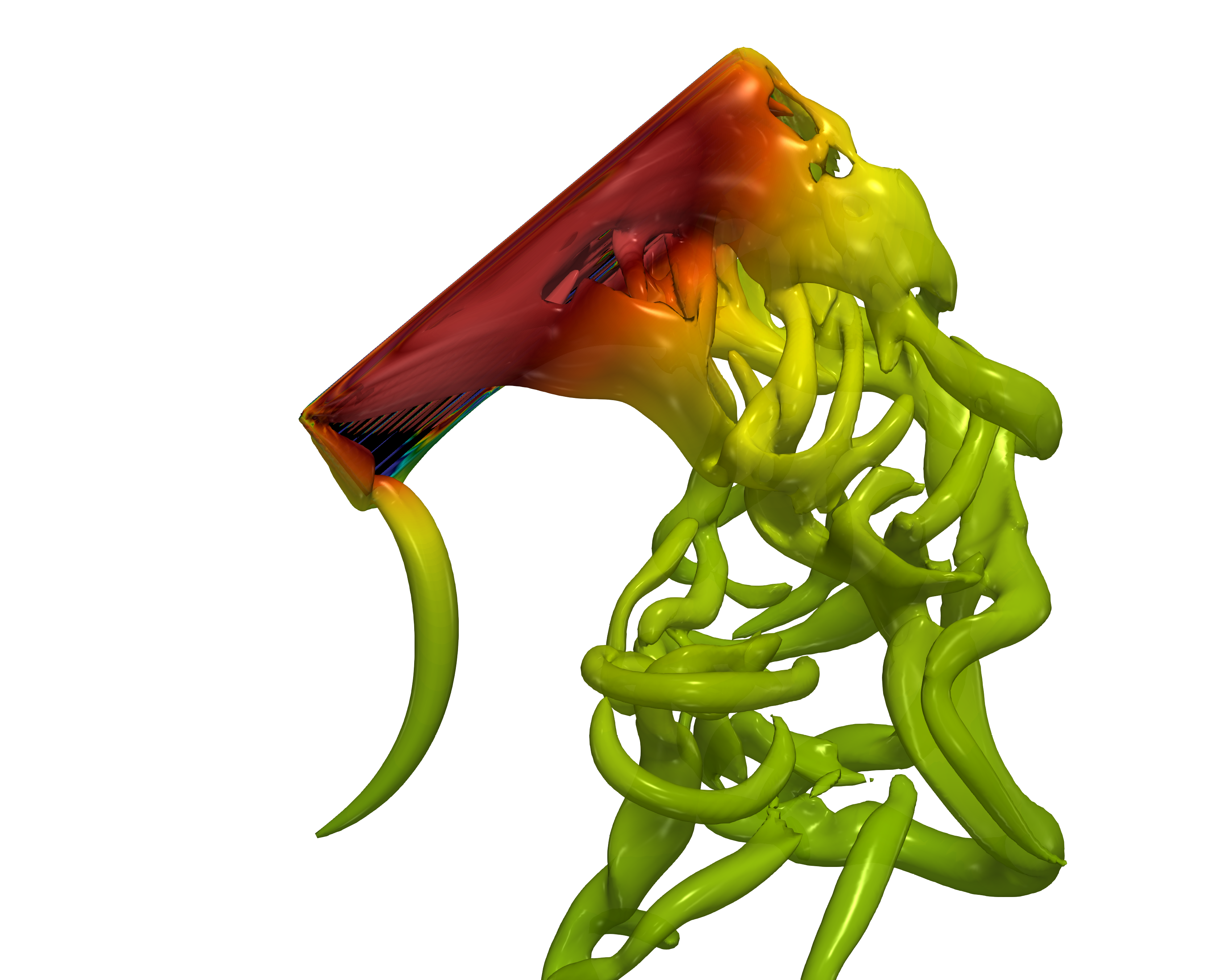}
         \caption{}
         \label{fig:rw:setup:e}
     \end{subfigure}
     \hfill
     \begin{subfigure}[b]{0.19\textwidth}
         \centering
         \includegraphics[width=\textwidth]{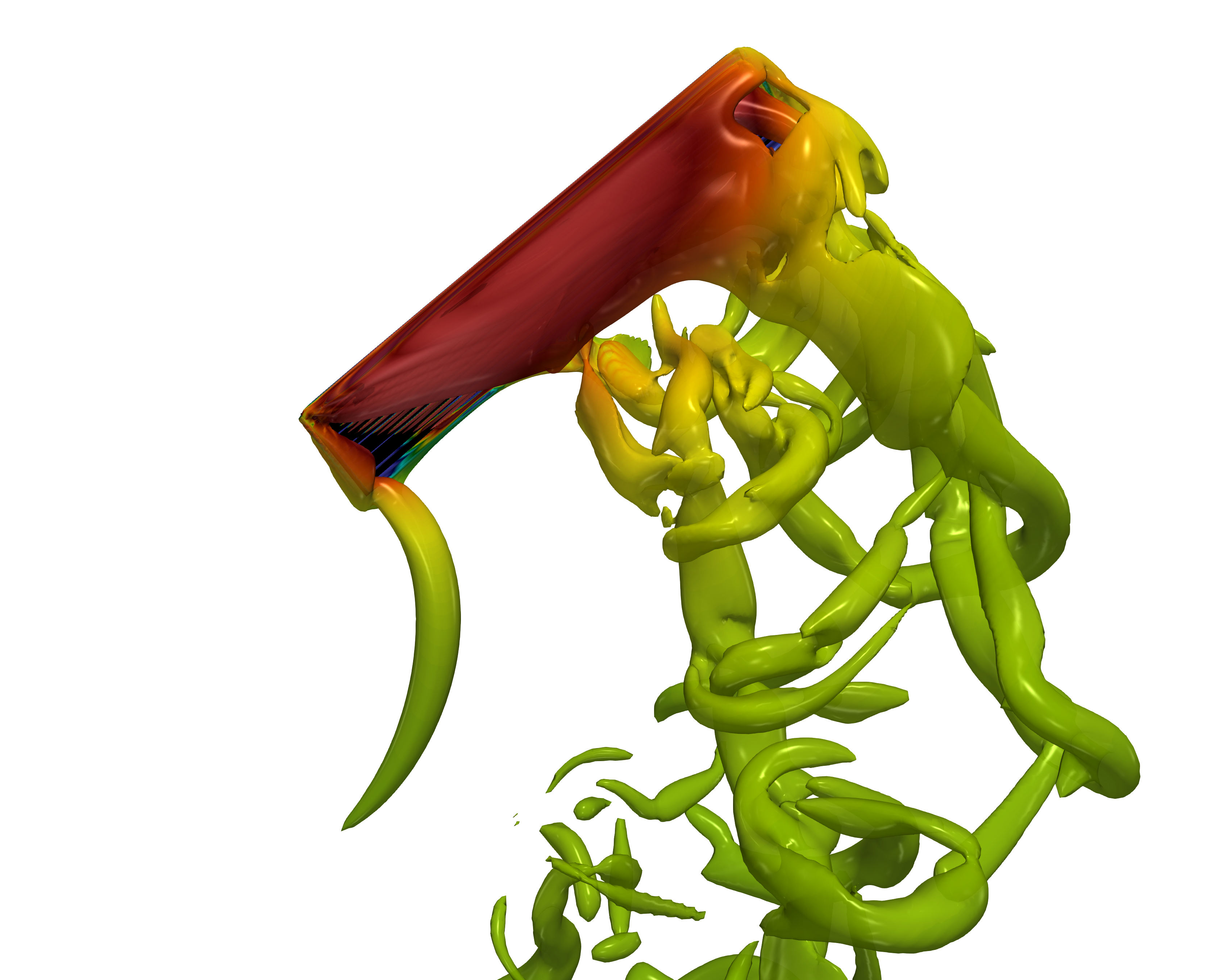}
         \caption{}
         \label{fig:rw:setup:f}
     \end{subfigure}
\hfill
     \begin{subfigure}[b]{0.19\textwidth}
      \centering
         \includegraphics[width=\textwidth]{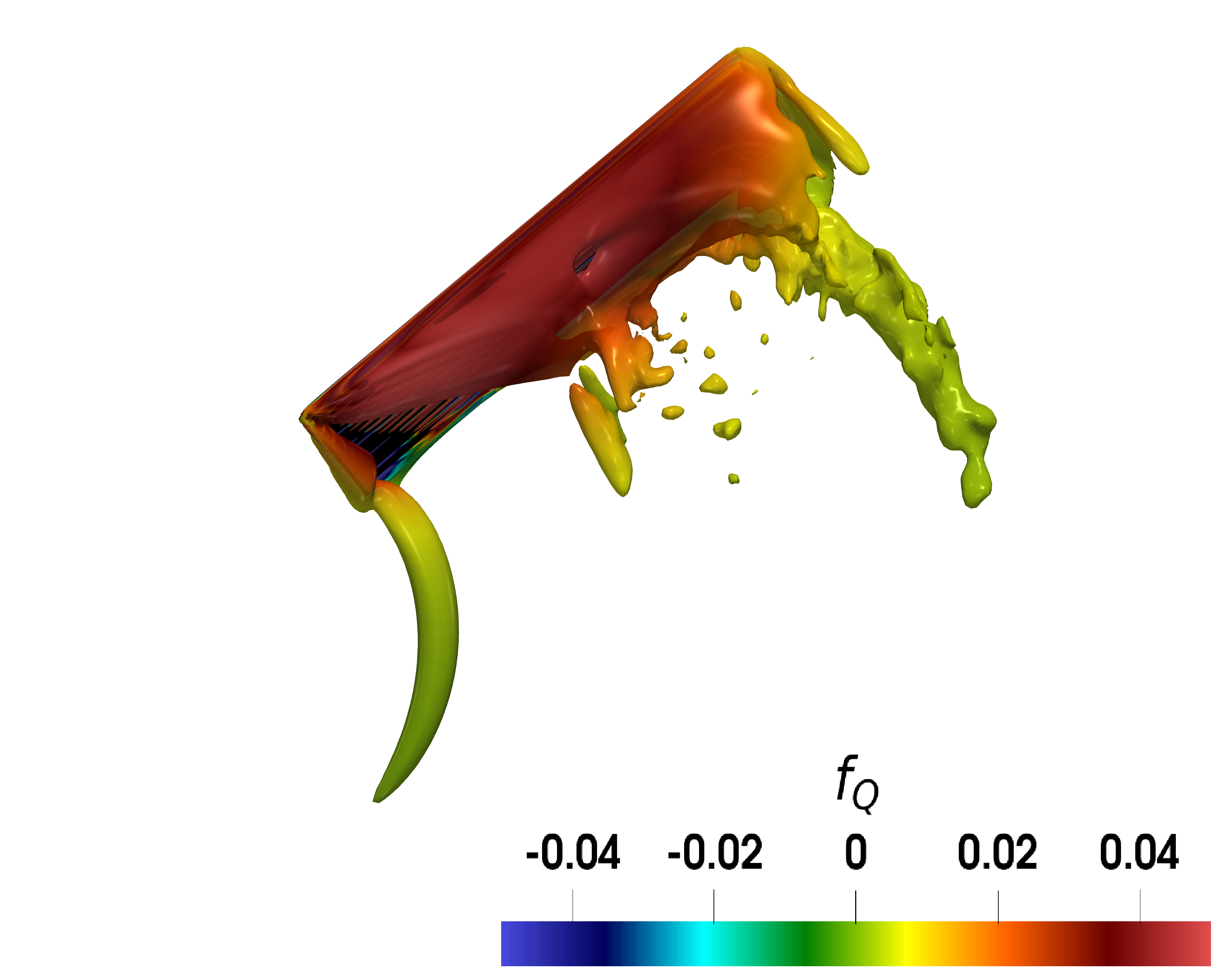}
         \caption{}
         \label{fig:rw:setup:g}
     \end{subfigure}
        \caption{(a) The computational domain and mesh for the rotor simulation with a zoomed image of the rotor with axis shown at the center of the revolution. (b) The lift coefficient (normalized using tip velocity and the rotor area) for the rotor with the vertical line showing the time instances where the flow vortex structures are shown. (c-f) Instantaneous flow fields showing the shedding of the vortices and (g) the mean flow. The vortices are shown using the $Q$-field and colored by the vortex-induced lift force.}
\label{fig:rw:setup}
\end{figure}

The flow simulation is carried out for 10 revolution cycles, and the time history of the lift coefficient is plotted in figure \ref{fig:rw:setup:b}.
The instantaneous flow fields are visualized by $Q$-field in figure \ref{fig:rw:setup}(c-f) at four time instances marked by the vertical dashed blue lines on the lift coefficient profile. In addition, the mean flow obtained from averaging between revolution cycles 5 to 10 is also shown in figure \ref{fig:rw:setup:g}. We see a smooth root vortex and leading-edge vortex that sheds in two locations along the span; one at around 50\% chord (visible in figure \ref{fig:rw:setup:d}) and one closer to the blade tip (visible in figures \ref{fig:rw:setup:c} and \ref{fig:rw:setup:f} where it merges with the blade-tip vortex. Indeed, figure \ref{fig:rw:setup:e} shows both these phenomena occurring at the same time.  Careful examination of the plots also indicates the formation and shedding of secondary leading-edge vortices. Vortices are also shed from the trailing-edge of the blade and all of these vortices interact in a complex way in the wake region of the blade.

The $Q$ associated with the mean flow shown in figure \ref{fig:rw:setup:g} shows only vortices near the root and the blade tip as well as the leading-edge but the vortex dynamics and associated aerodynamic loads and aeroacoustic sound are associated with a very complex behavior evident in the instantaneous snapshots. Thus, the task of determining the flow modes with dominant contributions to the aerodynamic loads and aeroacoustic sound is significantly more complex. Based on our discussion in the previous sections, we choose to directly decompose the $Q$-field to eliminate the inter-modal interactions.

\subsubsection{Modal FPM applied to POD of the $Q$-field}
\begin{figure}
    \centering
 \begin{subfigure}[b]{0.34\textwidth}
      \centering
         \includegraphics[width=\textwidth]{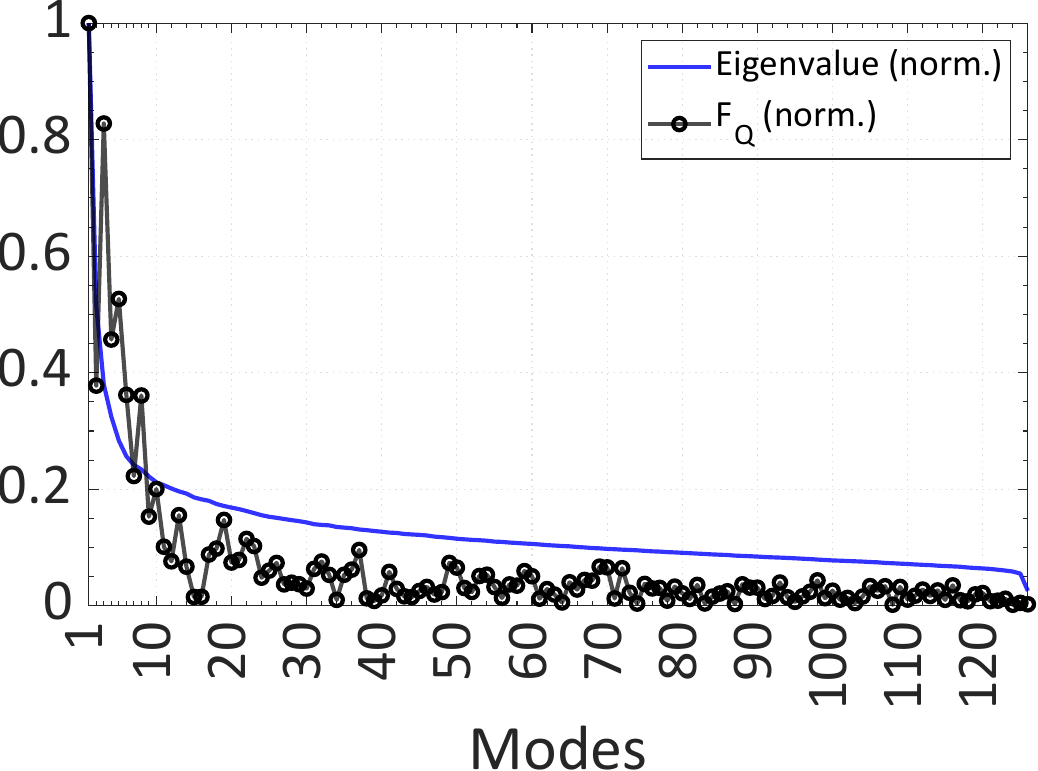}
         \caption{}
         \label{fig:rw:podQ:fqvst:a}
     \end{subfigure}
  \hfill
     \begin{subfigure}[b]{0.65\textwidth}
         \centering
         \includegraphics[width=\textwidth]{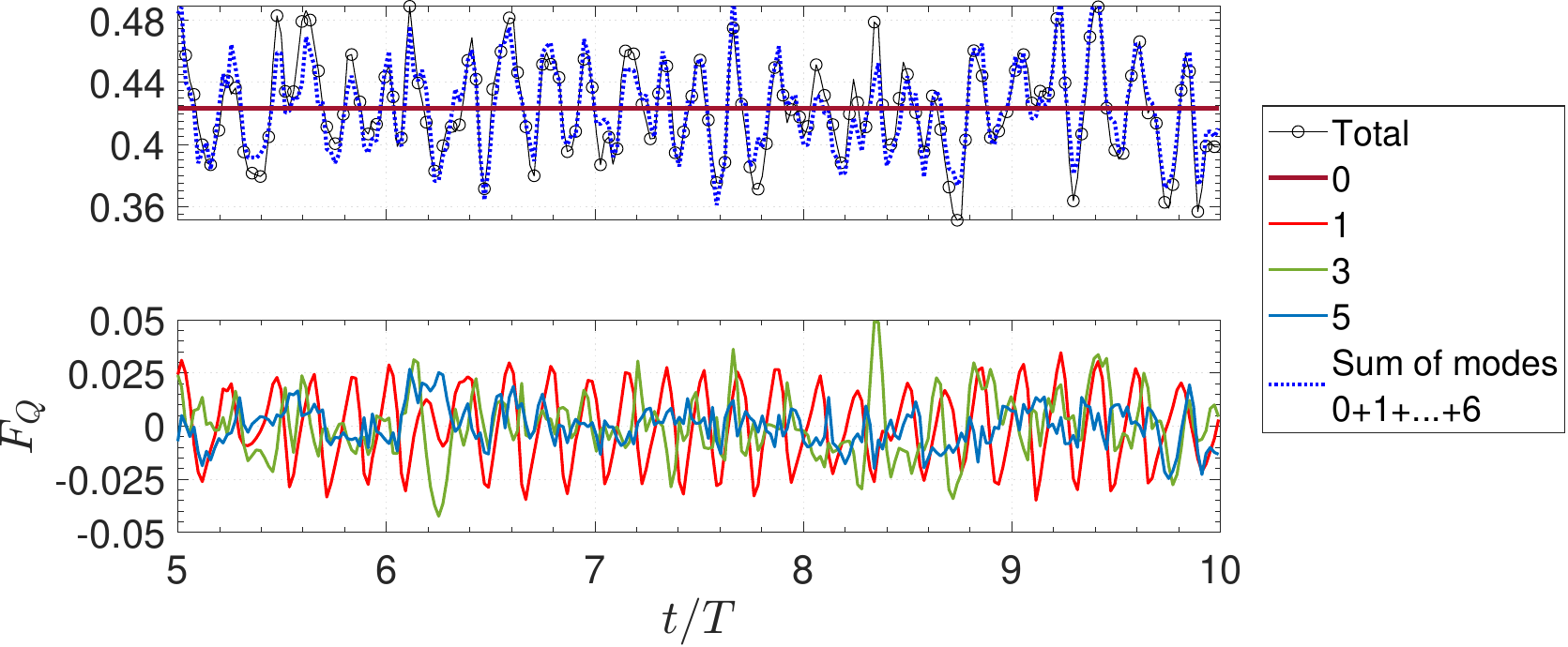}
         \caption{}
         \label{fig:rw:podQ:fqvst:b}
     \end{subfigure}
        \caption{(a): \oc{Eigenvalues corresponding to POD modes along with the vortex-induced lift force ($F_Q^{(2)}$) corresponding to each mode.} (b): vortex-induced lift force vs time corresponding to the POD applied on Q showing the dominant modes. The vortex induced lift force is normalized using tip velocity and the rotor area while the time is normalized using time period of each revolution ($T$).}
\label{fig:rw:podQ:fqvst}
\end{figure}

The POD is applied to the $Q$-field of the flow between revolution cycle, $t/T=$ 5 to 10, with 50 snapshots in each cycle and the eigenvalues for the modes are shown in the figure  \ref{fig:rw:podQ:fqvst:a} and 121 modes are required in this case to reconstruct 98\% of the original $Q$-field. The contribution of each mode to the vortex-induced lift force ($F_Q$) is also plotted in the figure. Note that, since the angle-of-attack of the blade is 45 degrees, the pressure lift is identically equal to the pressure drag for this case. The normalized value of $F_Q$ shows that the three dominant modes for the aerodynamic lift force are modes 1, 3 and 5. This also suggests that these 3 modes could possibly be the dominant sources of the aerodynamic loading noise as well, since the dipole loading noise for a rotor is generated by the pressure force fluctuation on the rotor blades. While we expect the POD mode of $Q$  to be arranged with decreasing ``energy'' content, the net force depends on the product of $Q$ and the influence field ($\phi$). With the $\phi$ decreasing rapidly away from the surface of the rotor, the modes containing larger vortical structures near the rotor will contribute more toward the aerodynamic force and hence we do not see a monotonic reduction in the net force generated by these modes.

We find that the cumulative value of the vortex-induced lift force converges very rapidly to the value of the total vortex-induced lift force. Furthermore, the first six modes capture 91.3\% of the total variance in the lift. The corresponding time histories of the vortex-induced lift forces are shown in the figure \ref{fig:rw:podQ:fqvst:b} and we see that most of the large-amplitude fluctuations being captured by the dominant modes (1, 3, and 5). We note that while Mode-1 captures the most periodic component of the force fluctuation, the intermittent, lower-frequency force fluctuations are represented by Mode-3. Mode-5 on the other hand, generates smaller-scale but highly stochastic variations in the force. The sum of the first six modes (1-6) and the mean are a good approximation of the total force, as shown in figure \ref{fig:rw:podQ:fqvst:b} although we see in figure \ref{fig:rw:podQ:fqvst:a} that several higher modes contribute to the total vortex-induced lift force. Thus, even though a large number of modes generate contributions to the $Q$-field and the vortex-induced lift force, just a few of the first few modes of the $Q$-field recover most of the aerodynamic force. Thus, this modal force partitioning approach provides a highly compact representation of the force-generating modes.

\begin{figure}
    \centering
 \begin{subfigure}[b]{0.19\textwidth}
      \centering
         \includegraphics[width=\textwidth]{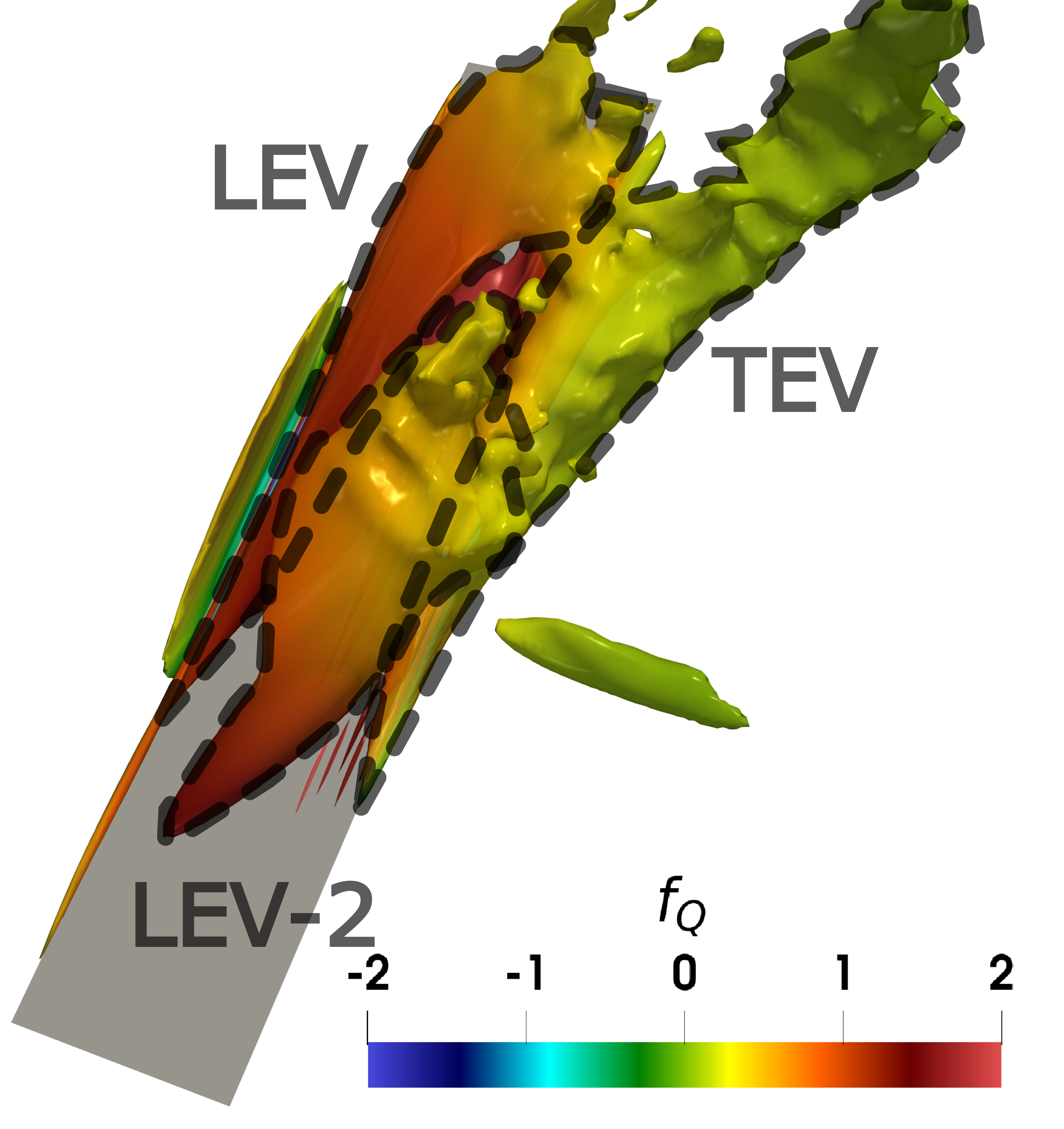}
         \caption{Mode-1}
         \label{fig:rw:podQ:US:Qisov2:a}
     \end{subfigure}
  \hfill
     \begin{subfigure}[b]{0.19\textwidth}
         \centering
         \includegraphics[width=\textwidth]{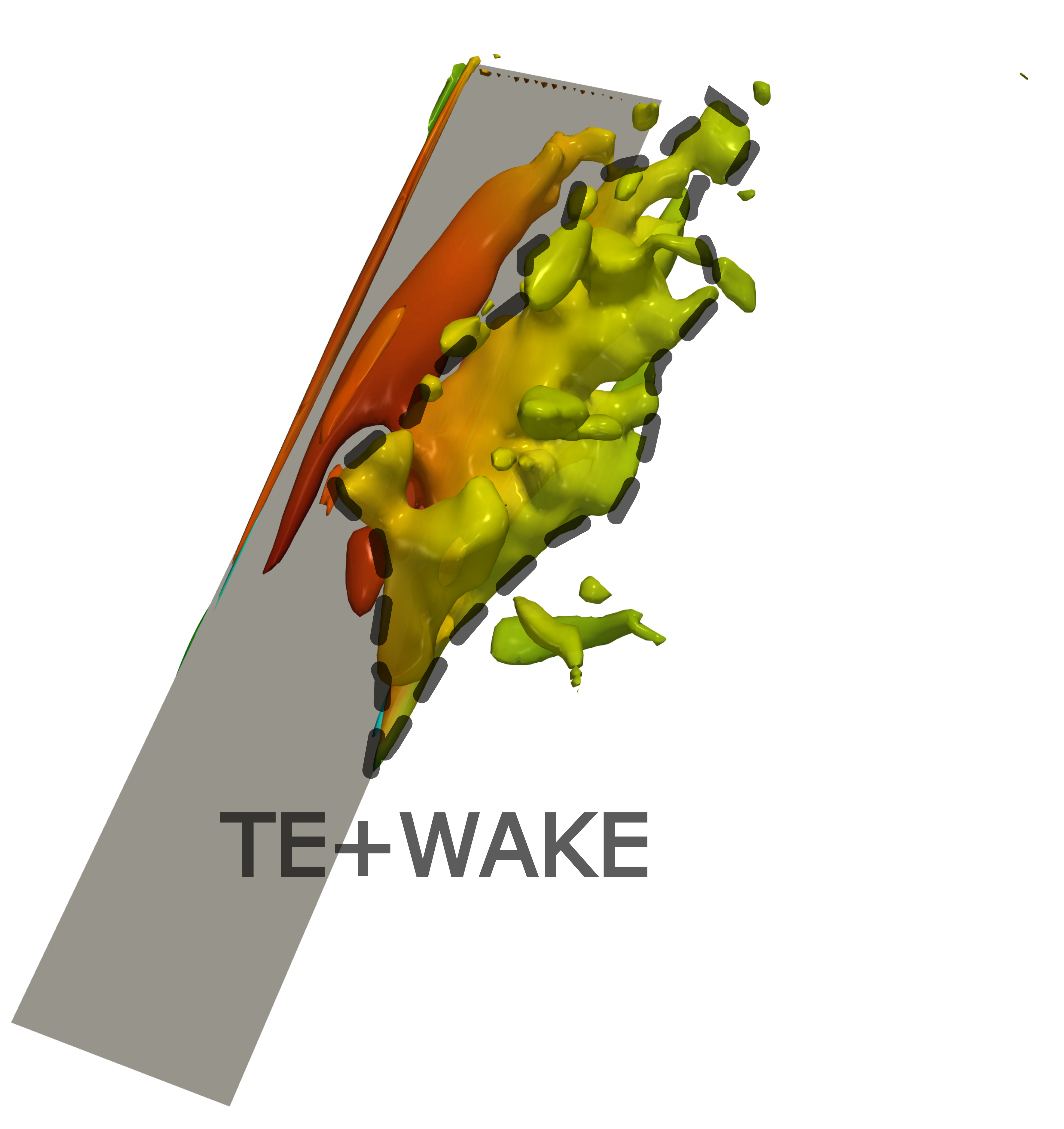}
         \caption{Mode-2}
         \label{fig:rw:podQ:US:Qisov2:b}
     \end{subfigure}
   \hfill
     \begin{subfigure}[b]{0.19\textwidth}
         \centering
         \includegraphics[width=\textwidth]{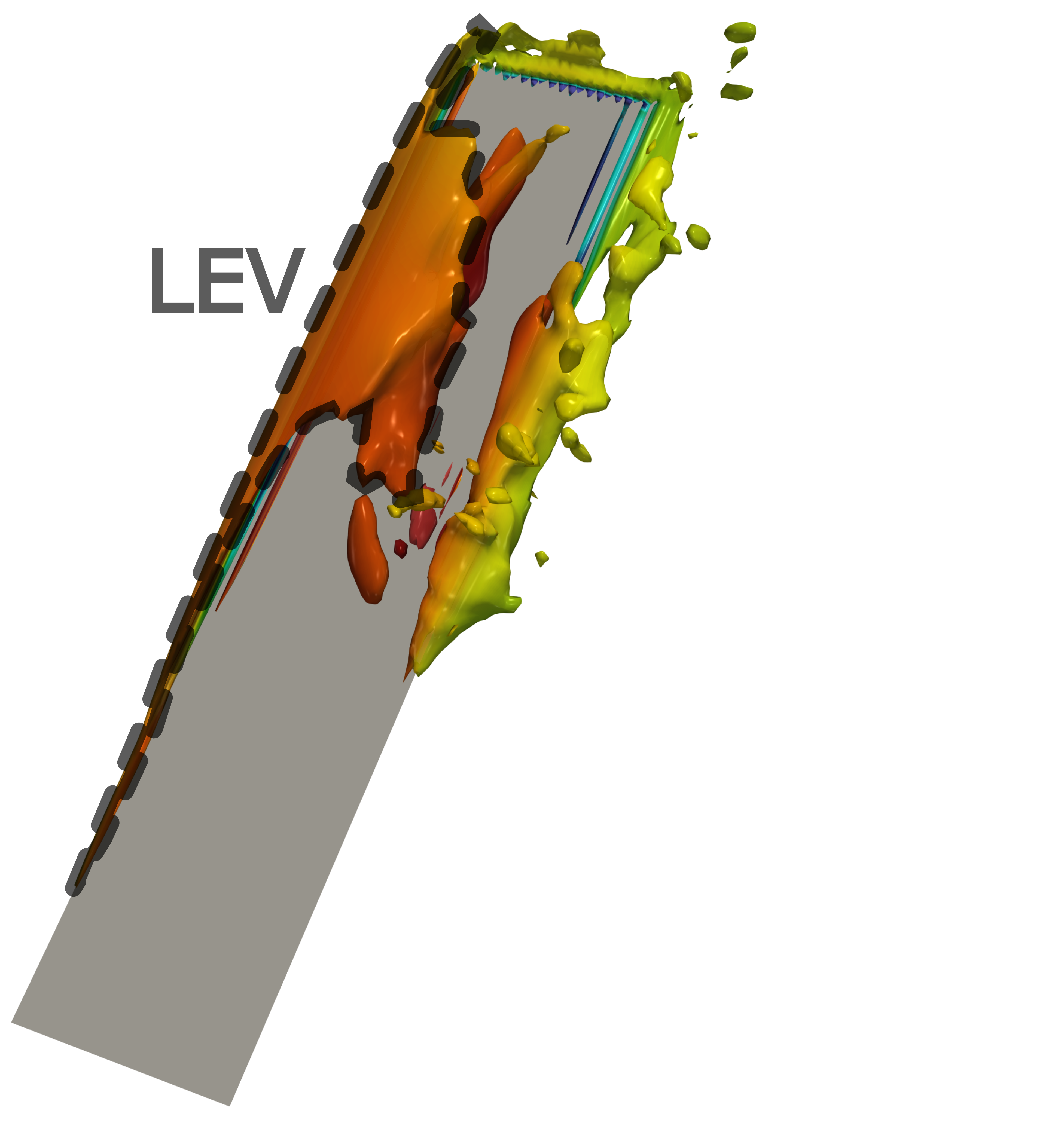}
         \caption{Mode-3}
         \label{fig:rw:podQ:US:Qisov2:c}
     \end{subfigure}
     \hfill
     \begin{subfigure}[b]{0.19\textwidth}
         \centering
         \includegraphics[width=\textwidth]{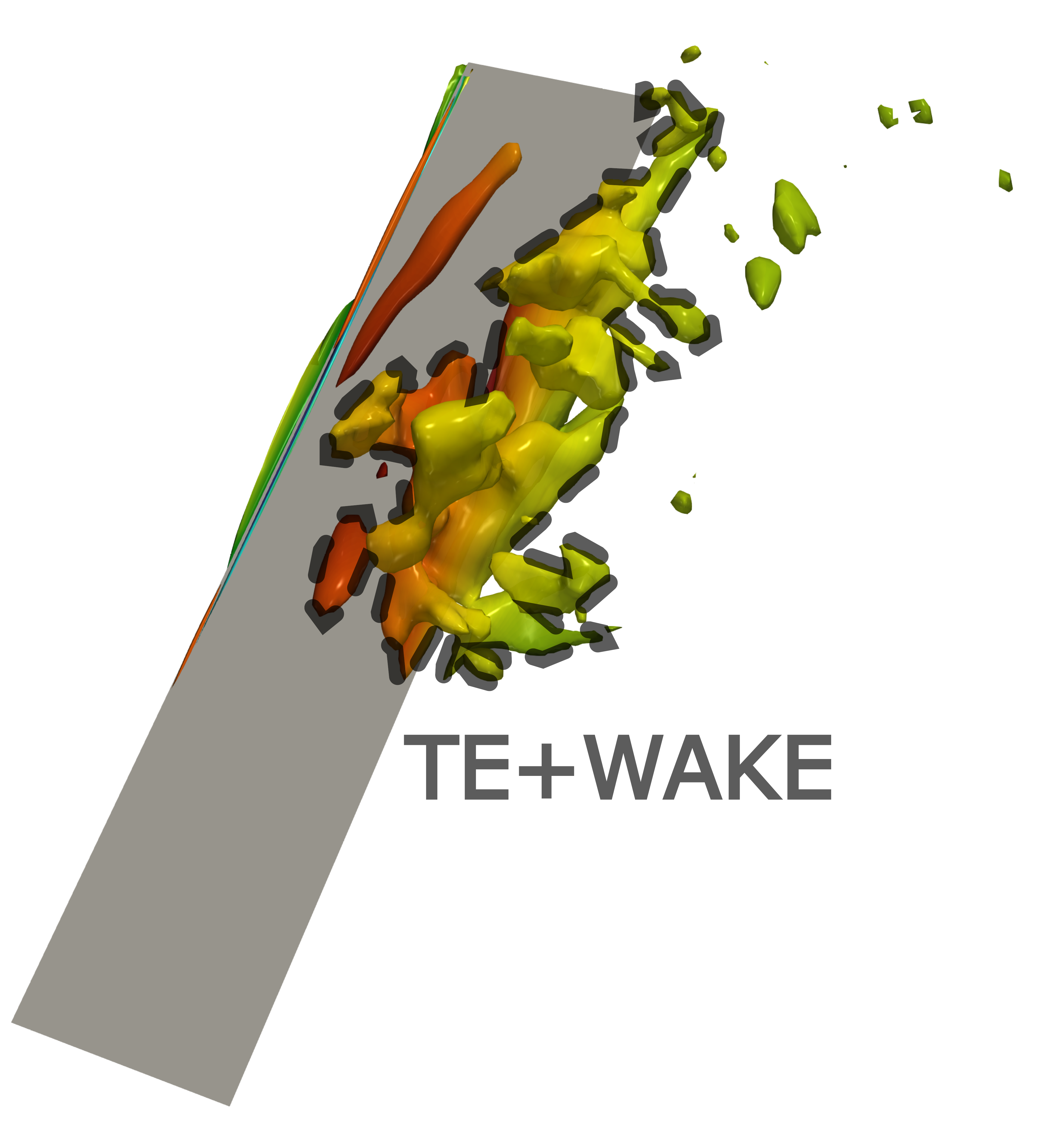}
         \caption{Mode-4}
         \label{fig:rw:podQ:US:Qisov2:d}
     \end{subfigure}
     \hfill
     \begin{subfigure}[b]{0.19\textwidth}
         \centering
         \includegraphics[width=\textwidth]{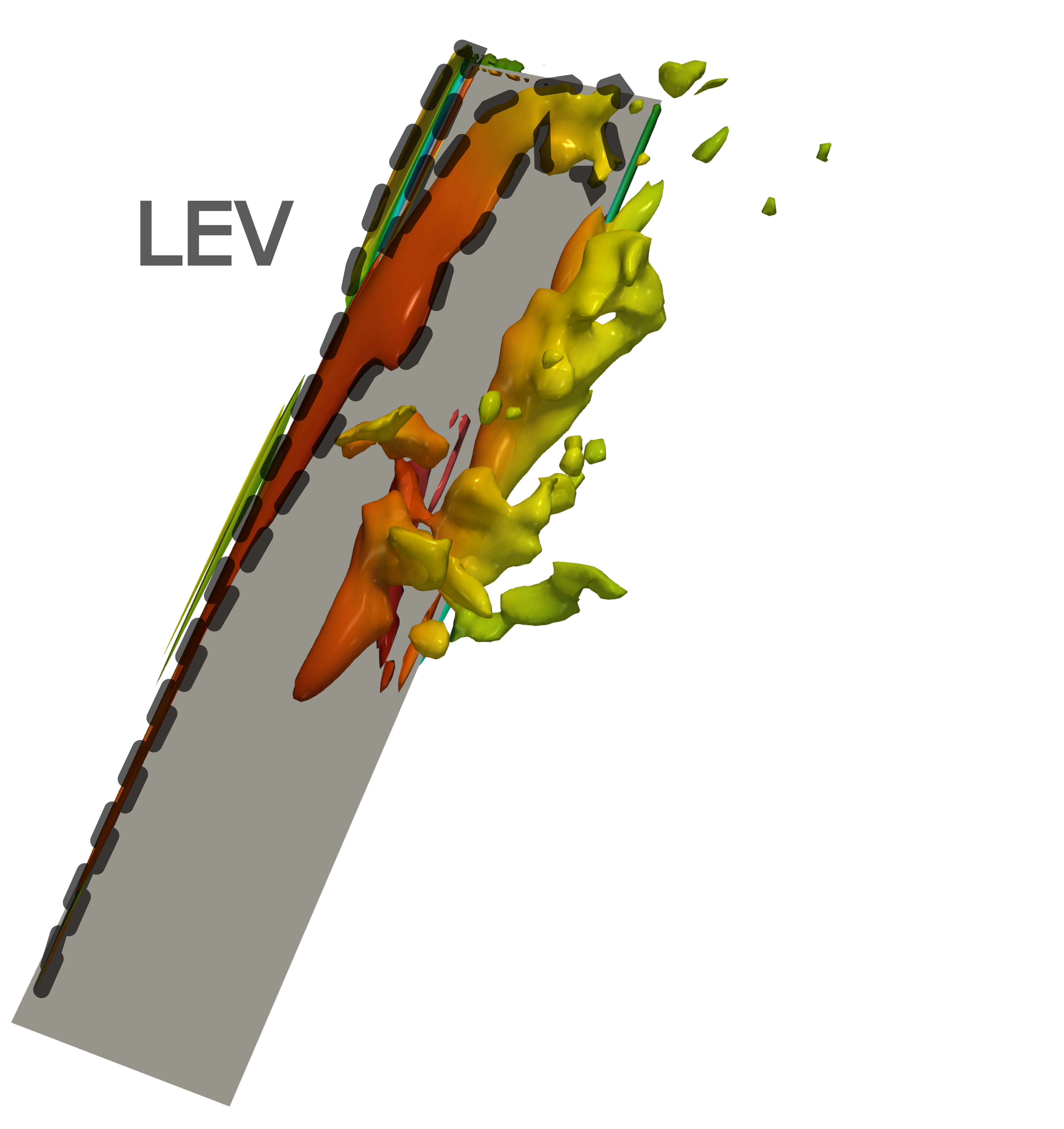}
         \caption{Mode-5}
         \label{fig:rw:podQ:US:Qisov2:e}
     \end{subfigure}
     \hfill
     \begin{subfigure}[b]{0.8\textwidth}
         \centering
         \includegraphics[width=\textwidth]{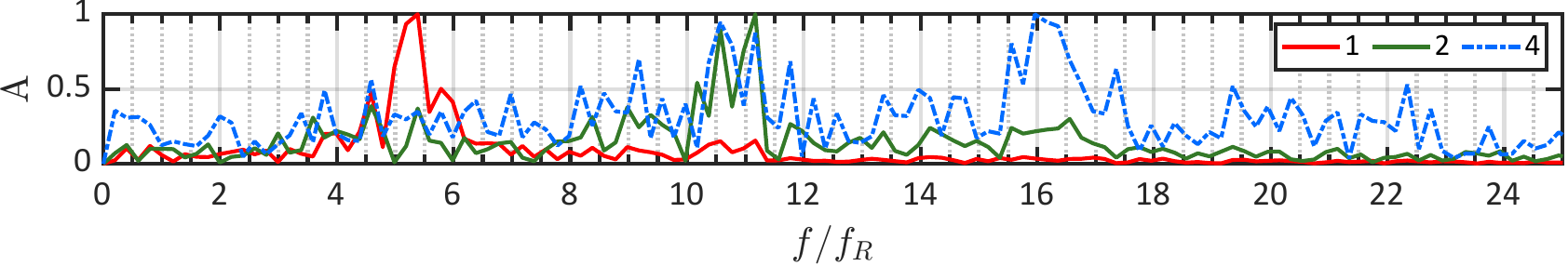}
         \caption{}
         \label{fig:rw:podQ:US:Qisov2:f}
     \end{subfigure}
     \hfill
     \begin{subfigure}[b]{0.8\textwidth}
         \centering
         \includegraphics[width=\textwidth]{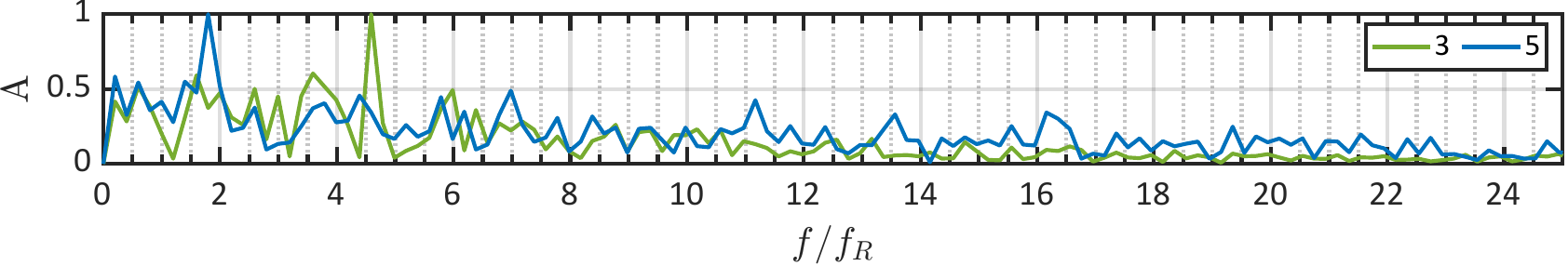}
         \caption{}
         \label{fig:rw:podQ:US:Qisov2:g}
     \end{subfigure}
      \hfill
     \begin{subfigure}[b]{0.8\textwidth}
         \centering
         \includegraphics[width=\textwidth]{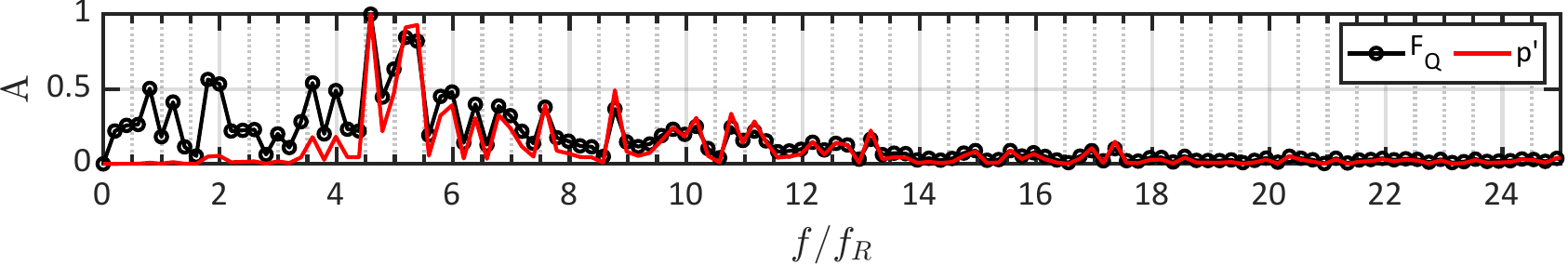}
         \caption{}
         \label{fig:rw:podQ:US:Qisov2:h}
     \end{subfigure}
        \caption{Results for modal force partitioning applied to POD of the $Q$-field. Plots of the scaled spatial eigenvectors ($U \Sigma$) of the rotor showing isosurface of $Q$-field colored by the vortex-induced lift force ($f_Q^{(2)})$ corresponding to the (a) Mode-1, (b) Mode-2, (c) Mode-3, (d) Mode-4 and (e) Mode-5. The frequency spectrum of the temporal eigenvector ($V^T$) of the POD for (f) Modes- 1,2,4 and (g) Modes- 3,5 (h) The frequency spectrum for the total vortex-induced lift force ($F_Q^{(2)}$) and the sound pressure ($p'$) at (200$R_c$, 90$^\circ$). The amplitudes for each mode is normalized by their respective maximum value to highlight the spectral content of each mode and the frequency is normalized by the revolution frequency \oc{($f_R$)}.}
\label{fig:rw:podQ:US:Qisov2}
\end{figure}

We now examine the spatial and temporal topology of the dominant force-generating modes by separating the spatial structure and temporal variation of these modes. The spatial structure of these dominant POD modes (1-5) is visualized via the spatial eigenvector times the corresponding eigenvalues ($U\Sigma$) in figure \ref{fig:rw:podQ:US:Qisov2}(a-e). While these modes have complex topologies, some useful observations can be made. First, Mode-1 is dominated by at least three distinct structures: an LEV that separates near the mid-span (denoted as ``LEV-1'' in the figure), an LEV that extends from the mid-span to the blade-tip (``LEV-2''), and a trailing-edge vortex (``TEV'') that also extends from the mid-span to the blade-tip. Modes-3 and 5 also clearly show leading-edge vortices (``LEV'' in the figures) whereas Modes-2 and 4 show structures associated with the trailing-edge and near wake vortices (``TE+wake'' in the figures). These topologies provide some sense of the vortical phenomenon that is represented by these modes and also provide a phenomenological connection to the force generation. For instance, since the leading-edge vortex is a major contributor to the aerodynamic lift, it is not surprising that Modes-1,3 and 5 provide most of the vortex-induced lift force (figure \ref{fig:rw:podQ:fqvst:a}).

Beyond the spatial structure, the temporal content of these modes is an important differentiator of these modes as well, and in figures \ref{fig:rw:podQ:US:Qisov2:f} and \ref{fig:rw:podQ:US:Qisov2:g} we show the frequency spectrum of the temporal eigenvectors contained within $V$ corresponding to these first five modes, and several observations can be made from these plots. First, Mode-1 is dominated by a fairly narrow band of frequencies centered around 5.2 (designated here by $f_0$). Mode-2 primarily contains $2f_0$, while Mode-4 is characterized by both $2f_0$ and $3f_0$. Thus, these three modes are likely connected with the periodic shedding of multiple vortices from the leading edge and the trailing-edge of the blade. Mode-3 has a clearly dominant frequency at 4.58 and this combined with the spatial structure suggests that this is the shedding of yet another LEV from near the 75\% span of the blade. Finally, Mode-5 corresponds to a low frequency (1.79) mode, which could be a subharmonic mode of one of the other dominant modes. Figure \ref{fig:rw:podQ:US:Qisov2:h} shows the spectra of the total lift force as well as that of the sound pressure at \oc{(200$R_c$, 90$^\circ$)} and we note that the dominant peaks in both these quantities are in the vicinity of a frequency of 5.0 and clearly connected with Modes-1 and 3. The lift force and sound pressure are in very good agreement for the higher frequencies. For the lower frequencies the sound pressure is of very small amplitude and this mismatch to the lift is due to the fact that the far-field sound originates from the time derivative of the force, and thus the amplitude in the low frequency range is reduced by its multiplication with the frequency.
\begin{figure}
    \centering
 \begin{subfigure}[b]{0.33\textwidth}
      \centering
         \includegraphics[width=\textwidth]{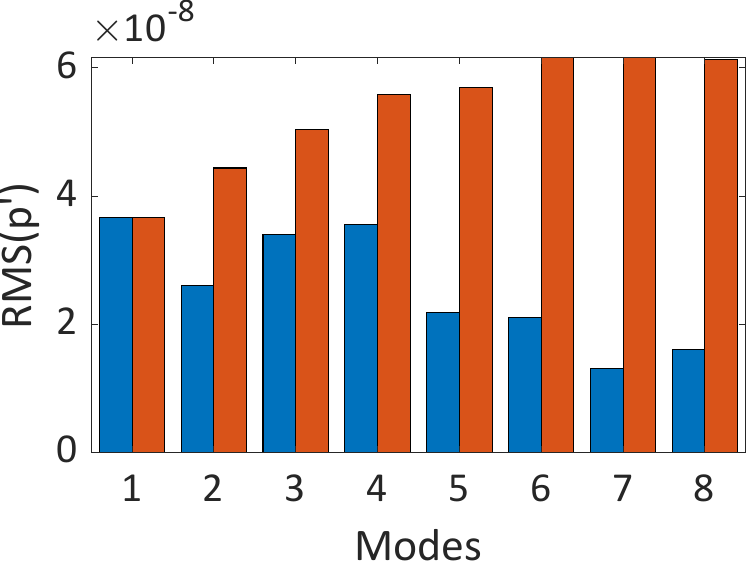}
         \caption{}
         \label{fig:rw:pod:Q:spl:a}
     \end{subfigure}
  \hfill
     \begin{subfigure}[b]{0.45\textwidth}
         \centering
        \includegraphics[width=\textwidth]{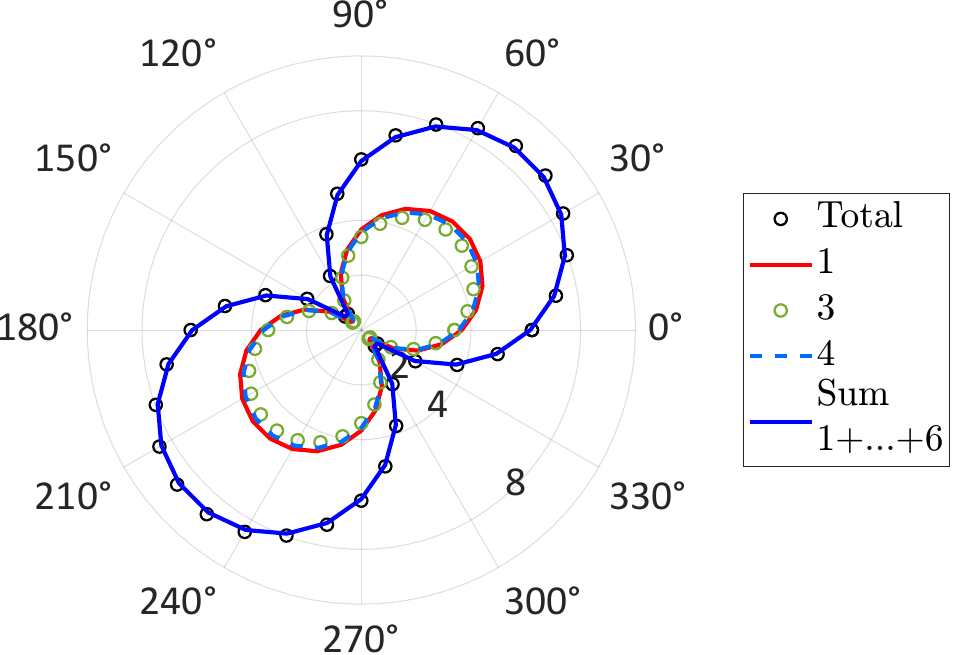}
         \caption{}
         \label{fig:rw:pod:Q:spl:b}
     \end{subfigure}
     \hfill
     \begin{subfigure}[b]{0.45\textwidth}
         \centering
        \includegraphics[width=\textwidth]{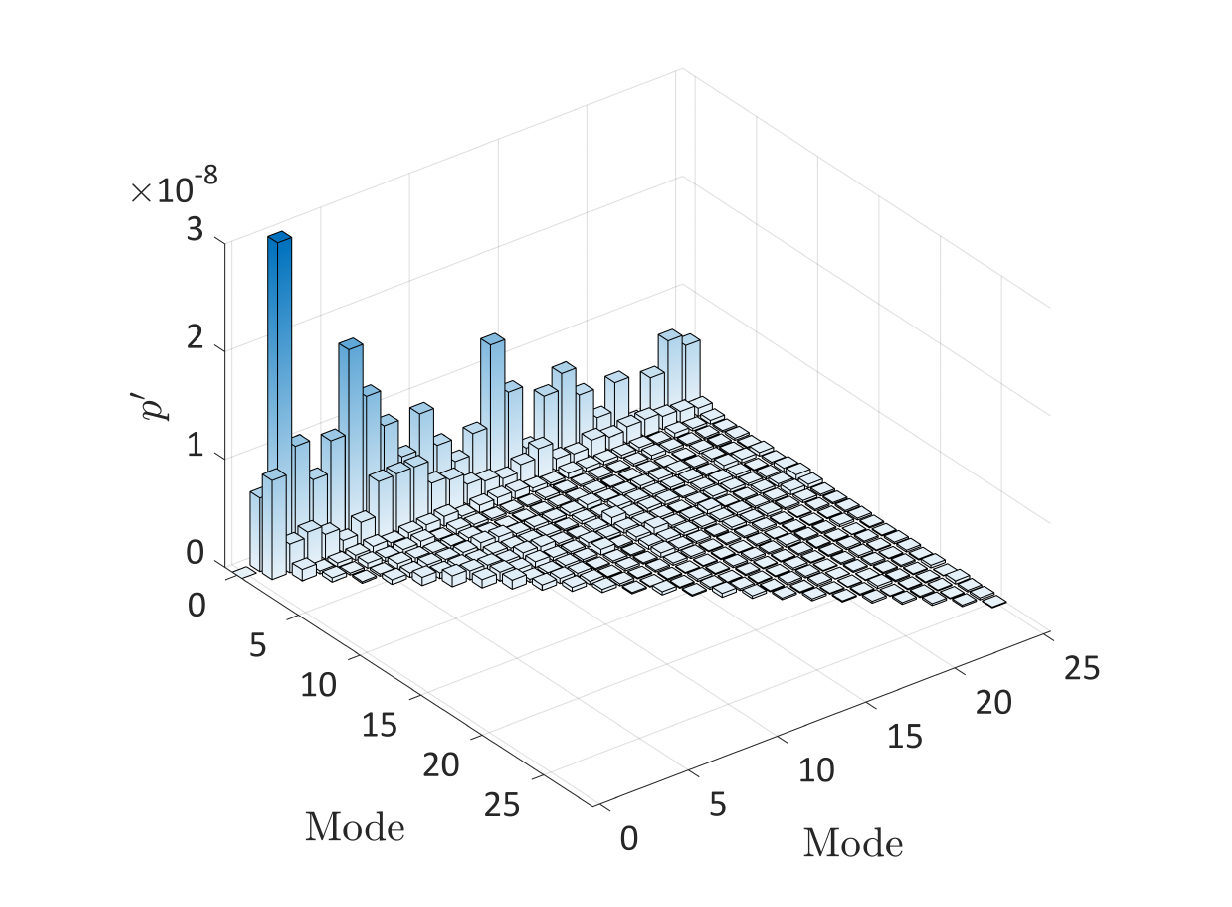}
         \caption{}
         \label{fig:rw:pod:Q:spl:c}
     \end{subfigure}
     \hfill
     \begin{subfigure}[b]{0.45\textwidth}
         \centering
        \includegraphics[width=\textwidth]{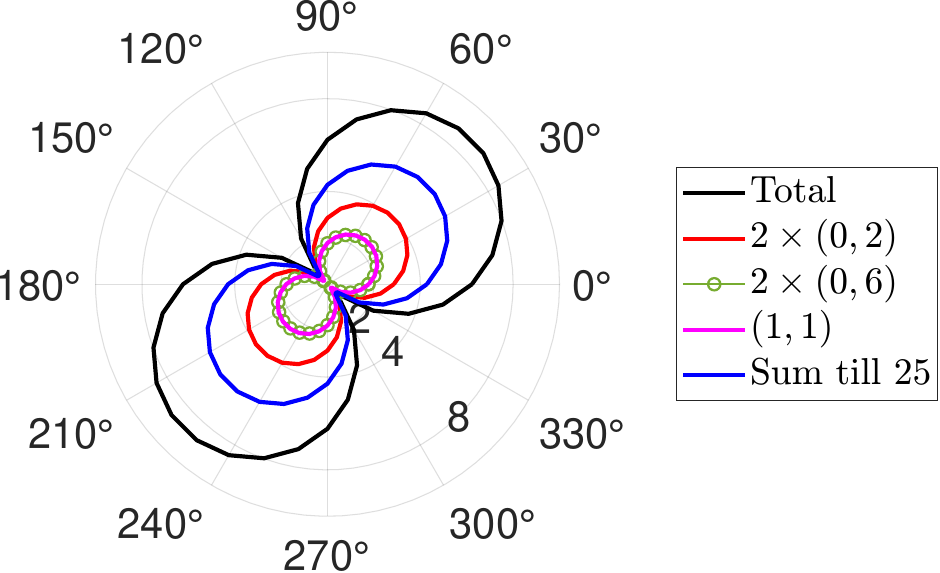}
         \caption{}
         \label{fig:rw:pod:Q:spl:d}
     \end{subfigure}
        \caption{Results from modal force and acoustic partitioning applied to the POD of the $Q$-field for the rotor blade (a) RMS value of the sound pressure for first 8 modes in blue and their cumulative values in red along with (b) the directivity \oc{($p'_{rms}[\times 10^{-8}]$)} for some dominant modes. Results from modal force and acoustic partitioning applied to the POD of the velocity field for the rotor blade showing (c) the RMS of sound pressure level for first 25 modal interaction and (d) the directivity \oc{($p'_{rms}[\times 10^{-8}]$)} corresponding to some dominant modes and modal interactions. The sound in all of these plots are measured at a distance \oc{200$R_c$ away and correspond to tip-velocity based Mach number of 0.25. The RMS values of the sound pressure level are recorded at a 90$^\circ$ angular orientation relative to the center of the rotor}}
\label{fig:rw:pod:Q:spl}
\end{figure}

Finally, the aeroacoustic sound generated by the revolving wing is calculated at a distance of \oc{200$R_c$ away from the revolution center and a tip-velocity-based Mach number of 0.25.} 
The RMS value of the sound pressure for each mode, recorded at an angular location of \oc{90$^\circ$ (this corresponds to a point vertically above the revolution center),} is shown in figure \ref{fig:rw:pod:Q:spl:a}. We note that similar to the vortex-induced lift force results, the sound pressure generated by these modes could easily be captured by a few dominant modes. Specifically, Modes-1,3 and 4 are found to be the top 3 dominant modes for the aeroacoustic noise. Figure \ref{fig:rw:pod:Q:spl:b} shows that by solely considering the first 6 modes, we almost fully match the sound directivity pattern for the rotor blade. 

For comparison, figures \ref{fig:rw:pod:Q:spl:c} and \ref{fig:rw:pod:Q:spl:d} show results from modal force and acoustic partitioning applied to the POD of the \emph{velocity field} for the rotor blade. The contributions of intra-modal and inter-modal interactions to the sound pressure are shown in \ref{fig:rw:pod:Q:spl:c} and we note that inter-modal interactions of Modes-0, 1, and 2 with all the other modes (we only show up to Mode-25) make significant contributions to the sound field. Indeed, the two largest contributions are from the (0,2) and (0,6) modes. We also note from figure \ref{fig:rw:pod:Q:spl:d} that even the addition of all the intra-modal and inter-modal contributions from Modes-1 to 25 modes does not provide a good estimate of the total sound directivity. \cite{Chiu2023} employed the ``force representation theory'' wherein, instead of the $Q$, the force density is based on the Lamb-vector as follows: $\left( \bf{u}\times \bf{\omega} \right) \cdot \nabla \phi$. They applied their method to a SPOD of the velocity flow field, and due to the non-linearity of the Lamb-vector, they also encountered inter-modal interaction between the SPOD modes, which complicated the interpretability of their results. Based on our experience, a direct SPOD of the Lamb-vector would alleviate this issue for their method as well.

\section{Conclusions}
In order to identify the flow structures (or modes) that contribute significantly to aerodynamic force and aeroacoustic sound generation in complex flows, we propose a modal force and acoustic decomposition method by combining the previously developed force and acoustic partitioning method (FAPM) \citep{zhang2015centripetal,menon2021a,Seo2022} with the modal decomposition of flows.
The modal FAPM is applied first to a canonical case of flow past a 2D circular cylinder at Re=300 and flow past a NACA 0015 airfoil at a higher Reynolds number of 2500 to examine the implications of applying FAPM to modes obtained from various decomposition techniques such as Reynolds decomposition, triple decomposition, and POD. Because the aerodynamic force is determined by the $Q$-field, which is a non-linear observable of the flow, modal decomposition of the velocity field leads to a situation where non-linear interactions between the decomposed modes become important in the determination of the aerodynamic loads and aeroacoustic sound. These inter-modal interactions make it difficult to interpret the contributions of individual modes to aerodynamic loading and aeroacoustic sound.

Based on this observation, we propose the direct modal decomposition of the $Q$-field. This approach eliminates inter-modal interactions from modal aerodynamic loads and aeroacoustic sound associated with the modes. The application of this approach to the circular cylinder flow shows that the modes and forces obtained from this decomposition exhibit certain symmetries that make it easier to interpret the significant of different modes. When triple decomposition is applied directly to the $Q$-field for the airfoil, we find, for instance, that the sound induced by the incoherent component has a larger contribution from the induced drag than for the coherent contribution. This has interpretable implications for the directivity of sound for these two modes. We also use modal acoustic partitioning to compute the distinct sound spectra for the coherent and incoherent modes. To our knowledge, this type of partitioning of sound sources has not been demonstrated before.

The final application of the modal force and acoustic partitioning is to a relatively complex case of a revolving rectangular wing at a tip Reynolds number of 3300. This configuration generates a complex three-dimensional flow that requires 119 POD velocity modes to reconstruct 98\% of the velocity field. Modal force and acoustic partitioning applied to the POD modes obtained from the velocity field exhibit very significant contributions from inter-modal interactions. In contrast, POD of the $Q$ field eliminates the inter-modal interactions and indicates that 3 dominant modes modes (Mode-1, 3 and 4) contribute 69.4\% of the total aeroacoustic sound (calculated based on the acoustic intensity) and the sum for the first 6 POD modes recovers nearly all the aeroacoustic sound contributions. Thus, POD applied to the $Q$ field demonstrates that even for this relatively complex three-dimensional flow, the aerodynamic loads and aeroacoustic sound can be represented by a set of modes that are significantly more compact than those required for the flow itself.

Modal decompositions are primarily considered tools for examining the kinematics of the flow and for developing low-dimensional representations of the flow field. However, the contributions of these modes to the aerodynamic loads, which are quite often key quantities of interest, are not usually considered because there has not been a systematic way to compute the aerodynamic loads for the decomposed flow modes. The modal force partitioning method proposed here provides this capability. The direct modal decomposition of $Q$ enables us to determine the modes that make the dominant contribution to the aerodynamic loads. This also suggests that $Q$, which has mostly been used for identification of vortices in flows, is a non-linear observable with exceptional importance in the analysis of aerodynamic loads.  In addition to the applications for determining the mechanisms for the generation of aerodynamic loads and aeroacoustic sound explored here, it is expected that modal force partitioning will find use in several other arenas, including flow-induced vibration/flutter and flow control. 

In closing we note that while the current paper had focused on Reynolds decomposition, triple-decomposition and POD, modal force partitioning can in principle be applied equally well to other modal decomposition techniques. \oc{In particular, new methods for dimensionality reduction based on machine-learning techniques and neural-networks are being developed \citep{sam_ae,vinuesa_ae} and it would be interesting to apply FPM in conjunction with these techniques. Finally, while the current study applies mFPM to results from DNS, FPM and mFPM could also be applied to data from large-eddy simulations where one would have to account for the contributions of the subgrid-scale stress term to the surface pressure force. These are useful directions to pursue in a future study.}

\section{Acknowledgments}
The authors acknowledge support from the Army Research Office
(Cooperative Agreement No. W911NF2120087) for this work. Computational resources for this work were provided by the Advanced Research Computing at Hopkins (ARCH) core facility (rockfish.jhu.edu) and high-performance computer time and resources from the DoD High Performance Computing Modernization Program.

\section{Declaration of Interests}
The authors report no conflict of interest.

\appendix
\section{Derivation of the Force Partitioning Method}
\label{appendix:fpm}
The derivation of FPM can be found in previous papers \citep{zhang2015centripetal,menon2021c,menon2022} but it included here for the sake of completeness. The gradient of the field of influence ($\phi$) (equation \ref{eqn:phi}) is projected onto the Navier-Stokes equation (equation \ref{momeqn}) which results in the following equation:
\begin{equation}
   \int_{V_f} \rho \left[ \frac{\partial {\bf u}}{\partial t} + {\bf u} \cdot \nabla {\bf u} \right]\cdot \nabla\phi^{(i)} dV= \int_{V_f} \left[ -\bold{\nabla} P +\mu \nabla^2 {\bf u} \right] \cdot \bold{\nabla} \phi^{(i)} dV .
\end{equation}
Rearranging the terms, we get:
\begin{equation}
    -\int_{V_f}\bold{\nabla} P \cdot \bold{\nabla} \phi^{(i)} dV= \int_{V_f} \left[\rho \frac{\partial {\bf u}}{\partial t}\cdot \nabla\phi^{(i)} + \rho ({\bf u} \cdot \nabla {\bf u})\cdot \nabla\phi^{(i)} - \mu \nabla^2 {\bf u} \cdot \nabla\phi^{(i)} \right] dV .
    \label{eqn:FPM:step1}
\end{equation}
Since $\nabla \cdot(P \nabla \phi^{(i)})=\nabla P\cdot \nabla \phi^{(i)} + P\cdot \nabla^2 \phi^{(i)}$, rearranging these terms, we see that $\nabla P\cdot \nabla \phi^{(i)}=\nabla \cdot(P \nabla \phi^{(i)})-P\cdot \nabla^2 \phi^{(i)}$. Thus the term on the left hand side of equation \ref{eqn:FPM:step1} can be rewritten as,
\begin{equation}
     -\int_{V_f}{\bf \nabla} P \cdot {\bf \nabla} \phi^{(i)} dV= -\int_{V_f}\nabla \cdot(P \nabla \phi^{(i)})dV + \int_{V_f}P\cdot \nabla^2 \phi^{(i)} dV \,\, ,
\end{equation}
Here, the last term is zero due to equation \ref{eqn:phi} and using the Gauss divergence theorem, $-\int_{V_f}\nabla \cdot(P \nabla \phi^{(i)})dV= -\int_{B+\Sigma} P {\bf n}\cdot \nabla \phi^{(i)}dS$ which can be further simplified using equation \ref{eqn:phi:bc} and we get,

\begin{equation}
     -\int_{V_f}{\bf \nabla} P \cdot {\bf \nabla} \phi^{(i)} dV=  -\int_{B} P n_i dS \,\, .
     \label{eqn:fpm:left1}
\end{equation}
We note that $\nabla \cdot \frac{\partial {\bf u}}{\partial t} \phi =0$ due to the continuity equation and therefore,
\begin{equation}
    \nabla \cdot (\frac{\partial {\bf u}}{\partial t} \phi^{(i)} ) = \nabla \cdot \frac{\partial {\bf u}}{\partial t}\phi^{(i)} + \frac{\partial {\bf u}}{\partial t} \cdot \nabla \phi^{(i)} =\frac{\partial {\bf u}}{\partial t} \cdot \nabla \phi^{(i)} \,\, ,
\end{equation}
The first term on the RHS of the equation \ref{eqn:FPM:step1} can then be rewritten as,
\begin{equation}
    \int_{V_f} \rho \frac{\partial {\bf u}}{\partial t}\cdot \nabla\phi^{(i)} dV = \int_{V_f} \rho\nabla \cdot (\frac{\partial {\bf u}}{\partial t} \phi^{(i)} ) dV= \int_{B+\Sigma} \rho{\bf n} \cdot (\frac{\partial {\bf u}}{\partial t} \phi^{(i)} ) dS \,\, .
    \label{eqn:fpm_term2_temp}
\end{equation}
Similarly, 
\begin{equation}
    \nabla \cdot (({\bf u}\cdot \nabla {\bf u}) \phi^{(i)}) = \nabla \cdot ({\bf u}\cdot \nabla {\bf u})\phi^{(i)} + ({\bf u}\cdot \nabla {\bf u})\cdot \nabla\phi^{(i)} \,\, ,
\end{equation}
and hence the second RHS term of equation \ref{eqn:FPM:step1} can be modified as,
\begin{equation}
    \int_{V_f} \rho ({\bf u} \cdot \nabla {\bf u})\cdot \nabla\phi^{(i)} dV=\int_{V_f} \rho \nabla \cdot ({\bf u}\cdot \nabla {\bf u}\phi^{(i)}) dV-\int_{V_f} \rho \nabla \cdot ({\bf u}\cdot \nabla {\bf u})\phi^{(i)} dV \,\, .
\end{equation}
The first RHS term in the above equation can again be converted to surface integral and thus we can write, $\int_{V_f}  \nabla \cdot ({\bf u}\cdot \nabla {\bf u}\phi^{(i)}) dV=\int_{B+\Sigma}  {\bf n} \cdot ({\bf u}\cdot \nabla {\bf u}\phi^{(i)}) dS$. To simplify the second RHS term in the above equation, we use equation \ref{Qcriteqn} and write $ \nabla \cdot ({\bf u}\cdot \nabla {\bf u})\phi= -2Q\phi^{(i)}$. Thus, 
\begin{equation}
    \int_{V_f} \rho ({\bf u} \cdot \nabla{\bf u})\cdot \nabla\phi^{(i)} dV=\int_{B+\Sigma} \rho {\bf n} \cdot ({\bf u}\cdot \nabla {\bf u}\phi^{(i)}) dS+\int_{V_f}  2\rho Q\phi^{(i)} dV \,\, .
    \label{eqn:fpm_term3_temp}
\end{equation}
The equations \ref{eqn:fpm_term2_temp} and \ref{eqn:fpm_term3_temp} can be combined to give:
\begin{equation}
    \int_{V_f} \rho \frac{\partial {\bf u}}{\partial t}\cdot \nabla\phi^{(i)} dV + \int_{V_f} \rho ({\bf u} \cdot \nabla {\bf u})\cdot \nabla\phi^{(i)} dV= \int_{B+\Sigma} \rho \frac{D{\bf u}}{Dt} \cdot {\bf n} \phi^{(i)} dS +\int_{V_f}  2\rho Q\phi^{(i)} dV
    \label{eqn:fpm_rhs_2_3}
\end{equation}

Next, we consider the last term of equation \ref{eqn:FPM:step1} and note that $\nabla \cdot (\nabla^2 {\bf u} \phi^{(i)})= \nabla \cdot (\nabla^2 {\bf u})\phi^{(i)} + \nabla^2 {\bf u} \cdot \nabla\phi^{(i)} $
and since $\nabla \cdot (\nabla^2 {\bf u})\phi^{(i)}= (\nabla^2 \nabla \cdot{\bf u})\phi^{(i)}=0$ due to the continuity equation, we can write
\begin{equation}
   \int_{V_f}\mu \nabla^2 {\bf u} \cdot \nabla\phi^{(i)} dV=  \int_{V_f} \mu\nabla \cdot (\nabla^2 {\bf u} \phi^{(i)}) dV \,\, ,
\end{equation} 
which is further simplified using the Gauss divergence theorem to give surface integral as,
\begin{equation}
    \int_{V_f} \mu \nabla \cdot (\nabla^2 {\bf u} \phi^{(i)}) dV = \int_{B+\Sigma} (\mu \nabla^2 {\bf u}\cdot {\bf n}) \phi^{(i)} dS
    \label{eqn:fpm:rhs_4}
\end{equation}
We see that combining equations \ref{eqn:fpm:left1}, \ref{eqn:fpm_rhs_2_3}, \ref{eqn:fpm:rhs_4} gives us the equation \ref{fpmEqn}.
\section{Grid Convergence}
\label{appendix:gc} 

\begin{figure}
    \centering
     \begin{subfigure}[b]{0.6\textwidth}
         \centering
         \includegraphics[width=\textwidth]{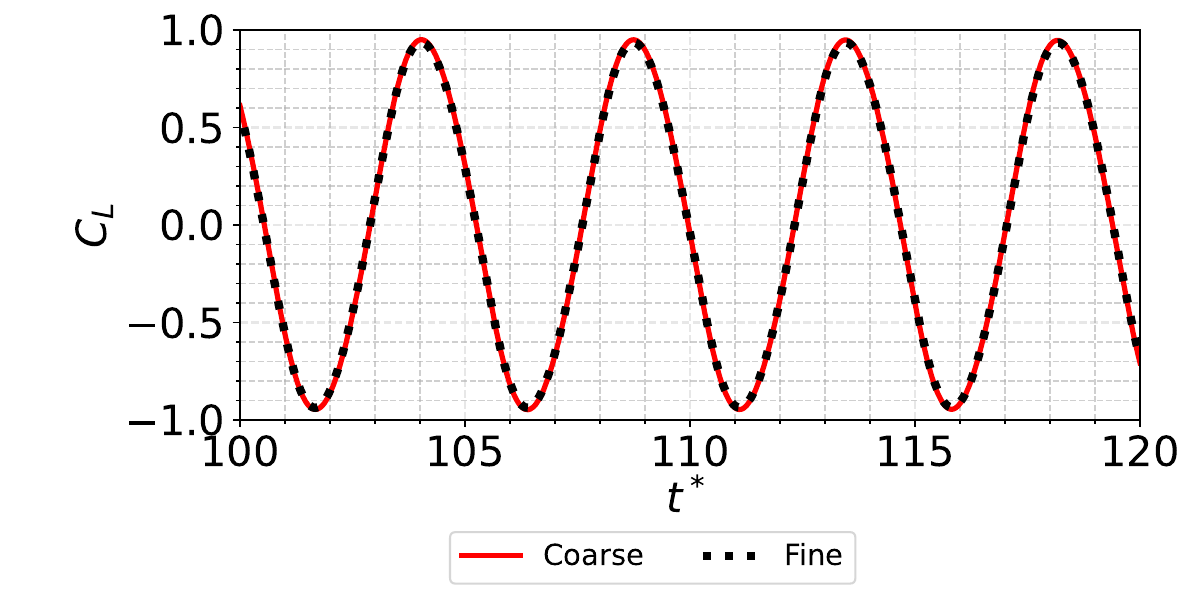}
         \caption{}
         \label{fig:app:grid_conv:a}
     \end{subfigure}
   \hfill
     \begin{subfigure}[b]{0.6\textwidth}
         \centering
         \includegraphics[width=\textwidth]{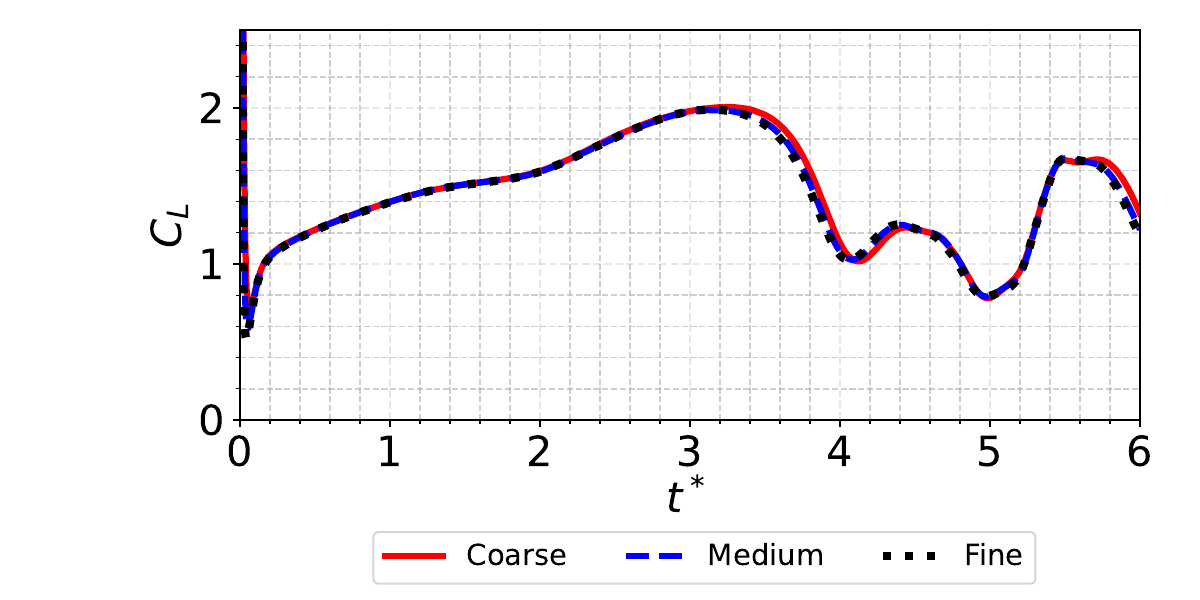}
         \caption{}
         \label{fig:app:grid_conv:b}
     \end{subfigure}
       \hfill
     \begin{subfigure}[b]{0.6\textwidth}
         \centering
         \includegraphics[width=\textwidth]{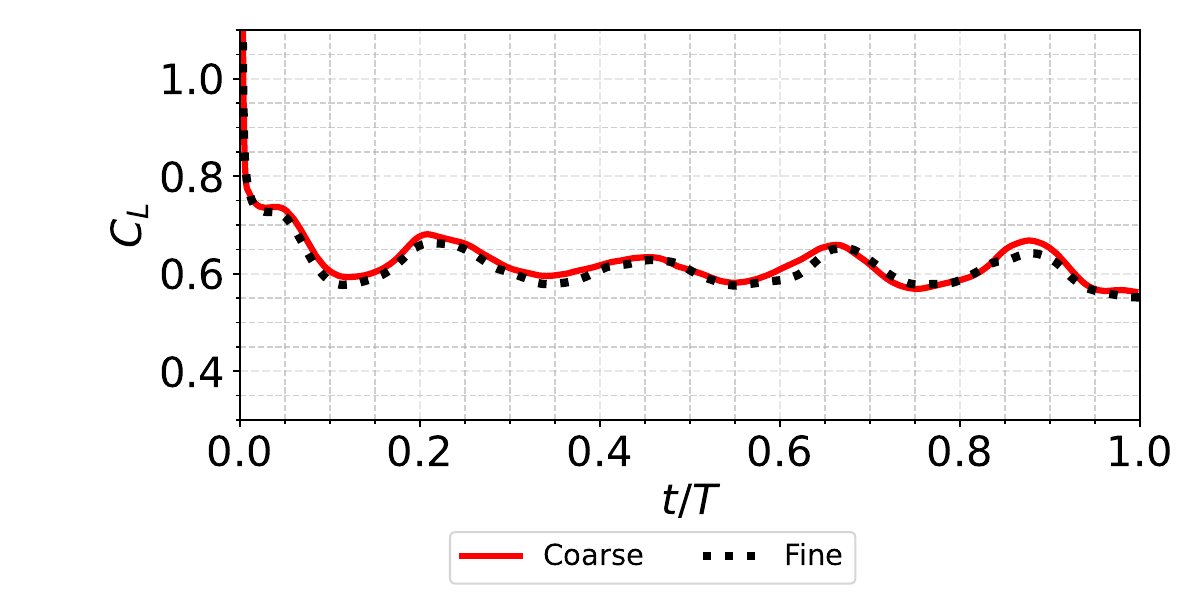}
         \caption{}
         \label{fig:app:grid_conv:c}
     \end{subfigure}
        \caption{ Grid convergence shown for (a) circular cylinder, (b) airfoil and (c) revolving wing.}
\label{fig:app:grid_conv}
\end{figure}

The grid convergence for the circular cylinder is shown in figure \ref{fig:app:grid_conv:a} with the coarse and fine mesh corresponding to 580K and 990K total grid points respectively. The time series for the fine mesh lift coefficient is shifted backwards to match the phase of the time series on the coarse mesh and the percentage difference between the RMS values of the lift coefficient is 1.34\%. The case with fine mesh is used for all the circular cylinder results presented in this paper.

The grid convergence for the airfoil is shown in figure \ref{fig:app:grid_conv:b} with the coarse, medium and fine mesh corresponding to 423K, 665K and 960K  grid points respectively. The percentage difference between the RMS values of the lift coefficients between the coarse and medium grid size is 1.26\% and this error reduces to 0.62\% between medium and fine grid size. The medium grid is used for all the airfoil results presented in this paper. 

The grid convergence for the revolving wing is shown in figure \ref{fig:app:grid_conv:c} with the coarse and fine mesh containing around 9.6M and 16M grid points respectively. The RMS values of the lift coefficient between these two mesh differ by 1.61\%. The fine mesh is used for all the results presented in this paper.


\begin{thebibliography}{56}
\providecommand{\natexlab}[1]{#1}
\providecommand{\url}[1]{\texttt{#1}}
\expandafter\ifx\csname urlstyle\endcsname\relax
  \providecommand{\doi}[1]{doi: #1}\else
  \providecommand{\doi}{doi: \begingroup \urlstyle{rm}\Url}\fi

\bibitem[Aghaei-Jouybari et~al.(2022)Aghaei-Jouybari, Seo, Yuan, Mittal, and Meneveau]{aghaei2022contributions}
M.~Aghaei-Jouybari, J.-H. Seo, J.~Yuan, R.~Mittal, and C.~Meneveau.
\newblock Contributions to pressure drag in rough-wall turbulent flows: insights from force partitioning.
\newblock \emph{Physical Review Fluids}, 7\penalty0 (8):\penalty0 084602, 2022.

\bibitem[Bailoor et~al.(2021)Bailoor, Seo, Schena, and Mittal]{bailoor_heart}
S.~Bailoor, J.-H. Seo, S.~Schena, and R.~Mittal.
\newblock Detecting aortic valve anomaly from induced murmurs: insights from computational hemodynamic models.
\newblock \emph{Frontiers in Physiology}, 12:\penalty0 734224, 2021.

\bibitem[Baj et~al.(2015)Baj, Bruce, and Buxton]{trip_decomp_ex}
P.~Baj, P.~J.~K. Bruce, and O.~R.~H. Buxton.
\newblock The triple decomposition of a fluctuating velocity field in a multiscale flow.
\newblock \emph{Physics of Fluids}, 27\penalty0 (7):\penalty0 075104, 07 2015.
\newblock ISSN 1070-6631.
\newblock \doi{10.1063/1.4923744}.

\bibitem[Berkooz et~al.(1993)Berkooz, Holmes, and Lumley]{lumleyPOD}
G.~Berkooz, P.~Holmes, and J.~L. Lumley.
\newblock The {Proper} {Orthogonal} {Decomposition} in the {Analysis} of {Turbulent} {Flows}.
\newblock \emph{Annual Review of Fluid Mechanics}, 25\penalty0 (1):\penalty0 539--575, Jan. 1993.
\newblock ISSN 0066-4189, 1545-4479.
\newblock \doi{10.1146/annurev.fl.25.010193.002543}.

\bibitem[Brentner and Farassat(1994)]{farassat_noise}
K.~S. Brentner and F.~Farassat.
\newblock Helicopter noise prediction: the current status and future direction.
\newblock \emph{Journal of Sound and Vibration}, 170\penalty0 (1):\penalty0 79--96, 1994.

\bibitem[Brentner and Farassat(1998)]{fwheqn2}
K.~S. Brentner and F.~Farassat.
\newblock Analytical comparison of the acoustic analogy and kirchhoff formulation for moving surfaces.
\newblock \emph{AIAA Journal}, 36\penalty0 (8):\penalty0 1379--1386, 1998.
\newblock \doi{10.2514/2.558}.

\bibitem[Candeloro et~al.(2022)Candeloro, Ragni, and Pagliaroli]{DroneNoiseRev}
P.~Candeloro, D.~Ragni, and T.~Pagliaroli.
\newblock Small-scale rotor aeroacoustics for drone propulsion: A review of noise sources and control strategies.
\newblock \emph{Fluids}, 7\penalty0 (8), 2022.
\newblock ISSN 2311-5521.
\newblock \doi{10.3390/fluids7080279}.

\bibitem[Canuto et~al.(2006)Canuto, Hussaini, Quarteroni, and Zang]{canuto_spectral_2006}
C.~Canuto, M.~Y. Hussaini, A.~Quarteroni, and T.~A. Zang.
\newblock \emph{Spectral {Methods}: {Fundamentals} in {Single} {Domains}}.
\newblock Scientific {Computation}. Springer Berlin Heidelberg, Berlin, Heidelberg, 2006.
\newblock ISBN 978-3-540-30725-9 978-3-540-30726-6.
\newblock \doi{10.1007/978-3-540-30726-6}.

\bibitem[Chang(1992)]{chang1992potential}
C.-C. Chang.
\newblock Potential flow and forces for incompressible viscous flow.
\newblock \emph{Proceedings of the Royal Society of London. Series A: Mathematical and Physical Sciences}, 437\penalty0 (1901):\penalty0 517--525, 1992.

\bibitem[Chatterjee(2000)]{chatterPOD}
A.~Chatterjee.
\newblock An introduction to the proper orthogonal decomposition.
\newblock \emph{Current Science}, 78\penalty0 (7):\penalty0 808--817, 2000.
\newblock ISSN 00113891.
\newblock URL \url{http://www.jstor.org/stable/24103957}.

\bibitem[Chiu et~al.(2023)Chiu, Tseng, Chang, and Chou]{Chiu2023}
T.-Y. Chiu, C.-C. Tseng, C.-C. Chang, and Y.-J. Chou.
\newblock Vorticity forces of coherent structures on the naca0012 aerofoil.
\newblock \emph{Journal of Fluid Mechanics}, 974:\penalty0 A52, 2023.
\newblock \doi{10.1017/jfm.2023.815}.

\bibitem[Ffowcs~Williams et~al.(1969)Ffowcs~Williams, Hawkings, and Lighthill]{FWHorigpaper}
J.~E. Ffowcs~Williams, D.~L. Hawkings, and M.~J. Lighthill.
\newblock Sound generation by turbulence and surfaces in arbitrary motion.
\newblock \emph{Philosophical Transactions of the Royal Society of London. Series A, Mathematical and Physical Sciences}, 264\penalty0 (1151):\penalty0 321--342, 1969.
\newblock \doi{10.1098/rsta.1969.0031}.

\bibitem[Fukami and Taira(2023)]{sam_ae}
K.~Fukami and K.~Taira.
\newblock Grasping extreme aerodynamics on a low-dimensional manifold.
\newblock \emph{Nature Communications}, 14\penalty0 (1):\penalty0 6480, 2023.

\bibitem[Glegg and Devenport(2017)]{lowmachaerobook}
S.~Glegg and W.~Devenport.
\newblock \emph{Aeroacoustics of Low Mach Number Flows: Fundamentals, Analysis, and Measurement}.
\newblock Elsevier Science, 2017.
\newblock ISBN 9780128097939.

\bibitem[Gururaj et~al.(2021)Gururaj, Moaven, Tan, Thurow, and Raghav]{45RotorExp}
A.~Gururaj, M.~Moaven, Z.~P. Tan, B.~Thurow, and V.~Raghav.
\newblock Rotating three-dimensional velocimetry.
\newblock \emph{Experiments in Fluids}, 62, 2021.
\newblock \doi{https://doi.org/10.1007/s00348-021-03241-4}.

\bibitem[Herrmann et~al.(2021)Herrmann, Baddoo, Semaan, Brunton, and McKeon]{brunton_resolvent}
B.~Herrmann, P.~J. Baddoo, R.~Semaan, S.~L. Brunton, and B.~J. McKeon.
\newblock Data-driven resolvent analysis.
\newblock \emph{Journal of Fluid Mechanics}, 918:\penalty0 A10, 2021.

\bibitem[Howe(1995)]{howe_fpm}
M.~Howe.
\newblock On the force and moment on a body in an incompressible fluid, with application to rigid bodies and bubbles at high and low reynolds numbers.
\newblock \emph{The Quarterly Journal of Mechanics and Applied Mathematics}, 48\penalty0 (3):\penalty0 401--426, 1995.

\bibitem[Howe(2002)]{Howe_book}
M.~S. Howe.
\newblock \emph{Theory of Vortex Sound}.
\newblock Cambridge Texts in Applied Mathematics. Cambridge University Press, 2002.

\bibitem[Hunt et~al.(1988)Hunt, Wary, and Moin]{Qcrit}
J.~C.~R. Hunt, A.~A. Wary, and P.~Moin.
\newblock Eddies, streams and convergence zones in turbulent flows.
\newblock \emph{Proc. Summer Program Center Turbulence Research, NASA Ames}, pages 193--207, 1988.

\bibitem[Hussain and Reynolds(1970{\natexlab{a}})]{hussain_mechanics_1970}
A.~K. M.~F. Hussain and W.~C. Reynolds.
\newblock The mechanics of an organized wave in turbulent shear flow.
\newblock \emph{Journal of Fluid Mechanics}, 41\penalty0 (2):\penalty0 241--258, Apr. 1970{\natexlab{a}}.
\newblock ISSN 0022-1120, 1469-7645.
\newblock \doi{10.1017/S0022112070000605}.

\bibitem[Hussain and Reynolds(1970{\natexlab{b}})]{tripDecompSource}
A.~K. M.~F. Hussain and W.~C. Reynolds.
\newblock The mechanics of an organized wave in turbulent shear flow.
\newblock \emph{Journal of Fluid Mechanics}, 41\penalty0 (2):\penalty0 241--258, 1970{\natexlab{b}}.

\bibitem[Ianniello et~al.(2013)Ianniello, Muscari, and Di~Mascio]{ship_acoustic}
S.~Ianniello, R.~Muscari, and A.~Di~Mascio.
\newblock Ship underwater noise assessment by the acoustic analogy. part i: nonlinear analysis of a marine propeller in a uniform flow.
\newblock \emph{Journal of marine Science and technology}, 18:\penalty0 547--570, 2013.

\bibitem[Laksham(2019)]{dronerescue}
K.~B. Laksham.
\newblock Unmanned aerial vehicle (drones) in public health: A swot analysis.
\newblock \emph{Journal of family medicine and primary care}, 8\penalty0 (2):\penalty0 342, 2019.

\bibitem[Lamptey and Serwaa(2020)]{ziplinecovid}
E.~Lamptey and D.~Serwaa.
\newblock The use of zipline drones technology for covid-19 samples transportation in ghana.
\newblock \emph{HighTech and Innovation Journal}, 1\penalty0 (2):\penalty0 67--71, 2020.

\bibitem[Lumley(1967)]{lumley1967structure}
J.~L. Lumley.
\newblock The structure of inhomogeneous turbulent flows.
\newblock \emph{Atmospheric turbulence and radio wave propagation}, pages 166--178, 1967.

\bibitem[Menon and Mittal(2021{\natexlab{a}})]{menon2021a}
K.~Menon and R.~Mittal.
\newblock Significance of the strain-dominated region around a vortex on induced aerodynamic loads.
\newblock \emph{Journal of Fluid Mechanics}, 918:\penalty0 R3, 2021{\natexlab{a}}.
\newblock \doi{10.1017/jfm.2021.359}.

\bibitem[Menon and Mittal(2021{\natexlab{b}})]{menon2021b}
K.~Menon and R.~Mittal.
\newblock On the initiation and sustenance of flow-induced vibration of cylinders: insights from force partitioning.
\newblock \emph{Journal of Fluid Mechanics}, 907:\penalty0 A37, 2021{\natexlab{b}}.
\newblock \doi{10.1017/jfm.2020.854}.

\bibitem[Menon and Mittal(2021{\natexlab{c}})]{menon2021c}
K.~Menon and R.~Mittal.
\newblock Quantitative analysis of the kinematics and induced aerodynamic loading of individual vortices in vortex-dominated flows: A computation and data-driven approach.
\newblock \emph{Journal of Computational Physics}, 443:\penalty0 110515, 2021{\natexlab{c}}.
\newblock ISSN 0021-9991.
\newblock \doi{https://doi.org/10.1016/j.jcp.2021.110515}.

\bibitem[Menon et~al.(2022)Menon, Kumar, and Mittal]{menon2022}
K.~Menon, S.~Kumar, and R.~Mittal.
\newblock Contribution of spanwise and cross-span vortices to the lift generation of low-aspect-ratio wings: Insights from force partitioning.
\newblock \emph{Phys. Rev. Fluids}, 7:\penalty0 114102, Nov 2022.
\newblock \doi{10.1103/PhysRevFluids.7.114102}.

\bibitem[Misiorowski et~al.(2019)Misiorowski, Gandhi, and Oberai]{droneflowstruct}
M.~Misiorowski, F.~Gandhi, and A.~A. Oberai.
\newblock Computational study on rotor interactional effects for a quadcopter in edgewise flight.
\newblock \emph{AIAA Journal}, 57\penalty0 (12):\penalty0 5309--5319, 2019.
\newblock \doi{10.2514/1.J058369}.

\bibitem[Mittal and Balachandar(1995)]{mittal1995effect}
R.~Mittal and S.~Balachandar.
\newblock Effect of three-dimensionality on the lift and drag of nominally two-dimensional cylinders.
\newblock \emph{Physics of Fluids}, 7\penalty0 (8):\penalty0 1841--1865, 1995.

\bibitem[Mittal and Seo(2023)]{mittal_origin_2023}
R.~Mittal and J.~H. Seo.
\newblock Origin and evolution of immersed boundary methods in computational fluid dynamics.
\newblock \emph{Physical Review Fluids}, 8\penalty0 (10):\penalty0 100501, Oct. 2023.
\newblock ISSN 2469-990X.
\newblock \doi{10.1103/PhysRevFluids.8.100501}.

\bibitem[Mittal et~al.(2003)Mittal, Simmons, and Najjar]{triple_ref}
R.~Mittal, S.~P. Simmons, and F.~Najjar.
\newblock Numerical study of pulsatile flow in a constricted channel.
\newblock \emph{Journal of Fluid Mechanics}, 485:\penalty0 337–378, 2003.
\newblock \doi{10.1017/S002211200300449X}.

\bibitem[Mittal et~al.(2008)Mittal, Dong, Bozkurttas, Najjar, Vargas, and {von Loebbecke}]{mittal2008}
R.~Mittal, H.~Dong, M.~Bozkurttas, F.~Najjar, A.~Vargas, and A.~{von Loebbecke}.
\newblock A versatile sharp interface immersed boundary method for incompressible flows with complex boundaries.
\newblock \emph{Journal of Computational Physics}, 227\penalty0 (10):\penalty0 4825--4852, 2008.
\newblock ISSN 0021-9991.
\newblock \doi{https://doi.org/10.1016/j.jcp.2008.01.028}.

\bibitem[Mittal et~al.(2021)Mittal, Seo, and Raghav]{apsrotor}
R.~Mittal, J.-H. Seo, and V.~Raghav.
\newblock Evolution of the leading-edge vortex on a revolving wing and its effect on aerodynamic loading.
\newblock \emph{APS Division of Fluid Dynamics Meeting Abstracts}, 2021.

\bibitem[Mittal et~al.(2025)Mittal, Seo, Turner, Kumar, Prakhar, and Zhou]{new_ibm}
R.~Mittal, J.-H. Seo, J.~Turner, S.~Kumar, S.~Prakhar, and J.~Zhou.
\newblock Freeman scholar lecture (2021)—sharp-interface immersed boundary methods in fluid dynamics.
\newblock \emph{Journal of Fluids Engineering}, 147\penalty0 (3):\penalty0 030801, 01 2025.
\newblock ISSN 0098-2202.
\newblock \doi{10.1115/1.4067385}.

\bibitem[Nedunchezian et~al.(2019)Nedunchezian, kwon Kang, and Aono]{zoo_acoustic}
K.~Nedunchezian, C.~kwon Kang, and H.~Aono.
\newblock Effects of flapping wing kinematics on the aeroacoustics of hovering flight.
\newblock \emph{Journal of Sound and Vibration}, 442:\penalty0 366--383, 2019.
\newblock ISSN 0022-460X.
\newblock \doi{https://doi.org/10.1016/j.jsv.2018.11.014}.

\bibitem[Nekkanti and Schmidt(2021)]{schmidt_spod}
A.~Nekkanti and O.~T. Schmidt.
\newblock Frequency--time analysis, low-rank reconstruction and denoising of turbulent flows using spod.
\newblock \emph{Journal of Fluid Mechanics}, 926:\penalty0 A26, 2021.

\bibitem[Quartapelle and Napolitano(1983)]{quartapelle1983force}
L.~Quartapelle and M.~Napolitano.
\newblock Force and moment in incompressible flows.
\newblock \emph{AIAA journal}, 21\penalty0 (6):\penalty0 911--913, 1983.

\bibitem[Reynolds(1895)]{Reynolds_1895}
O.~Reynolds.
\newblock {On} the dynamical theory of incompressible viscous fluids and the determination of the criterion.
\newblock \emph{Philosophical Transactions of the Royal Society of London. (A.)}, 186:\penalty0 123--164, Dec. 1895.
\newblock ISSN 0264-3820, 2053-9231.
\newblock \doi{10.1098/rsta.1895.0004}.

\bibitem[Rochuon et~al.(2006)Rochuon, Trébinjac, and Billonnet]{rochuon_pod}
N.~Rochuon, I.~Trébinjac, and G.~Billonnet.
\newblock An extraction of the dominant rotor-stator interaction modes by the use of {Proper} {Orthogonal} {Decomposition} ({POD}).
\newblock \emph{Journal of Thermal Science}, 15\penalty0 (2):\penalty0 109--114, June 2006.
\newblock ISSN 1003-2169, 1993-033X.
\newblock \doi{10.1007/s11630-006-0109-4}.

\bibitem[Schmid(2010)]{schmid_dmd}
P.~J. Schmid.
\newblock Dynamic mode decomposition of numerical and experimental data.
\newblock \emph{Journal of fluid mechanics}, 656:\penalty0 5--28, 2010.

\bibitem[Seo and Mittal(2011)]{seo2011}
J.~H. Seo and R.~Mittal.
\newblock A sharp-interface immersed boundary method with improved mass conservation and reduced spurious pressure oscillations.
\newblock \emph{Journal of Computational Physics}, 230\penalty0 (19):\penalty0 7347--7363, 2011.
\newblock ISSN 0021-9991.
\newblock \doi{https://doi.org/10.1016/j.jcp.2011.06.003}.

\bibitem[Seo et~al.(2022)Seo, Menon, and Mittal]{Seo2022}
J.-H. Seo, K.~Menon, and R.~Mittal.
\newblock {A method for partitioning the sources of aerodynamic loading noise in vortex dominated flows}.
\newblock \emph{Physics of Fluids}, 34\penalty0 (5), 05 2022.
\newblock ISSN 1070-6631.
\newblock \doi{10.1063/5.0094697}.
\newblock 053607.

\bibitem[Seo et~al.(2023)Seo, Zhang, Mittal, and Cattafesta]{Seo2023}
J.-H. Seo, Y.~Zhang, R.~Mittal, and L.~N. Cattafesta.
\newblock {Vortex-induced sound prediction of slat noise from time-resolved particle image velocimetry data}.
\newblock \emph{Experiments in Fluids}, 64\penalty0 (99), 05 2023.
\newblock \doi{10.1007/s00348-023-03636-5}.

\bibitem[Sirovich(1987)]{sirovich}
L.~Sirovich.
\newblock Turbulence and the dynamics of coherent structures. i. coherent structures.
\newblock \emph{Quarterly of applied mathematics}, 45\penalty0 (3):\penalty0 561--571, 1987.

\bibitem[Solera-Rico et~al.(2024)Solera-Rico, Sanmiguel~Vila, G{\'o}mez-L{\'o}pez, Wang, Almashjary, Dawson, and Vinuesa]{vinuesa_ae}
A.~Solera-Rico, C.~Sanmiguel~Vila, M.~G{\'o}mez-L{\'o}pez, Y.~Wang, A.~Almashjary, S.~T. Dawson, and R.~Vinuesa.
\newblock $\beta$-variational autoencoders and transformers for reduced-order modelling of fluid flows.
\newblock \emph{Nature Communications}, 15\penalty0 (1):\penalty0 1361, 2024.

\bibitem[Taira et~al.(2017)Taira, Brunton, Dawson, Rowley, Colonius, McKeon, Schmidt, Gordeyev, Theofilis, and Ukeiley]{st_modal}
K.~Taira, S.~L. Brunton, S.~T.~M. Dawson, C.~W. Rowley, T.~Colonius, B.~J. McKeon, O.~T. Schmidt, S.~Gordeyev, V.~Theofilis, and L.~S. Ukeiley.
\newblock Modal {Analysis} of {Fluid} {Flows}: {An} {Overview}.
\newblock \emph{AIAA Journal}, 55\penalty0 (12):\penalty0 4013--4041, Dec. 2017.
\newblock ISSN 0001-1452, 1533-385X.
\newblock \doi{10.2514/1.J056060}.

\bibitem[Van~Kan(1986)]{vankan}
J.~Van~Kan.
\newblock A second-order accurate pressure-correction scheme for viscous incompressible flow.
\newblock \emph{SIAM journal on scientific and statistical computing}, 7\penalty0 (3):\penalty0 870--891, 1986.

\bibitem[Wang et~al.(2016)Wang, McBee, and Iliescu]{parPODformula}
Z.~Wang, B.~McBee, and T.~Iliescu.
\newblock Approximate partitioned method of snapshots for pod.
\newblock \emph{Journal of Computational and Applied Mathematics}, 307:\penalty0 374--384, 2016.
\newblock ISSN 0377-0427.
\newblock \doi{https://doi.org/10.1016/j.cam.2015.11.023}.
\newblock 1st Annual Meeting of SIAM Central States Section, April 11–12, 2015.

\bibitem[Weiss()]{PodTutorial}
J.~Weiss.
\newblock A tutorial on the proper orthogonal decomposition.
\newblock \doi{10.2514/6.2019-3333}.

\bibitem[Wu(1981)]{wu_fpm}
J.~C. Wu.
\newblock Theory for aerodynamic force and moment in viscous flows.
\newblock \emph{AIAA Journal}, 19\penalty0 (4):\penalty0 432--441, 1981.

\bibitem[Zhang et~al.(2015)Zhang, Hedrick, and Mittal]{zhang2015centripetal}
C.~Zhang, T.~L. Hedrick, and R.~Mittal.
\newblock Centripetal acceleration reaction: an effective and robust mechanism for flapping flight in insects.
\newblock \emph{PloS one}, 10\penalty0 (8):\penalty0 e0132093, 2015.

\bibitem[Zhao et~al.(2019)Zhao, Zhao, Liu, and Du]{zhao_pod}
M.~Zhao, Y.~Zhao, Z.~Liu, and J.~Du.
\newblock Proper {Orthogonal} {Decomposition} {Analysis} of {Flow} {Characteristics} of an {Airfoil} with {Leading} {Edge} {Protuberances}.
\newblock \emph{AIAA Journal}, 57\penalty0 (7):\penalty0 2710--2721, July 2019.
\newblock ISSN 0001-1452, 1533-385X.
\newblock \doi{10.2514/1.J058010}.

\bibitem[Zhu et~al.(2023)Zhu, Lee, Kumar, Menon, Mittal, and Breuer]{zhu2023force}
Y.~Zhu, H.~Lee, S.~Kumar, K.~Menon, R.~Mittal, and K.~Breuer.
\newblock Force moment partitioning and scaling analysis of vortices shed by a 2d pitching wing in quiescent fluid.
\newblock \emph{Experiments in Fluids}, 64\penalty0 (10):\penalty0 158, 2023.

\bibitem[Zorumski and Weir(1986)]{zorumski_compact_noise}
W.~E. Zorumski and D.~S. Weir.
\newblock Aircraft noise prediction program theoretical manual: Propeller aerodynamics and noise.
\newblock \emph{Technical Memorandum No. NASA-TM-83199-PT-1 (NASA Langley Research Center, Hampton, VA)}, 1986.

\end{thebibliography}
\end{document}